\documentclass[reprint,twocolumn,superscriptaddress,groupedaddress,aps]{revtex4-2} 
\usepackage{graphicx} 
\usepackage{dcolumn}  
\usepackage{bm}     
\usepackage[pdftex,breaklinks,colorlinks=true,citecolor=blue,linkcolor=blue,urlcolor=blue]{hyperref}
\usepackage{amssymb} 
\usepackage{amsmath}
\usepackage{subfigure}
\usepackage{bbold}
\usepackage{blindtext}
\usepackage[utf8]{inputenc}
\usepackage{float}
\usepackage{color}
\usepackage{mwe}
\usepackage{multirow}
\usepackage{ulem} 
\usepackage{algorithm}
\usepackage{algpseudocode}
\usepackage[dvipsnames]{xcolor}
\usepackage{natbib}

\newcommand{\CUAaddress}{Harvard-MIT Center for Ultracold Atoms, Cambridge, Massachusetts 02138, USA}
\newcommand{\HarvardPhysicsAddress}{Department of Physics, Harvard University, Cambridge, Massachusetts 02138, USA}
\newcommand{\HarvardChemistryaddress}{Department of Chemistry and Chemical Biology, Harvard University, Cambridge, Massachusetts 02138, USA}

\newcommand*{\TT}{\textcolor{Blue}} 

\newcommand{\threejm}[6]{ \left(\begin{array}{ccc} #1 & #3 & #5\\
                                              #2 & #4 & #6
                                \end{array}
                          \right)}
\newcommand{\sixj}[6]{ \left\{\begin{array}{ccc} #1 & #3 & #5\\
                                              #2 & #4 & #6
                                \end{array}
                          \right\}}

\begin{document}

\title{Hyperfine-to-rotational energy transfer in ultracold atom-molecule collisions}

\author{Yi-Xiang Liu}
\thanks{These authors contributed equally to this work.}

\affiliation{\HarvardChemistryaddress} 
\affiliation{\HarvardPhysicsAddress}
\affiliation{\CUAaddress}

\author{Lingbang Zhu}
\thanks{These authors contributed equally to this work.}
\affiliation{\HarvardChemistryaddress}
\affiliation{\HarvardPhysicsAddress}
\affiliation{\CUAaddress}

\author{Jeshurun~Luke}
\affiliation{\HarvardChemistryaddress}
\affiliation{\HarvardPhysicsAddress}
\affiliation{\CUAaddress}

\author{Mark~C.~Babin}
\affiliation{\HarvardChemistryaddress}
\affiliation{\HarvardPhysicsAddress}
\affiliation{\CUAaddress}

\author{Timur~V.~Tscherbul}
\affiliation{Department of Physics, University of Nevada, Reno, Nevada, 89557, USA}

\author{Marcin Gronowski}
\affiliation{Faculty of Physics, University of Warsaw, Pasteura 5, 02-093 Warsaw, Poland}
\author{Hela Ladjimi}
\affiliation{Faculty of Physics, University of Warsaw, Pasteura 5, 02-093 Warsaw, Poland}
\author{Micha{\l} Tomza}
\affiliation{Faculty of Physics, University of Warsaw, Pasteura 5, 02-093 Warsaw, Poland}

\author{John L. Bohn}
\affiliation{JILA, NIST, and Department of Physics, University of Colorado, Boulder, Colorado 80309-0440, USA
}

\author{Kang-Kuen Ni}
\email{ni@chemistry.harvard.edu}
\affiliation{\HarvardChemistryaddress}
\affiliation{\HarvardPhysicsAddress}
\affiliation{\CUAaddress}

\date{\today}

\begin{abstract}
Energy transfer between different mechanical degrees of freedom in atom-molecule collisions has been widely studied and largely understood. However, systems involving spins remain less explored, especially with a state-to-state precision.
Here, we directly observed the energy transfer from atomic hyperfine to molecular rotation in the 
$^{87}$Rb ($|F_a,M_{F_a}\rangle = |2,2\rangle$) + $^{40}$K$^{87}$Rb (in the rovibronic ground state $N=0$) $\longrightarrow$ Rb ($ |1,1\rangle$) + KRb ($N=0,1,2$)  
exothermic collision. We probed the quantum states of the collision products using resonance-enhanced multi-photon ionization followed by time-of-flight mass spectrometry. 
We also carried out state-of-the-art quantum scattering calculations, which rigorously take into account the coupling between the spin and rotational degrees of freedom at short range, and assume that the KRb monomer can be treated as a rigid rotor moving on a single potential energy surface.
The calculated product rotational state distribution deviates from the observations even after extensive tuning of the atom-molecule potential energy surface, suggesting that vibrational degrees of freedom and conical intersections play an important part in ultracold Rb~+~KRb collisions. Additionally, our {\it ab initio} calculations indicate that spin-rotation  coupling is dramatically enhanced near a conical intersection, which is energetically accessible at short range.
The observations confirm that spin is coupled to mechanical rotation at short range and establish a benchmark for future theoretical studies.
\end{abstract}

\maketitle

\section{Introduction} 
Advances in joint experimental and theoretical tools have allowed scattering experiments to probe molecular interactions during collisions, revealing reaction mechanisms, dynamics, and the structure of collisional complexes~\cite{Herschbach-Nobel,lee1987molecular,Neumark2018,zhao2017state}. 
After decades of work, sophisticated and accurate quantum dynamical calculations on atom-molecule collisions are now feasible \cite{althorpe2003quantum},
 especially for collisions and reactions involving only vibrational, rotational, and electronic states~\cite{rebentrost1978,yuan2018observation,klein2017directly,dong2010transition,DeJonghScience2020}.

Adding spin degrees of freedom  complicates the interactions and poses challenges to experiments and a full understanding of the dynamics. 
Yet, collisions involving spins are of interest to many areas, such as buffer gas cooling~\cite{Maussang2005}, sympathetic cooling~\cite{Lara:06,Morita:18,son2020collisional,Jurgilas2021}, spin-exchange optical pumping~\cite{Walker:97,tscherbul2011anisotropic}, Feshbach resonances~\cite{wang2021magnetic,park2023spectrum,bird2023tunable,park2023feshbach}, 
and the magneto-association of triatomic molecules~\cite{yang2022creation}.   
{Another notable example is the spin-orbit interaction, which plays a significant role in the paradigmatic chemical reaction F~+~H$_2$ $\to$ HF + H, the only known source of HF in the interstellar medium \cite{Tizniti:14,Yang:19}.}

Besides their fundamental significance, the spin-dependent  interactions could play an essential role in the formation of long-lived  complexes in ultracold atom-molecule and molecule-molecule collisions  \cite{mayle2012statistical,mayle2013scattering,Christianen2019}. 
The complexes are commonly described using the statistical Rice-Ramsperger-Kassel-Marcus (RRKM) theory, which gives the complex lifetime as $\tau= 2\pi\hbar \rho/N_o$, where $\rho$ is the density of states, and $N_o$ is the number of open channels \cite{mayle2012statistical,mayle2013scattering}. While the lifetimes predicted by the RRKM theory agree well with some experiments such as KRb + KRb~\cite{liu2020photo} and RbCs + RbCs~\cite{gregory2020loss}, this is not the case for other complexes~\cite{Gersema2021,Bause2021collisions}. 
A striking example is the Rb-KRb collision complex formed by the Rb atom in the ground hyperfine state $|F_a,M_{F_a} \rangle = |1,1 \rangle$  and KRb in its rovibronic ground state, whose experimentally measured complex lifetime is five orders of magnitude longer than RRKM estimates treating spins as ``spectators"~\cite{nichols2022detection}.
At present, this discrepancy remains unsolved~\cite{bause2023ultracold}. 
Recently, {\it ab initio} evidence for  strong spin-dependent {hyperfine} interactions has been reported for the Rb-KRb complex \cite{jachymski2022collisional}. Because sufficiently strong spin-dependent  interactions can couple electron and nuclear spins to molecular rotations at short range, spins can  dramatically increase the density of states of ultracold collision complexes and hence their RRKM lifetimes~\cite{mayle2012statistical,Christianen2019}, potentially explaining the large discrepancy.

These crucial spin-dependent hyperfine interactions have never been experimentally observed directly or rigorously explored in quantum scattering calculations. 
On the experiment side, detecting these interactions would require state-to-state measurements of collisional processes specifically mediated by spin-dependent interactions.  
On the theory side, enormous rotational and spin basis sets have prevented rigorous atom-molecule quantum scattering calculations~\cite{Tscherbul:23}.

\begin{figure}
\centering
\includegraphics[width=0.48\textwidth]{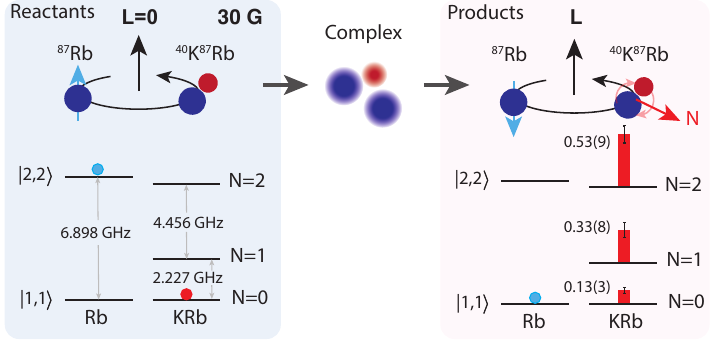}
\caption{\textbf{Schematic of hyperfine-to-rotational energy transfer in $^{87}$Rb + $^{40}$K$^{87}$Rb collisions.} Left panel: Rb atoms were prepared in the $|F_a=2,M_{F_a}=2\rangle$ excited hyperfine state of the 5S$_{1/2}$ manifold, while KRb molecules were prepared in the ground rotational state $|N =0\rangle$ of their vibronic ground state, X$^1\Sigma^+, \nu = 0$. $L$ is the collision partial wave between the atom and molecule. Right panel: Following collision, Rb atoms relaxed to the $F_{a} = 1$ manifold, predominantly to the $|1,1\rangle$ state, while molecules populated multiple rotational states up to $N = 2$. Red bars represent the experimentally measured branching ratio. This process occurred due to the coupling between atomic spin and molecular rotation in the KRb$_2$ collision complex. This collision was studied under an external magnetic field B = 30 G and an electric field E = 17 V/cm, where B and E were  perpendicular to one another.}
\label{fig:scheme}
\end{figure}

\begin{figure*}
\centering
\includegraphics[width=\textwidth]{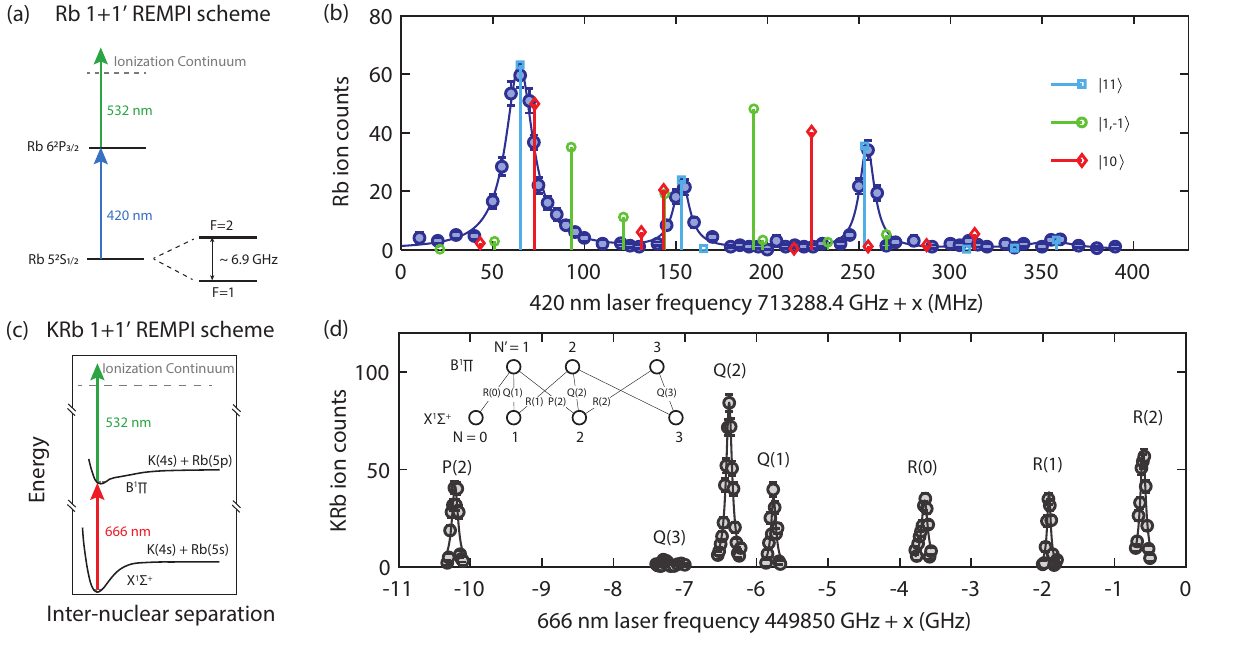}
\caption{\textbf{REMPI detection of collision products.} (a) Rb REMPI detection scheme. State selective ionization of Rb was performed using a 1 + 1' REMPI scheme, with the 420 nm laser pulse resonant with the 5S$_{1/2}$, F = 1 to 6P$_{3/2}$ transition, followed by the 532 nm pulsed laser (10 ns) ionizing atoms in the 6P state.
(b) Rb REMPI spectroscopy on the $|2,2\rangle $ + KRb collision products. Solid blue circles are data points. Solid dark blue line represents a fit to Lorentzian line shapes. We extract the 420 nm laser polarization composition through the relative heights of the Lorentzian peaks. Light blue (open squares) vertical lines are predicted line positions for Rb atoms in the $|1,1\rangle$ state in 5S$_{1/2}$ targeting different hyperfine states in 6P$_{3/2}$. Similarly, red (open diamonds) and green (open circles) vertical lines are predicted line positions for $|1,0\rangle$ and $|1,-1\rangle$ in 5S$_{1/2}$, respectively.  The vertical lines were simulated using the fitted polarization and assuming equal population among the various $M_{F_a}$ levels, scaled such that $|1,1\rangle$ population matches data. 
(c) KRb REMPI detection scheme. Rotational-state selective ionization of KRb was performed using a 1 + 1' REMPI scheme, where the 666 nm laser pulse was resonant with one of the X$^1\Sigma^+$, $\nu=0$, $N$ = 0, 1, and 2 to B$^1\Pi,\nu'=0$ rovibronic transitions, followed by a 532 nm pulsed laser directly ionizing KRb in the B$^1\Pi$ state. (d) REMPI spectra of KRb collision products. The 666 nm laser frequency was scanned over 11 GHz, revealing a number of features corresponding to P ($\Delta N = N'-N = -1$), Q ($\Delta N = 0$), and R ($\Delta N = 1$) transitions of the $N$ = 0, 1, and 2 levels of X$^1\Sigma^+$ KRb (circles). The solid lines are Lorentzian fits. Notably, no molecules were observed in the $N$ = 3 level. (Inset) Herzberg diagram of relevant transitions between the KRb ground and excited states~\cite{Herzberg}. 
}
\label{fig:spectroscopy}
\end{figure*}

Here, we directly probed the spin-dependent interactions in Rb + KRb collisions by activating the spin degree of freedom with the Rb atom prepared in an excited hyperfine state $|F_a,M_{F_a} \rangle = |2,2\rangle$ [see Fig.~\ref{fig:scheme}]. We also calculated the underlying potential energy surfaces and
developed the formalism for rigorous quantum scattering calculations to model these collisions. 
As the KRb + Rb $\rightarrow$ K + Rb$_2$ reaction is endothermic~\cite{nichols2022detection}, only inelastic collision outcomes are allowed at ultracold temperatures. 
We directly measure the final states of the atoms and the molecules, finding the excitation of molecules to all energetically allowed rotational levels commensurate with a relaxation of the atoms to their ground hyperfine states.  This observation confirms the coupling between spin and mechanical rotation in the KRb$_2$ complex.

\section{Results \& Discussion}
The experiment started with a cloud of $\sim10^4$ $^{40}$K$^{87}$Rb and $\sim 10^4$ of $^{87}$Rb atoms co-trapped in an optical dipole trap (ODT) at 1064 nm.
The molecules were prepared in the rovibronic ground state X$^1\Sigma^+$ $|N=0, M_N=0, M_{I,\text{K}}=-4, M_{I,\text{Rb}}=1/2\rangle$, where $N$ is the rotational quantum number, $M_N$ is the rotation projection onto the magnetic field axis, $M_{I,\text{K}}$ and $M_{I,\text{Rb}}$ are the nuclear spin projection of K ($I_\mathrm{K}=4$) and Rb ($I_\mathrm{Rb}=3/2$) on the magnetic field axis, at 0.4 $\mu$K following the procedure described previously~\cite{ni2008high,hu2019direct}. The Rb atoms were residuals from the Feshbach association process and were in the $|F_a,M_{F_a}\rangle$ = $|1,1\rangle$ state at 0.8 $\mu$K. 
Following molecule production, the magnetic field was ramped down from 543.5 to 30 G in 20 ms, after which the Rb atoms were transferred from the $|1,1\rangle$ to the $|2,2\rangle$ state via an adiabatic rapid passage (ARP) microwave pulse. At this point, the atom-molecule collisions of interest began. The excess energy in this inelastic collision is carried away as kinetic energy of the products, allowing for their escape from the ODT.
The products were then probed via a state-selective 1+1' resonance enhanced multi-photon ionization (REMPI) scheme [see Appendix].  In the experiment we protected the reactant cloud from REMPI beams using a dark mask [see Appendix], and only probed the escaping products of this collision.

\subsection{REMPI detection of collision products} 

To detect the hyperfine states of the product Rb atoms, we applied a 420 nm laser resonant to 5S$_{1/2}, F_a=1$ to 6P$_{3/2}$ transition and then photoionized the 6P$_{3/2}$ Rb atoms using a 532 nm pulsed laser [see Fig.~\ref{fig:spectroscopy}(a)]. The resulting Rb REMPI spectrum is shown in Fig. \ref{fig:spectroscopy}(b) along with theoretical stick spectra for transitions resulting from various $M_{F_a}$ levels within $F_a = 1$. The height of the theoretical lines were calculated assuming equal population in each $M_{F_a}$ states. This spectrum indicates that the atoms predominately ($>95\%$) populate the $|1,1\rangle$ state following collisions. We similarly performed F = 2 to 6P$_{3/2}$ spectroscopy and found that the population of $|2,1\rangle$ is less than 5\% of the total Rb population {and negligible population in other  $M_{F_a}$ levels}. 

Similarly, we probed the rotational state of KRb molecules after collisions with a REMPI scheme shown in Fig.~\ref{fig:spectroscopy}(c). By scanning the 666 nm laser frequency, we identified KRb molecules in the $N$ = 0, 1, and 2 states following a collision, as shown in Fig. \ref{fig:spectroscopy}(d). We found no evidence for KRb excited to $N$ = 3 level, suggesting these rotationally excited KRb species originated from a single collisional event, as a secondary collision with $|2,2\rangle$ Rb atom would have sufficient energy required to reach the $N$ = 3 level ($\sim6.683$ GHz from $N$ = 2).

\begin{figure}
\centering
	\includegraphics[width=0.45\textwidth]{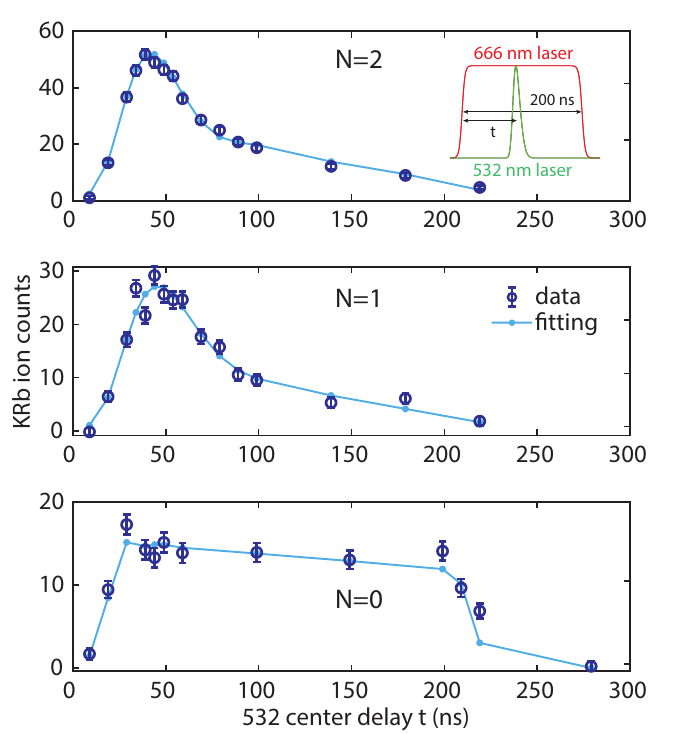}
  \caption{\textbf{Extracting KRb branching ratio.}  The relative time delay between the 666 nm and 532 nm REMPI pulses was scanned. The 666 nm laser was resonant with the Q(2) (top panel), Q(1) (middle) and R(0) (bottom panel) transitions respectively. The pulse length of the 666 nm laser was fixed at 200 ns. The inset shows powers of the REMPI beams as a function of time. The open circles are data. Error bars represent shot noise. The solid dots connected by lines are fits to the Monte Carlo simulation. From the line shape we can extract the Rabi rate of the three transitions and the population of molecules in $N$ = 2, 1, and 0 states post collision.}
\label{fig:calibration}
\end{figure}

To further obtain a quantitative population distribution of the KRb rotational states, we scanned the 532 nm pulsed laser delay $t$ with respect to the beginning of the 666 nm excitation pulse [see Fig. \ref{fig:calibration}].
With a pulse delay time $t$, the ion counts is proportional to the population at $B^1\Pi$ after the first transition of the REMPI [see Fig.~\ref{fig:spectroscopy}(c)], which mainly depends on the molecule population in $N$, the 666 nm Rabi rate for each transition, and the excited state lifetime.
For $N$ = 2 and 1, we observed a damped oscillation. 
The lineshape for the $N$ = 0 scan showed little dependence on the delay $t$ because the pulsed 532 nm beam induced a light shift on R(0) transition [see Appendix] and the 666 nm pulse was resonant only for a fixed duration when the 532 nm beam was on. 
 
Because the exoergicity of this inelastic collision is small, product molecules are relatively slow-moving and can experience multiple ionization pulses before they leave the detection region, while the REMPI efficiency for each ionization event varies as the ionization beam intensity decays following a Gaussian profile. 
To account for all these factors, we performed a Monte Carlo simulation of the multi-pulse ionization process and fitted to the data in Fig.~\ref{fig:calibration} to extract the branching ratio [see Appendix for details].
 
From the fitting, we found the number of molecules in each rotational state and obtained the branching ratio of $N = 2, 1$ and $0$ to be $0.53(9):0.33(8):0.13(3)$. The error bars were calculated by propagating a 68\% confidence interval from the fitted molecule numbers and assuming a 10\% overall fluctuation in the initial atom and molecule numbers. 
We also studied the $|2,2\rangle$ Rb colliding with KRb in $N=1$ and found that the product KRb rotational states population followed a ratio $N = 2,1,0$ to be $0.59(13):0.32(9):0.09(2)$ [see Appendix Table~\ref{table:ratio}]. This indicates that the scattering process is relatively insensitive to the details of the interaction potential.

A pure chemical exchange channel, wherein the incoming Rb atom swaps with the Rb atom in the KRb molecule, could qualitatively explain our observations. In the $|0,0,-4,1/2\rangle $ KRb + $|2,2\rangle$ Rb collision, the nuclear spin states of the two Rb atoms are different, allowing the Rb atom to end up in the $|1,1\rangle$ state post-exchange.
To test whether chemical exchange alone could explain our results, we also studied KRb in $|0,0,-4,3/2\rangle$ and Rb in $|2,2\rangle$, where both Rb atoms have the same nuclear spin states $M_{I,\mathrm{Rb}}=3/2$. We observed KRb in $N$ = 2 states post collision, indicating the coupling between atomic spin and mechanical rotation in the KRb$_2$ system, as a pure chemical exchange process would effectively result in an elastic collision.

\subsection{Statistical aspects of Rb+KRb scattering}
One way to interpret the observed Rb spin flip is to use a statistical theory of collision complexes and a time-dependent perturbation theory applied to the spins~\cite{mayle2012statistical}.  We modelled the complex as a Rb atom repeatedly bouncing off the KRb molecule at a time interval estimated by averaging over classical trajectories in a pure $C_6$ potential. Given a typical GHz spin-dependent interaction strength~\cite{jachymski2022collisional} and a 0.54(10)~ns measured complex lifetime [see Supplementary Materials (SM)], a single Rb spin flip probability is found to be near unity while a second spin flip to $|1,0\rangle$ is unlikely [see Appendix], matching the observation. 

The fact that the relative populations of the rotational states $N$ are in proportion to their degeneracies $2N+1$ invites a statistical interpretation of the branching ratios before diving into rigorous quantum scattering calculations. We calculated the expected KRb branching ratio assuming that the collision process is ergodic and all states of the collision complex, as well as all exit channels, are populated with equal probability.  In total, there are six angular momenta in the system, 
$\hat{\bf N}$  = molecular rotation, $\hat{\bf L}$  = collision partial wave, $\hat{\bf I}_a$  = Rb atom nuclear spin, $\hat{\bf S}$ = Rb electronic spin, $\hat{\bf I}_{1}$ = K molecular nuclear spin, and $\hat{\bf I}_2 $ = Rb molecular nuclear spin. As our detection can only resolve the atom spins and molecular rotation, the molecular nuclear spins and partial wave go in the degeneracies of $N = 0,1,2$ product channels. When the external magnetic field is weak (our measurement at 2 G yields the same ratio as at 30 G [see Appendix Table~\ref{table:ratio}]), the total angular momentum $J$ is conserved. The total angular momentum $\hat{\bf J}$ is the vector sum of the angular momentum for total mechanical rotation $\hat{\bf J}_{r} = \hat{\bf N}  + \hat{\bf L}$ and the spin angular momentum $\hat{\bf F} = \hat{\bf S} + \hat{\bf I}_a + \hat{\bf I}_1 + \hat{\bf I}_2$.  

If we assume that all the degrees of freedom are scrambled, the degeneracies of $N=2,1,0$ channels have a ratio of $0.544:0.340:0.116$ following the statistical model [see Appendix], in agreement with the measured ratio. 

We acknowledge that spin-dependent interaction terms in the Hamiltonian are weak, in which case $J_r$ would be conserved in the collision.  Assuming conservation of   $J_r=0$ throughout the process, the statistical model predicts the ratio of $N=2:1:0$ to be $1:1:1$, which deviates from the experimental finding.
  Thus the statistical hypothesis can explain the observed distribution of rotational levels, provided $J_r$ is able to change freely during the collision.

\begin{figure}
\centering
\includegraphics[width=\columnwidth]{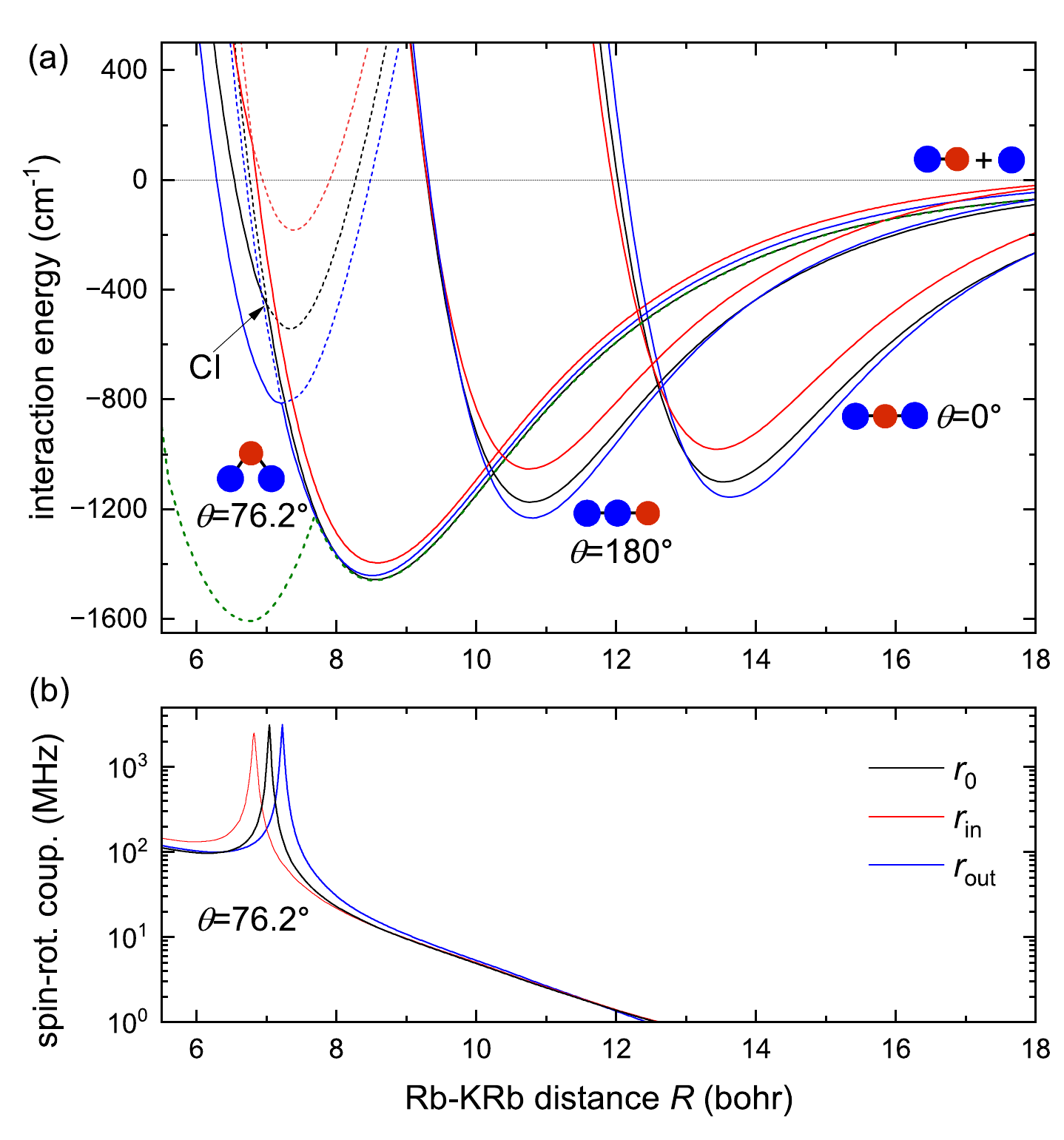}
\caption{\textbf{Potential energy surfaces and spin-rotation coupling of the Rb+KRb system.} (a) One-dimensional cuts through the three-dimensional electronic interaction potentials $V(R,r,\theta)$ for the ground (solid lines) and the first excited (dashed lines) electronic states. Three orientations $\theta$ are presented: two linear configurations ($\theta=0^{\circ}$ and $180^{\circ}$) and one bend arrangement for which the conical intersection (indicated as CI) exists ($\theta_\text{CI}\approx 76.2^{\circ}$). Three intramolecular distances in KRb $r$ are selected: vibrationally averaged value ($r_0$) and inner and outer classical turning points ($r_\text{in}$ and $r_\text{out}$) of the ground rovibronic level of KRb. The excited state for linear geometries is outside the range of the plot. The green dotted line shows the ground-state interaction energy for $\theta_\text{CI}$ and minimized over $r$.
(b) Isotropic part of the spin-rotation coupling tensor $|\text{Tr}[\bm{\epsilon}(R,r,\theta)]/3|$ corresponding to the geometries presented in panel~(a). Couplings for the linear geometries around their minima overlap with the ones for $\theta\approx 76.2^{\circ}$.}
\label{fig:PESs}
\end{figure}

\subsection{Rb+KRb interactions}
\label{sec:interaction}

As an alternative hypothesis to explain the distribution of rotational levels, we now proceed to describe a full coupled-channel calculation.

We begin by using \textit{ab initio} quantum-chemical methods to obtain the interaction potentials $V(R,r,\theta)$ in the Jacobi coordinates [see SM], which dominate the atom-molecule collisions but do not couple spins.
Examples of these potential energy surfaces (PESs) for selected geometries are presented in Fig.~\ref{fig:PESs}(a).    

These surfaces reveal various important facts about the interaction: (i) the Rb+KRb PESs are very anisotropic;  
(ii) the conical intersection (CI) between the ground and first excited electronic states is energetically accessible for ultracold Rb+KRb collisions; (iii) the excited electronic state is also energetically accessible at short range; 
(iv) the excited-state PES depends dramatically on the internuclear distance in KRb (that is its vibration), while the ground-state PES shows little dependence on it; and (v) the exchange of Rb atoms can proceed without any electronic barrier. 

In addition, we contemplate spin-dependent interactions that are active in the short range and that take the form

\begin{multline}\label{V_sd}
    \hat{V}_\mathrm{sd}  = \sum_{m=1,2,a} \bigl{[} A^{\text{FC}}_m(R,r,\theta)  \hat{\mathbf{S}} \cdot \hat{\mathbf{I}}_m 
+\hat{\mathbf{S}}\cdot \bm{A}^\text{ahf}_m(R,r,\theta) \cdot \hat{\mathbf{I}}_m\bigr{]} \\
+ \hat{\mathbf{S}}\cdot \bm{\epsilon}(R,r,\theta) \cdot (\hat{\mathbf{N}}+ \hat{\mathbf{L}}).
\end{multline}
The intermolecular hyperfine interactions between $\hat{\mathbf{{S}}}$ and $\hat{\mathbf{{I}}}_{m}$ consist of the scalar Fermi contact couplings $A^{\text{FC}}_m(R,r,\theta)$ and tensor  couplings  $\bm{A}^\text{ahf}_m(R,r,\theta)$ \cite{SchweigerBook}, which depend on the atom-molecule distance $R$, the internuclear distance in KRb $r$, and the orientation angle $\theta$.  Finally, the spin-rotation interaction between the overall rotation of the collision complex and its electronic spin is described by the  tensor $\bm{\epsilon}(R,r,\theta)$. 
All of the terms in Eq.~\eqref{V_sd} depend on $\theta$ and couple spins to the mechanical rotation, which qualitatively explain the hyperfine-to-rotational energy transfer. 

The Fermi contact interaction is generally the dominant spin-dependent interactions at short range, with typical values of GHz~\cite{jachymski2022collisional} 
but conserves  $J_r$. 
By contrast,  the intermolecular spin-rotation and tensor hyperfine interactions in Eq.~\eqref{V_sd} couple the states with different $J_r$, and so does the intramolecular nuclear electric quadrupole interaction in KRb, which is part of the KRb molecular Hamiltonian $\hat{H}_\text{mol}$ [Eq.~\eqref{eq:H} in SM]. However, the spin-rotation coupling has a typical short-range value of around 10 MHz and is enhanced to over GHz at the CI only [see Fig.~\ref{fig:PESs}(b)]. The typical strength of tensor hyperfine is on the order of 10 MHz for the Rb-KRb complex, while the strength of  nuclear electric quadrupole interaction in KRb is less than 1 MHz~\cite{AldegundePRA2017}.

\subsection{Coupled-channel calculations}

A complete quantum scattering calculation incorporating both electronic energy surfaces, and the molecular vibration, together with nuclear spins, molecular rotations, and external fields is not possible at present.  
  
We therefore make the following assumptions [see SM]: First, the KRb fragment is assumed to be rigid, which is necessary to limit the number of scattering channels, and hence the computational complexity, to a tractable level. Second, the CI between the ground and the first excited electronic states of Rb-KRb is neglected for the same reason.
Under these assumptions, we developed a coupled-channel (CC) model of  Rb~+~KRb collisions in an external magnetic field based on the {\it ab initio} PES~\cite{jachymski2022collisional} and spin-dependent interactions [Eq.~\eqref{V_sd}].  
These approximations enable us to rigorously account for all rotational and spin degrees of freedom of Rb-KRb in the presence of an external magnetic field. Our approach complements   previous calculations on ultracold alkali dimer-atom collisions, which did not include the spin degrees of freedom and Zeeman interactions, but rigorously treated the rovibrational modes and CIs \cite{croft2017universality,Kendrick:21}.

To handle the computational complexity, we use a recently developed CC basis set composed of the eigenstates of the total mechanical rotational angular momentum of the atom-molecule system $\hat{\mathbf{J}}_r=\hat{\mathbf{N}} + \hat{\mathbf{L}}$ \cite{Tscherbul:23}. 
As shown in SM, our CC model, which rigorously includes all six angular momenta in Rb-KRb and  the couplings between them, predicts that $J_r$ is nearly perfectly conserved in the rigid-rotor approximation. In fact, the calculations  show that, in order to break the conservation of $J_r$, the spin-rotation interaction, which is enhanced at CI, would have to be enhanced by another factor of $\ge$100.

The fact that the CC model conserves $J_r$ greatly reduces the computational cost. Using  extensive basis sets including as many as 180 rotational states of KRb, we demonstrate numerical convergence of Rb~+~KRb collision  cross sections  in the rigid rotor approximation. This is the first time that such convergence is reached for a heavy, strongly anisotropic atom-molecule collision system in a magnetic field including the rotational and  spin degrees of freedom. 

\begin{figure}
\centering
	\includegraphics[width=0.48\textwidth]{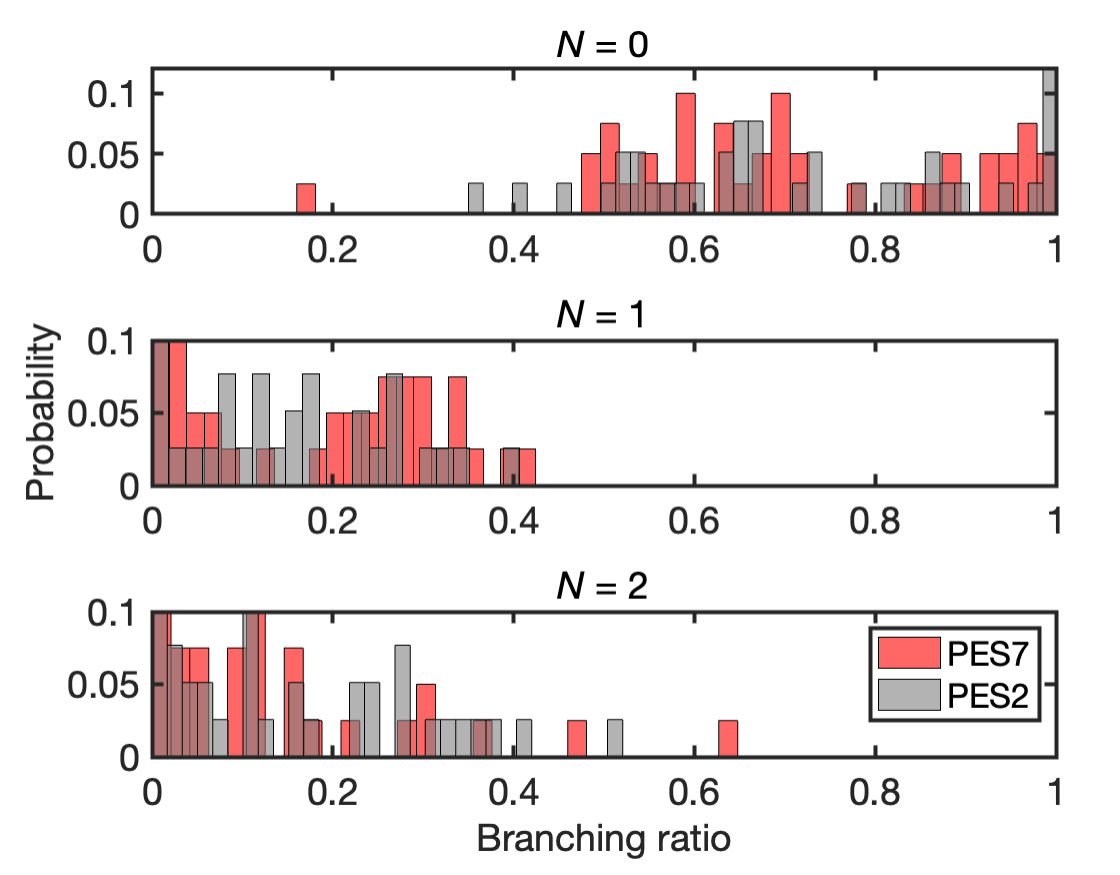}
  \caption{{\bf Histograms of product KRb rotational state
  distributions.} We show the KRb rotational state population distribution following the ultracold Rb ($|2,2\rangle$)~+~KRb ($|0,0,-4,1/2\rangle$) collisions obtained from converged CC calculations that include 181 lowest rotational states of KRb and $J_r=0$.
  The distributions are defined as state-to-state inelastic cross sections to a given final rotational state of KRb ($N$) summed over all  hyperfine sublevels in the $N$-th  manifold and normalized to the total inelastic cross section. The distributions are averaged over 
  over 40 samples of PES2 (grey bars) and PES7 (red bars). The PES samples are obtained by scaling the corresponding PESs by a constant factor $\lambda$ as described in SM.
  }
\label{fig:hist}
\end{figure}

To account for the variation of the final product state distributions with respect to fine details of the PES, we average the calculated distributions over 40 PES samples \cite{Morita:19}. Fig.~\ref{fig:hist} shows the resulting histograms, which show that the product KRb rotational states tend to peak at the lowest final rotational state $N=0$, in disagreement with the experiment. The corresponding Rb product state distribution is largely dominated by the $|F_a=1,M_{F_a}=1\rangle$ and $|F_a=2,M_{F_a}=1\rangle$  hyperfine states, consistent with the selection rules for spin-dependent interactions in Eq.~\eqref{V_sd}.

Although it is possible to bring our converged $J_r=0$ rigid-rotor CC calculations in agreement with the observed product state distributions by fine tuning the strength of the Rb-KRb interaction, the resulting PES is extremely unlikely [see SM], and the observed total inelastic rate could still not be reproduced.

The lack of complete agreement between experiment and our  scattering calculations using only the ground state PES and the rigid-rotor approximation strongly suggests the importance of the KRb vibration and excited PES which are the main missing elements in our theory model. 
This is supported by the presented \textit{ab initio} calculations  which show that, at large $R$, the rigid rotor assumption is a good approximation (the green and black lines overlap, Fig.~\ref{fig:PESs}a) while at short range, it is not due to the excited electronic state of Rb-KRb becoming energetically accessible.
The related strong enhancement of spin-rotation coupling at the CI  is shown in Fig.~\ref{fig:PESs}(b).
This enhancement correlates with the electron g-factor reaching zero, which can potentially lead to Majorana-type spin-flip transitions~\cite{Majorana32} at the CI. 
 The CI could potentially introduce strong mixing between different $J_r$ and explain the statistical distribution of the KRb rotational states.

\section{Conclusion}

Our experiment demonstrated energy transfer from the atomic spin degree of freedom to molecular rotation in Rb ($|2,2\rangle$) and KRb ($|0,0,-4,1/2\rangle$) collisions. 
Following collision, Rb predominantly populated a single hyperfine state ($|1,1\rangle$), and KRb populated its various energetically allowed rotational levels up to $N=2$. 
Our results represent, to our knowledge, the first direct experimental observation of inter-species spin-dependent {hyperfine} interactions.
 Intriguingly, our observed result matches well with the statistical model under the assumption of $J_r$-nonconservation. We developed a state-of-the-art {quantum scattering methodology} in order to understand the results.
However, under rigid-rotor assumption and using only the ground state PES, $J_r$-nonconservation cannot be justified and the quantitative branching ratio as well as the scattering cross sections do not match experiment after extensive efforts. 
In electronic structure calculations, we identified the enhancement of spin-rotation coupling near the CI, which is energetically accessible and could play an important role in the collision dynamics. We note that fully converged scattering calculations including vibrational modes and CIs is beyond the current state-of-the-art~\cite{croft2017universality,Kendrick:21,morita2023non}.  Our findings also shed new light on a previously measured exceptionally long lifetime of ground-state Rb-KRb complex~\cite{nichols2022detection}, which may have a similar origin. This joint experimental and theoretical efforts manifest the complexity of atom-molecule dynamics and call for more advanced theoretical and computational tools to study quantum scattering. {One possible direction for future research could involve combining the CC methodology developed in this work with rigorous quantum reactive scattering calculations, which accurately account for the electronic, vibrational, and rotational degrees of freedom of the collision partners \cite{croft2017universality,Kendrick:21,morita2023non}}.

\section*{Acknowledgement}
 We thank Matthew Frye, Yu Liu,  Jessie Zhang, Zhi Li, and Arfor Houwman for helpful discussions. The experimental work is  supported by the U.S. Department of Energy (DOE), Office of Science, Basic Energy Sciences (BES), under Award No. DE-SC0024087 (molecule state detection), 
 the Center for Ultracold Atoms (an NSF Physics Frontiers Center,  PHY-2317134) (atom state detection), AFOSR DURIP FA9550-23-1-0122 (instrument upgrade). Warsaw team acknowledges the European Union (ERC, 101042989 – QuantMol) and the PL-Grid Infrastructure (Grant No.~PLG/2023/016115). 
{T.V.T. gratefully acknowledges support from the NSF CAREER program (grant No. PHY-2045681).} JLB acknowledges support from the JILA Physics Frontier Center, PHY-2317149.

\section*{Appendix} 
\renewcommand{\tablename}{Table.}
\renewcommand{\thetable}{A\arabic{table}}

\setcounter{figure}{0}
\renewcommand{\figurename}{Fig.}
\renewcommand{\thefigure}{A\arabic{figure}}

\setcounter{equation}{0}
\renewcommand{\theequation}{A\arabic{equation}}

\subsection{Rb spin flip rate based on a statistical model}
We modelled the complex as a Rb atom repeatedly bouncing off the KRb molecule at average time intervals of $\Delta t$.  This time was estimated to be $\Delta t = 1.0 \times 10^{-10}$ s in Ref.~\cite{mayle2012statistical} {by averaging over classical trajectories in a pure $C_6$ potential}.  While these mini-collisions are going on, the hyperfine constant of the Rb atom changes due to the presence of the other atoms.  The change from the native Rb hyperfine constant at large $R$, to that at small $R$, is given by {$\Delta A = A^{\text{FC}}$
~\cite{jachymski2022collisional}}; it vanishes at large $R$, and approaches a value $\Delta A \approx 1$ GHz.  Thus the Rb atom sees its hyperfine constant fluctuate on a scale $\Delta A$, at time intervals of roughly $\Delta t$.
Ref.~\cite{mayle2012statistical} estimates the probability that a hyperfine transition may occur during a time $T$ spent in complex:
\begin{align}
P =  \pi^2  \left( \frac{ \Delta A T}{  h } \right) \left( \frac{ \Delta A \Delta t }{ h } \right),
\end{align}
where $h$ is Planck's constant. {Experimentally,  we measured the intermediate complex lifetime to be $T$ = 0.54(10) ns [see SM], and thus the spin-flip probability is near unity.} This back-of-envelope estimation suggests that the transition can indeed be made on the observed timescale, while suggesting that a second transition, to $|10\rangle$, is less likely. 

\subsection{Statistical model of KRb branching ratio}

We propose a statistical model as follows.  We choose, in principle, a Jacobi coordinate system for the triatomic encounter.   Let ${\bf r}$ be the vector joining the K and Rb atoms in the molecule, and $\bf{ R}$ the vector between the molecule's center of mass and the incident Rb atom.  Similar to the coupled-channel calculations, we assume the magnitude $r$ is fixed, or more properly, this degree of freedom is ignored.  It may contribute to the lifetime of the complex or the density of resonant states, but is unlikely to directly affect rotational branching ratios.  Likewise, the magnitude $R$ is not treated explicitly, except as follows: When $R$ is large, the states of the system are given in terms of atom + molecule quantum numbers that define asymptotic $|a \rangle$  states, to be specified further below.  When $R$ is small, the triatomic system is described by states $|c \rangle$ of the complex, also specified below.  

Transitions between asymptotic states and states of the complex are handled in the spirit of a frame transformation.  That is, the probability of a given incident asymptotic channel $|a \rangle$ to find itself in state $|c \rangle$ of the complex is $|\langle a | c \rangle|^2$, the matrix element taken over all degrees of freedom except $R$.  Likewise, the same kind of matrix element gives  the probability of  complex channel $|c \rangle$ to decay into asymptotic channel $|a \rangle$.

For convenience, here we again identify the various angular momentum operators as follows.  
\begin{align}
\hat{{\bf N}} & = \mathrm{molecular}\;\; \mathrm{rotation} \\
\hat{\bf L} & = \mathrm{partial} \;\; \mathrm{wave} \\
\hat{\bf I}_a & = \mathrm{Rb} \;\;  \mathrm{atom} \;\; \mathrm{nuclear} \;\;  \mathrm{spin} \\
\hat{\bf S} &= \mathrm{Rb} \;\; \mathrm{electronic} \;\; \mathrm{spin} \\
\hat{\bf I}_{1} &= \mathrm{K} \;\; \mathrm{moleceular} \;\; \mathrm{nuclear} \;\; \mathrm{spin} \\
\hat{\bf I}_2 &= \mathrm{Rb} \;\; \mathrm{molecular} \;\; \mathrm{nuclear} \;\; \mathrm{spin} 
\end{align}
 Using these ingredients, the separated atom-molecule states are given by
\begin{align}
|a \rangle = |NM_N \rangle |LM_L \rangle | I_1 M_{I_1} I_2 M_{I_2} \rangle |F_a M_{F_a} \rangle
\end{align}
where $|F_a M_{F_a} \rangle$ are the eigenstates of $\hat{F}_a^2$ and $\hat{F}_{a_z}$, and  
 $\hat{\bf F}_a = \hat{\bf S} + \hat{\bf I}_{a}$ is the total angular momentum of the atom.  The complex is assumed to be a state of good total angular momentum $|JM_J \rangle$, in spite of the presence of small magnetic and electric fields.  
To describe states of the complex we choose the following angular momentum coupling scheme.  
\begin{align}
\hat{\bf I} &= \hat{\bf I}_{1} + \hat{\bf I}_{2}  = \mathrm{molecular} \;\; \mathrm{nuclear} \;\; \mathrm{spin} \\
\hat{\bf F} &= \hat{\bf I} + \hat{\bf F }_a  =\mathrm{total} \;\; \mathrm{spin}    \\
\hat{\bf J}_r &=  \hat{\bf N} + \hat{\bf L}  = \mathrm{total} \;\; \mathrm{mechanical} \;\; \mathrm{rotation} \\
 \hat{\bf J} &=\hat{\bf J}_r + \hat{\bf F} = \mathrm{total} \;\; \mathrm{angular} \;\; \mathrm{momentum}.
\end{align}
 A state of the complex is therefore denoted
\begin{align}
| c \rangle = \Big| \Big[ (NL)J_r ; [(I_{1}I_{2})I;F_a] F \Big] JM_J \rangle.
\end{align}
 One further constraint on both basis sets is that parity be conserved, that is, $(-1)^{N+L}=1$, as it is in the initial state with $N=L=0$.  
 
 To determine overlap integrals for these states, we expand the complex state explicitly as
 \begin{align}
 |c \rangle &= 
 \sum_{{\cal M}} 
  |NM_N \rangle |LM_L \rangle | I_1 M_{I_1} I_2 M_{I_2} \rangle |F_a M_{F_a} \rangle \\
  & \hphantom{{}= \sum_{{\cal M}} } 
  \times \langle NM_N LM_L | J_r M_r \rangle \langle I_1 M_{I_1} I_2 M_{I_2} | I M_I \rangle \\
   & \hphantom{{}= \sum_{{\cal M}} } 
  \times  \langle IM_I F_a M_{F_a} | F M_F \rangle \langle J_r M_rFM_F | J M_J \rangle.
  \end{align}
  Here ${\cal M} = \{ M_r,M_F,M_N,M_L,M_I,M_{F_a},M_{I_1},M_{I_2} \}$ denotes, formally, the set of magnetic quantum numbers in this expansion.  
 
 The projection of asymptotic states onto states of the complex therefore takes the form
 \begin{align}
 \langle a  | c \rangle &=
  \langle NM_N LM_L | J_r M_r \rangle \langle I_1 M_{I_1} I_2 M_{I_2} | I M_I \rangle \\
   & \hphantom{{} \langle a  | c \rangle  } 
     \times  \langle IM_I F_a M_{F_a} | F M_F \rangle \langle J_r M_rFM_F | J M_J \rangle.
   \end{align}
Some of the quantum numbers are specified in the basis sets; others are determined by angular momentum conservation, thus $M_r = M_N +M_L$, $M_I = M_{I_1} + M_{I_2}$, $M_F = M_J - M_r$.  

In the initial state with $N=L=0$, the overlap simplifies, since $J_r=0$ and $J=F$.  The probability for the initial complex to find itself in a state with total angular momentum $J$ is the sum of probabilities to arrive in any of the states with a given total $J$ and all values of any other quantum numbers, in this case $I$.  We denote this probability
\begin{align}
P(J) = \sum_I \Big| \langle I_1 M_{I_1} I_2 M_{I_2} | I M_I \rangle
\langle IM_I F_a M_{F_a} | J M_J \rangle \Big|^2,
\end{align}
Here $M_{I_1}=-4$, $M_{I_2}=1/2$, $F_a=M_{F_a}=2$ are specified by the initial condition.  

It is likely that only certain substates of the complex, with a given value of $JM_J$ are populated by the initial collision, for example, only those with $N=L=0$.  In the {\it full statistical approximation}, it is assumed that the complex thoroughly scrambles all the states that are allowed by angular momentum conservation.  That is, during the lifetime of the complex, all such states become equally populated.  This assumption can be restricted later, based on other physical considerations.  

In this circumstance, consider the probability that the complex with total angular momentum $J$ decays into an asymptotic channel where the molecule has rotation $N$.  This is the sum of the probabilities $|\langle a | c \rangle|^2$ for all states of the complex  with total angular momentum $J$, and all final asymptotic channels consistent with $N$, summed over all the other quantum numbers.  Such a probability is denoted
\begin{widetext}
 \begin{align}
 D(N,J) &= \sum_{J_r,I,F} \sum_{M_N,L,M_L,M_{I_1},M_{I_2} }
  \Big| \langle NM_N LM_L | J_r M_r \rangle \langle I_1 M_{I_1} I_2 M_{I_2} | I M_I \rangle \\
   & \hphantom{{} = \sum_{J_r,I,F} \sum_{M_N,L,M_L,M_{I_1},M_{I_2} }  } 
     \times  \langle IM_I F_a M_{F_a} | F M_F \rangle \langle J_r M_rFM_F | J M_J \rangle \Big|^2.
 \end{align}
 \end{widetext}
This expression requires $F_a=M_{F_a}=1$ in the final state that is measured.  The quantum numbers $M_{I_1}$, $M_{I_2}$ are no longer constrained, but run over all allowed values.

The probability for the incident state to produce, after the collision, a state of rotation $N$ of the molecule is therefore the probability that a complex of total angular momentum $J$ so decays, weighted by the probability that the collision populates such a complex in the first place:
\begin{align}
D(N) = \sum_J P(J) D(N,J).
\end{align}
As noted above, the full statistical model would allow for all possible states of the complex. In particular, the mechanical rotation $J_r$ could range over many values.  In this case, the branching ratio for $N=2:1:0$ are $0.544:0.340:0.116$, in agreement with the measured results.  However, as noted above, the available spin-rotation and tensor hyperfine interaction that can actually change $J_r$ in a collision is quite weak, whereby we must also consider incomplete statistical models that restrict the value of $J_r$.  If $J_r$ is completely restricted to its initial value of $J_r=0$ (consistent with $N=L=0$), then the unsuitable branching ratio $1:1:1$ results.  However, if this weak perturbation can drive the complex from $J_r=0$ to even $J_r=2$, the corresponding statistical model restores the branching ratio to $0.50:0.33:0.17$.  This suggests that only a weak scrambling is required to tip the complex into something like its full statistical behavior.  

\begin{figure*}
\centering
	\includegraphics[width=0.9\textwidth]{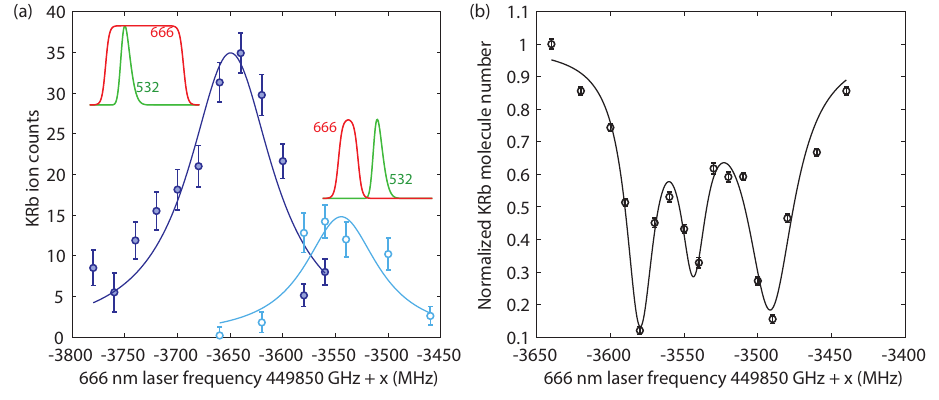}
  \caption{{\bf Light shift and hyperfine structure observed in the R(0) transition.} (a) Pulsed 532 nm light induced light shift in the R(0) transition. The dark blue solid circles represent the R(0) transition with an overlapping 666 nm pulse and 532 nm pulse. The light blue open circles represent the R(0) transition without overlapping 666 nm and 532 nm pulses. The solid lines are fittings to Lorentzian lineshapes. The insets are the corresponding REMPI pulse timing diagrams. (b) Depletion spectroscopy of $N$ = 0 KRb molecules from absorption imaging near R(0) transitions. The spectrum shows hyperfine structure of the N = 1 level in B$^1\Pi$ $\nu' = 0$ manifold.  }
\label{fig:R0_ls}
\end{figure*}

\subsection{1+1' REMPI detection of collisional products}

Our scheme of the 1+1' REMPI detection of KRb utilized two lasers. One was a continuous wave external cavity diode laser operating at 666 nm and the other was a pulsed frequency-doubled Nd:YVO$_4$ laser at 532 nm with a pulse duration of 10 ns. The 666 nm laser was resonant to the KRb X$^1\Sigma^+$ ($\nu = 0$) and B$^1\Pi$ ($\nu' = 0$) vibronic band transitions as shown in Fig. \ref{fig:spectroscopy}. The power of the 666 nm light was 8.7 mW with a $1/e^{2}$ beam radius of 1.0 mm. The pulse energy of the 532 nm laser was set to 189 $\mu$J with a $1/e^{2}$ beam radius of 1.0 mm.
Similarly, the 1+1' REMPI detection of Rb utilized  a continuous wave external cavity diode laser operating at 420 nm and the same 532 nm pulsed laser. The 420 nm laser was resonant to the  Rb 5S$_{1/2}, F=1$ to 6P$_{3/2}$ transitions as shown in Fig. \ref{fig:spectroscopy}. The power of the 420 nm light was 1.9 mW with a $1/e^{2}$ beam radius of 1.0 mm. 
The stabilization of the 666 nm and 420 nm laser frequency was achieved by locking to a wavelength meter (Bristol 871A) inside a pressure-stable chamber. We further calibrated the wavelength meter readings to an external cavity diode laser at 970 nm locked to a high-finesse stable cavity, which enabled us to improve the frequency accuracy to 5 MHz. 

 To protect the reactant KRb molecules in the ODT from scattering of the repeatedly fired REMPI pulses, we placed a copper wire (American Wire Gauge 36, 127 $\mu$m diameter) in the REMPI beam path and cast its shadow on the plane containing KRb molecules with a f = 250 mm achromatic plano-convex lens. The shadow of the copper wire had a width of 250 $\mu$m and was centered around the location of the molecular cloud.

The 666 nm and 420 nm laser pulses were fired at a repetition rate of 10 kHz that temporally overlapped with the 532 nm pulses. The pulse lengths of the 666 nm and 420 nm light and the relative timings to the 532 nm pulses were controlled and optimized using a delay function generator.  The resulting ions are accelerated onto an ion detector under a constant electric field (E =  17 V/cm) generated by a series of plate electrodes and counted~\cite{hu2019direct,liu2020probing}. The ionization sequence lasts for 1 s, where the optical dipole trap (ODT) intensity, the magnetic field, and the electric field (perpendicular to the magnetic field) are held constant for the entire duration. The atoms and molecules are depleted after the ionization sequence.

From our KRb REMPI spectra (Fig.~\ref{fig:spectroscopy} in the main text), we can extract the rotational constants of KRb in the X$^1\Sigma^+$ and B$^1\Pi$ ($\nu' = 0$) states as $2\pi\times 1112(3)$ MHz and $2\pi~\times 961(1)$ MHz, which are in good agreement with previously measured value~\cite{ni2008high} (B = $2\pi~\times$ 1.1139(1) GHz) and theory prediction~\cite{borsalino2014efficient} (B = $2\pi~ \times$  953 MHz). The $\Lambda-$doublet splitting of the B$^1\Pi$ excited state is $2\pi \times$ 7(5) MHz. Additionally, we observed that the frequency of the R(0) transition is offset by $2\pi~ \times$ 100 MHz from prediction using the rotational constants, due to the light shift caused by the 532 nm light [see Fig.~\ref{fig:R0_ls}]. It is important to take this into account to extract the population of post-collision N=0 molecules. 

\begin{figure}
\centering
	\includegraphics[width=0.4\textwidth]{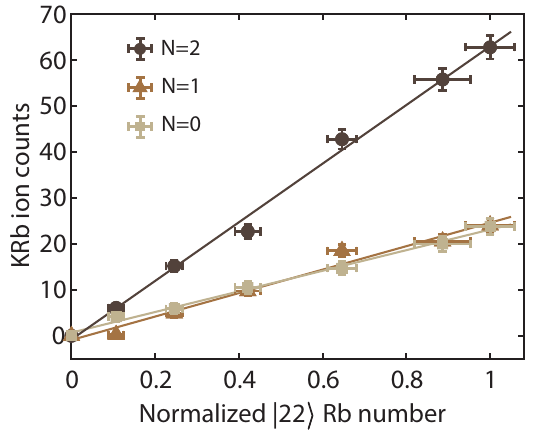}
  \caption{ \textbf{Rb atom number dependence of KRb ion counts.} KRb ion counts were measured with the 666 nm laser resonant with the Q(2) (circles), Q(1) (triangles) and R(0) (squares) transitions as a function of the initial $|2,2\rangle$ Rb atom number, which was controlled by varying the duration of the microwave ARP pulse. To account for imperfections in the dark mask that give rise to signal from $|1,1\rangle$ atoms in the reactant cloud, the same measurements were performed without a microwave ARP pulse to establish a background. The results with background subtraction are shown here.  X-axis error bars indicate the standard deviation of relative atom number of the five measurements, while Y-axis error bars represent shot noise. Solid lines are linear fits to the data.}
\label{fig:number}
\end{figure}

We additionally check the linear correlation between the product KRb molecule population and the initial $|2,2\rangle$ Rb atom number, as shown in Fig.~\ref{fig:number}, validating that the $N = 0$  KRb with sufficient kinetic energy to escape the ODT and the excited KRb molecules in $N$ = 1 and 2 originate from the inelastic atom-molecule collisions.

\subsection{Monte Carlo simulation of REMPI}

To obtain a branching ratio between different product rotational states, we created a model that accurately reflected the ionization efficiency for each state to convert the experimental signal ion counts to the corresponding molecule population. 
The model addressed several not-so-straightforward aspects, including the fact that the kinetic energy release from the collision is small and the products could experience multiple ionization events in the detection volume. Additionally, the 666 nm transition Rabi rate was comparable to the excited state decay rate. 
As a consequence, a single ionization event could not deplete the ground state population of the product molecules and later ionization events could also contribute to the total ion signal.
Moreover, due to the Gaussian intensity profile of our REMPI beams, ionization events exhibited different dynamics depending on their distances away from the trap center.

To get an accurate branching ratio, we modelled all ionization processes using Monte Carlo (MC) simulation and fitted the data shown in Fig.~\ref{fig:calibration} in the main text to our simulation.

In the simulation, both the 666 nm and 532 nm beams had Gaussian profiles with a $1/e^{2}$ beam radius of 1 mm. The width of the dark mask was 250 $\mu$m, covering the center of the ionization beams. These geometric parameters were chosen to be consistent with our experimental values. Meanwhile, product KRb molecules in three different rotational states had velocities $v_0=4.18$ m/s, $v_1$ = 3.54 m/s, and $v_2 = $ 0.74 m/s for $N=0,1,2$ respectively. The simulation was initialized with 1000 product molecules for each rotational state, all located at the center of the REMPI beams. Each molecule followed a uniform spherical velocity distribution and an initial starting time drawn from a uniform distribution between 0 and 100~$\mu$s, which was the time difference between two ionization events to model the asynchronous nature of product formation.

For every product molecule created with its initial time and velocity vector, subsequent ionization process became deterministic. We calculated the coordinates of the molecules when each ionization pulse was turned on, from which we obtained the REMPI beam intensities for each ionization event. Each ionization event was modelled using a two-level system with decay in the excited state~\cite{liu2021precision}. The dynamics of the ground state $|g\rangle$ in the X$^1\Sigma^+$ manifold and the intermediate state $|e\rangle$ in the B$^1\Pi$ could be described using the following rate equations
\begin{align}
\frac{d}{dt} \rho_{gg} & = -\frac{i}{2}(\Omega_{ge}\rho_{eg}-c.c.), \\
\frac{d}{dt} \rho_{ee} & = -(\Gamma+\Gamma_{\text{ion}})\rho_{ee} + \frac{i}{2} (\Omega_{ge}\rho_{eg}-c.c.), \\
\frac{d}{dt} \rho_{eg} & = -\frac{1}{2}(\Gamma + \Gamma_{\text{ion}}) \rho_{eg} + i \Delta \rho_{eg} + \frac{i}{2} \Omega_{eg}(\rho_{ee}-\rho_{gg}), 
\end{align}
where $c.c.$ stands for complex conjugate, $\Omega_{ge}$ is the Rabi rate of the 666 nm bound-to-bound transition, $\Delta$ is the 666 nm laser detuning, $\Gamma$ is the decay rate of the excited state $|e\rangle$, and $\Gamma_{\text{ion}}$ is the ionization rate of the 532 nm bound-to-continuum transition.  We characterized the decay rate of the excited state $\Gamma = 2\pi \times 15.6(3.2)$ MHz using absorption imaging and fixed this parameter in the MC simulation.  The ion counts were calculated using $\int dt\,\rho_{ee}(t)\Gamma_{\text{ion}}(t)$, where $\Gamma_{\text{ion}}(t)$ is proportional to the 532 nm pulse intensity, with peak value $\Gamma_{\text{ion}} = 2\pi \times$  5 MHz~\cite{liu2021precision} at the center of the Gaussian beam.  

For each ionization event, we solved the above rate equations and obtained the ion counts and remaining ground state population $\rho_{gg}$, which rolls over to the next ionization event. We summed up the ion counts from all product molecules and all ionization pulses each molecule experienced, from which we extracted the branching ratio. 
We note that the choice of peak 532 nm ionization rate did not affect the final branching ratio. Doppler shifts were taken into account in our simulation. 

When fitting the data to our MC simulation, we set the initial molecule number, 666 nm Rabi rate, and 666 nm laser detuning, as free fitting parameters. Meanwhile, we fixed the excited state decay rate, temporal profile of the REMPI pulses, the peak 532 nm laser ionization rate at the center of the beam, and the geometries of the REMPI beams with the dark line from experimentally measured values.

For the R(0) branch, we additionally took into account of the light shift caused by the pulsed 532 nm light observed in the experiment by introducing a time-dependent detuning. We also considered the excited state hyperfine structure in R(0) transition [see Fig.~\ref{fig:R0_ls}] by extending the two-level system model to include three independent excited states with the same Rabi rates and decay rates. We did not observe significant light shifts on the Q(2) and Q(1) transitions caused by the 532 nm light. The hyperfine structure in the excited state for Q(2) and Q(1) transitions are similar to R(0), and ignoring the hyperfine structure would cause at most a 5\% overestimation of the excited state population.

The fitted Rabi rates for Q(2), Q(1), and R(0) were $2\pi\times$ 10.2(2) MHz, $2\pi\times$ 10.0(5) MHz, and $2\pi\times$ 13.7(1.0) MHz, respectively, with the error bars representing a 68\% confidence interval. The extracted R(0) Rabi rate from the fitting agreed well with the Rabi rate measured using depletion spectroscopy, which is 2$\pi\times$ 14(4)MHz. Additionally, the ratio between the three Rabi rates showed good agreement with the H\"onl-London factors~\cite{hansson2005comment} for X$^1\Sigma^+$ to B$^1\Pi$ transitions, assuming that the $M_N$ distribution was randomized. In addition, no polarization dependence of KRb ion counts was observed across all transitions, indicating that the $M_N$ levels were scrambled after the collision. 

\begin{table}[h!]
\centering
 \begin{tabular}{||c | c c c||} 
 \hline
Condition & $N=2$ & $N=1$ & $N=0$ \\ [0.5ex] 
 \hline\hline
 2 G, $|2,2\rangle$ Rb + N=0 KRb & 0.54(9) & 0.36(7) & 0.10(2) \\ 
 \hline
 30 G, $|2,2\rangle$ Rb + N=0 KRb & 0.53(9) & 0.33(8) & 0.13(3) \\ 
 \hline
 30 G, $|2,2\rangle$ Rb + N=1 KRb & 0.59(13) & 0.32(9) & 0.09(2) \\[1ex] 
 \hline
 \end{tabular}
 \caption{Table of product KRb branching ratio under different conditions. We presented 30 G, $|2,2\rangle$ Rb + N=0 KRb data in the main text. Electric fields for all three datasets are the same (17 V/cm and perpendicular to the magnetic field).  } \label{table:ratio}
\end{table}


\begin{thebibliography}{76}%
\makeatletter
\providecommand \@ifxundefined [1]{%
 \@ifx{#1\undefined}
}%
\providecommand \@ifnum [1]{%
 \ifnum #1\expandafter \@firstoftwo
 \else \expandafter \@secondoftwo
 \fi
}%
\providecommand \@ifx [1]{%
 \ifx #1\expandafter \@firstoftwo
 \else \expandafter \@secondoftwo
 \fi
}%
\providecommand \natexlab [1]{#1}%
\providecommand \enquote  [1]{``#1''}%
\providecommand \bibnamefont  [1]{#1}%
\providecommand \bibfnamefont [1]{#1}%
\providecommand \citenamefont [1]{#1}%
\providecommand \href@noop [0]{\@secondoftwo}%
\providecommand \href [0]{\begingroup \@sanitize@url \@href}%
\providecommand \@href[1]{\@@startlink{#1}\@@href}%
\providecommand \@@href[1]{\endgroup#1\@@endlink}%
\providecommand \@sanitize@url [0]{\catcode `\\12\catcode `\$12\catcode
  `\&12\catcode `\#12\catcode `\^12\catcode `\_12\catcode `\%12\relax}%
\providecommand \@@startlink[1]{}%
\providecommand \@@endlink[0]{}%
\providecommand \url  [0]{\begingroup\@sanitize@url \@url }%
\providecommand \@url [1]{\endgroup\@href {#1}{\urlprefix }}%
\providecommand \urlprefix  [0]{URL }%
\providecommand \Eprint [0]{\href }%
\providecommand \doibase [0]{https://doi.org/}%
\providecommand \selectlanguage [0]{\@gobble}%
\providecommand \bibinfo  [0]{\@secondoftwo}%
\providecommand \bibfield  [0]{\@secondoftwo}%
\providecommand \translation [1]{[#1]}%
\providecommand \BibitemOpen [0]{}%
\providecommand \bibitemStop [0]{}%
\providecommand \bibitemNoStop [0]{.\EOS\space}%
\providecommand \EOS [0]{\spacefactor3000\relax}%
\providecommand \BibitemShut  [1]{\csname bibitem#1\endcsname}%
\let\auto@bib@innerbib\@empty
\bibitem [{\citenamefont {Herschbach}(1987)}]{Herschbach-Nobel}%
  \BibitemOpen
  \bibfield  {author} {\bibinfo {author} {\bibfnamefont {D.~R.}\ \bibnamefont
  {Herschbach}},\ }\bibfield  {title} {\bibinfo {title} {Molecular dynamics of
  elementary chemical reactions ({Nobel} lecture)},\ }\href
  {https://doi.org/https://doi.org/10.1002/anie.198712211} {\bibfield
  {journal} {\bibinfo  {journal} {Angewandte Chemie International Edition in
  English}\ }\textbf {\bibinfo {volume} {26}},\ \bibinfo {pages} {1221}
  (\bibinfo {year} {1987})}\BibitemShut {NoStop}%
\bibitem [{\citenamefont {Lee}(1987)}]{lee1987molecular}%
  \BibitemOpen
  \bibfield  {author} {\bibinfo {author} {\bibfnamefont {Y.~T.}\ \bibnamefont
  {Lee}},\ }\bibfield  {title} {\bibinfo {title} {Molecular beam studies of
  elementary chemical processes ({Nobel} lecture)},\ }\href
  {https://doi.org/https://doi.org/10.1002/anie.198709393} {\bibfield
  {journal} {\bibinfo  {journal} {Angewandte Chemie International Edition in
  English}\ }\textbf {\bibinfo {volume} {26}},\ \bibinfo {pages} {939}
  (\bibinfo {year} {1987})}\BibitemShut {NoStop}%
\bibitem [{\citenamefont {Wang}\ \emph {et~al.}(2018)\citenamefont {Wang},
  \citenamefont {Yang}, \citenamefont {Xiao}, \citenamefont {Sun},
  \citenamefont {Zhang}, \citenamefont {Yang}, \citenamefont {Weichman},\ and\
  \citenamefont {Neumark}}]{Neumark2018}%
  \BibitemOpen
  \bibfield  {author} {\bibinfo {author} {\bibfnamefont {T.}~\bibnamefont
  {Wang}}, \bibinfo {author} {\bibfnamefont {T.}~\bibnamefont {Yang}}, \bibinfo
  {author} {\bibfnamefont {C.}~\bibnamefont {Xiao}}, \bibinfo {author}
  {\bibfnamefont {Z.}~\bibnamefont {Sun}}, \bibinfo {author} {\bibfnamefont
  {D.}~\bibnamefont {Zhang}}, \bibinfo {author} {\bibfnamefont
  {X.}~\bibnamefont {Yang}}, \bibinfo {author} {\bibfnamefont {M.~L.}\
  \bibnamefont {Weichman}},\ and\ \bibinfo {author} {\bibfnamefont {D.~M.}\
  \bibnamefont {Neumark}},\ }\bibfield  {title} {\bibinfo {title} {Dynamical
  resonances in chemical reactions},\ }\href
  {https://doi.org/10.1039/C8CS00041G} {\bibfield  {journal} {\bibinfo
  {journal} {Chemical Society Reviews}\ }\textbf {\bibinfo {volume} {47}},\
  \bibinfo {pages} {6744} (\bibinfo {year} {2018})}\BibitemShut {NoStop}%
\bibitem [{\citenamefont {Zhao}\ and\ \citenamefont
  {Guo}(2017)}]{zhao2017state}%
  \BibitemOpen
  \bibfield  {author} {\bibinfo {author} {\bibfnamefont {B.}~\bibnamefont
  {Zhao}}\ and\ \bibinfo {author} {\bibfnamefont {H.}~\bibnamefont {Guo}},\
  }\bibfield  {title} {\bibinfo {title} {State-to-state quantum reactive
  scattering in four-atom systems},\ }\href
  {https://doi.org/https://doi.org/10.1002/wcms.1301} {\bibfield  {journal}
  {\bibinfo  {journal} {Wiley Interdisciplinary Reviews: Computational
  Molecular Science}\ }\textbf {\bibinfo {volume} {7}},\ \bibinfo {pages}
  {e1301} (\bibinfo {year} {2017})}\BibitemShut {NoStop}%
\bibitem [{\citenamefont {Althorpe}\ and\ \citenamefont
  {Clary}(2003)}]{althorpe2003quantum}%
  \BibitemOpen
  \bibfield  {author} {\bibinfo {author} {\bibfnamefont {S.~C.}\ \bibnamefont
  {Althorpe}}\ and\ \bibinfo {author} {\bibfnamefont {D.~C.}\ \bibnamefont
  {Clary}},\ }\bibfield  {title} {\bibinfo {title} {Quantum scattering
  calculations on chemical reactions},\ }\href
  {https://doi.org/https://doi.org/10.1146/annurev.physchem.54.011002.103750}
  {\bibfield  {journal} {\bibinfo  {journal} {Annual review of physical
  chemistry}\ }\textbf {\bibinfo {volume} {54}},\ \bibinfo {pages} {493}
  (\bibinfo {year} {2003})}\BibitemShut {NoStop}%
\bibitem [{\citenamefont {Rebentrost}(1978)}]{rebentrost1978}%
  \BibitemOpen
  \bibfield  {author} {\bibinfo {author} {\bibfnamefont {F.}~\bibnamefont
  {Rebentrost}},\ }\bibfield  {title} {\bibinfo {title}
  {Electronic-to-rotational energy transfer in the system {F + D$_2$}},\ }\href
  {https://doi.org/https://doi.org/10.1016/0009-2614(78)80308-X} {\bibfield
  {journal} {\bibinfo  {journal} {Chemical Physics Letters}\ }\textbf {\bibinfo
  {volume} {58}},\ \bibinfo {pages} {18} (\bibinfo {year} {1978})}\BibitemShut
  {NoStop}%
\bibitem [{\citenamefont {Yuan}\ \emph {et~al.}(2018)\citenamefont {Yuan},
  \citenamefont {Guan}, \citenamefont {Chen}, \citenamefont {Zhao},
  \citenamefont {Yu}, \citenamefont {Luo}, \citenamefont {Tan}, \citenamefont
  {Xie}, \citenamefont {Wang}, \citenamefont {Sun} \emph
  {et~al.}}]{yuan2018observation}%
  \BibitemOpen
  \bibfield  {author} {\bibinfo {author} {\bibfnamefont {D.}~\bibnamefont
  {Yuan}}, \bibinfo {author} {\bibfnamefont {Y.}~\bibnamefont {Guan}}, \bibinfo
  {author} {\bibfnamefont {W.}~\bibnamefont {Chen}}, \bibinfo {author}
  {\bibfnamefont {H.}~\bibnamefont {Zhao}}, \bibinfo {author} {\bibfnamefont
  {S.}~\bibnamefont {Yu}}, \bibinfo {author} {\bibfnamefont {C.}~\bibnamefont
  {Luo}}, \bibinfo {author} {\bibfnamefont {Y.}~\bibnamefont {Tan}}, \bibinfo
  {author} {\bibfnamefont {T.}~\bibnamefont {Xie}}, \bibinfo {author}
  {\bibfnamefont {X.}~\bibnamefont {Wang}}, \bibinfo {author} {\bibfnamefont
  {Z.}~\bibnamefont {Sun}}, \emph {et~al.},\ }\bibfield  {title} {\bibinfo
  {title} {Observation of the geometric phase effect in the {H+HD→H$_2$+D
  }reaction},\ }\href {https://doi.org/10.1126/science.aav1356} {\bibfield
  {journal} {\bibinfo  {journal} {Science}\ }\textbf {\bibinfo {volume}
  {362}},\ \bibinfo {pages} {1289} (\bibinfo {year} {2018})}\BibitemShut
  {NoStop}%
\bibitem [{\citenamefont {Klein}\ \emph {et~al.}(2017)\citenamefont {Klein},
  \citenamefont {Shagam}, \citenamefont {Skomorowski}, \citenamefont
  {{\.Z}uchowski}, \citenamefont {Pawlak}, \citenamefont {Janssen},
  \citenamefont {Moiseyev}, \citenamefont {van~de Meerakker}, \citenamefont
  {van~der Avoird}, \citenamefont {Koch} \emph {et~al.}}]{klein2017directly}%
  \BibitemOpen
  \bibfield  {author} {\bibinfo {author} {\bibfnamefont {A.}~\bibnamefont
  {Klein}}, \bibinfo {author} {\bibfnamefont {Y.}~\bibnamefont {Shagam}},
  \bibinfo {author} {\bibfnamefont {W.}~\bibnamefont {Skomorowski}}, \bibinfo
  {author} {\bibfnamefont {P.~S.}\ \bibnamefont {{\.Z}uchowski}}, \bibinfo
  {author} {\bibfnamefont {M.}~\bibnamefont {Pawlak}}, \bibinfo {author}
  {\bibfnamefont {L.~M.}\ \bibnamefont {Janssen}}, \bibinfo {author}
  {\bibfnamefont {N.}~\bibnamefont {Moiseyev}}, \bibinfo {author}
  {\bibfnamefont {S.~Y.}\ \bibnamefont {van~de Meerakker}}, \bibinfo {author}
  {\bibfnamefont {A.}~\bibnamefont {van~der Avoird}}, \bibinfo {author}
  {\bibfnamefont {C.~P.}\ \bibnamefont {Koch}}, \emph {et~al.},\ }\bibfield
  {title} {\bibinfo {title} {Directly probing anisotropy in atom--molecule
  collisions through quantum scattering resonances},\ }\href
  {https://doi.org/https://doi.org/10.1038/nphys3904} {\bibfield  {journal}
  {\bibinfo  {journal} {Nature Physics}\ }\textbf {\bibinfo {volume} {13}},\
  \bibinfo {pages} {35} (\bibinfo {year} {2017})}\BibitemShut {NoStop}%
\bibitem [{\citenamefont {Dong}\ \emph {et~al.}(2010)\citenamefont {Dong},
  \citenamefont {Xiao}, \citenamefont {Wang}, \citenamefont {Dai},
  \citenamefont {Yang},\ and\ \citenamefont {Zhang}}]{dong2010transition}%
  \BibitemOpen
  \bibfield  {author} {\bibinfo {author} {\bibfnamefont {W.}~\bibnamefont
  {Dong}}, \bibinfo {author} {\bibfnamefont {C.}~\bibnamefont {Xiao}}, \bibinfo
  {author} {\bibfnamefont {T.}~\bibnamefont {Wang}}, \bibinfo {author}
  {\bibfnamefont {D.}~\bibnamefont {Dai}}, \bibinfo {author} {\bibfnamefont
  {X.}~\bibnamefont {Yang}},\ and\ \bibinfo {author} {\bibfnamefont {D.~H.}\
  \bibnamefont {Zhang}},\ }\bibfield  {title} {\bibinfo {title}
  {Transition-state spectroscopy of partial wave resonances in the {F+ HD}
  reaction},\ }\href {https://doi.org/10.1126/science.1185694} {\bibfield
  {journal} {\bibinfo  {journal} {Science}\ }\textbf {\bibinfo {volume}
  {327}},\ \bibinfo {pages} {1501} (\bibinfo {year} {2010})}\BibitemShut
  {NoStop}%
\bibitem [{\citenamefont {de~Jongh}\ \emph {et~al.}(2020)\citenamefont
  {de~Jongh}, \citenamefont {Besemer}, \citenamefont {Shuai}, \citenamefont
  {Karman}, \citenamefont {van~der Avoird}, \citenamefont {Groenenboom},\ and\
  \citenamefont {van~de Meerakker}}]{DeJonghScience2020}%
  \BibitemOpen
  \bibfield  {author} {\bibinfo {author} {\bibfnamefont {T.}~\bibnamefont
  {de~Jongh}}, \bibinfo {author} {\bibfnamefont {M.}~\bibnamefont {Besemer}},
  \bibinfo {author} {\bibfnamefont {Q.}~\bibnamefont {Shuai}}, \bibinfo
  {author} {\bibfnamefont {T.}~\bibnamefont {Karman}}, \bibinfo {author}
  {\bibfnamefont {A.}~\bibnamefont {van~der Avoird}}, \bibinfo {author}
  {\bibfnamefont {G.~C.}\ \bibnamefont {Groenenboom}},\ and\ \bibinfo {author}
  {\bibfnamefont {S.~Y.~T.}\ \bibnamefont {van~de Meerakker}},\ }\bibfield
  {title} {\bibinfo {title} {Imaging the onset of the resonance regime in
  low-energy {NO-He} collisions},\ }\href
  {https://doi.org/10.1126/science.aba3990} {\bibfield  {journal} {\bibinfo
  {journal} {Science}\ }\textbf {\bibinfo {volume} {368}},\ \bibinfo {pages}
  {626} (\bibinfo {year} {2020})}\BibitemShut {NoStop}%
\bibitem [{\citenamefont {Maussang}\ \emph {et~al.}(2005)\citenamefont
  {Maussang}, \citenamefont {Egorov}, \citenamefont {Helton}, \citenamefont
  {Nguyen},\ and\ \citenamefont {Doyle}}]{Maussang2005}%
  \BibitemOpen
  \bibfield  {author} {\bibinfo {author} {\bibfnamefont {K.}~\bibnamefont
  {Maussang}}, \bibinfo {author} {\bibfnamefont {D.}~\bibnamefont {Egorov}},
  \bibinfo {author} {\bibfnamefont {J.~S.}\ \bibnamefont {Helton}}, \bibinfo
  {author} {\bibfnamefont {S.~V.}\ \bibnamefont {Nguyen}},\ and\ \bibinfo
  {author} {\bibfnamefont {J.~M.}\ \bibnamefont {Doyle}},\ }\bibfield  {title}
  {\bibinfo {title} {Zeeman relaxation of caf in low-temperature collisions
  with helium},\ }\href {https://doi.org/10.1103/PhysRevLett.94.123002}
  {\bibfield  {journal} {\bibinfo  {journal} {Physical Review Letters}\
  }\textbf {\bibinfo {volume} {94}},\ \bibinfo {pages} {123002} (\bibinfo
  {year} {2005})}\BibitemShut {NoStop}%
\bibitem [{\citenamefont {Lara}\ \emph {et~al.}(2006)\citenamefont {Lara},
  \citenamefont {Bohn}, \citenamefont {Potter}, \citenamefont {Sold\'an},\ and\
  \citenamefont {Hutson}}]{Lara:06}%
  \BibitemOpen
  \bibfield  {author} {\bibinfo {author} {\bibfnamefont {M.}~\bibnamefont
  {Lara}}, \bibinfo {author} {\bibfnamefont {J.~L.}\ \bibnamefont {Bohn}},
  \bibinfo {author} {\bibfnamefont {D.}~\bibnamefont {Potter}}, \bibinfo
  {author} {\bibfnamefont {P.}~\bibnamefont {Sold\'an}},\ and\ \bibinfo
  {author} {\bibfnamefont {J.~M.}\ \bibnamefont {Hutson}},\ }\bibfield  {title}
  {\bibinfo {title} {Ultracold {Rb-OH} collisions and prospects for sympathetic
  cooling},\ }\href {https://doi.org/10.1103/PhysRevLett.97.183201} {\bibfield
  {journal} {\bibinfo  {journal} {Physical Review Letters}\ }\textbf {\bibinfo
  {volume} {97}},\ \bibinfo {pages} {183201} (\bibinfo {year}
  {2006})}\BibitemShut {NoStop}%
\bibitem [{\citenamefont {Morita}\ \emph {et~al.}(2018)\citenamefont {Morita},
  \citenamefont {Kosicki}, \citenamefont {\ifmmode~\dot{Z}\else
  \.{Z}\fi{}uchowski},\ and\ \citenamefont {Tscherbul}}]{Morita:18}%
  \BibitemOpen
  \bibfield  {author} {\bibinfo {author} {\bibfnamefont {M.}~\bibnamefont
  {Morita}}, \bibinfo {author} {\bibfnamefont {M.~B.}\ \bibnamefont {Kosicki}},
  \bibinfo {author} {\bibfnamefont {P.~S.}\ \bibnamefont {\ifmmode~\dot{Z}\else
  \.{Z}\fi{}uchowski}},\ and\ \bibinfo {author} {\bibfnamefont {T.~V.}\
  \bibnamefont {Tscherbul}},\ }\bibfield  {title} {\bibinfo {title}
  {Atom-molecule collisions, spin relaxation, and sympathetic cooling in an
  ultracold spin-polarized
  {$\mathrm{Rb}(^{2}\mathrm{S})\ensuremath{-}\mathrm{SrF}(^{2}\mathrm{\ensuremath{\Sigma}}^{+})$}
  mixture},\ }\href {https://doi.org/10.1103/PhysRevA.98.042702} {\bibfield
  {journal} {\bibinfo  {journal} {Physical Review A}\ }\textbf {\bibinfo
  {volume} {98}},\ \bibinfo {pages} {042702} (\bibinfo {year}
  {2018})}\BibitemShut {NoStop}%
\bibitem [{\citenamefont {Son}\ \emph {et~al.}(2020)\citenamefont {Son},
  \citenamefont {Park}, \citenamefont {Ketterle},\ and\ \citenamefont
  {Jamison}}]{son2020collisional}%
  \BibitemOpen
  \bibfield  {author} {\bibinfo {author} {\bibfnamefont {H.}~\bibnamefont
  {Son}}, \bibinfo {author} {\bibfnamefont {J.~J.}\ \bibnamefont {Park}},
  \bibinfo {author} {\bibfnamefont {W.}~\bibnamefont {Ketterle}},\ and\
  \bibinfo {author} {\bibfnamefont {A.~O.}\ \bibnamefont {Jamison}},\
  }\bibfield  {title} {\bibinfo {title} {Collisional cooling of ultracold
  molecules},\ }\href
  {https://doi.org/https://doi.org/10.1038/s41586-020-2141-z} {\bibfield
  {journal} {\bibinfo  {journal} {Nature}\ }\textbf {\bibinfo {volume} {580}},\
  \bibinfo {pages} {197} (\bibinfo {year} {2020})}\BibitemShut {NoStop}%
\bibitem [{\citenamefont {Jurgilas}\ \emph {et~al.}(2021)\citenamefont
  {Jurgilas}, \citenamefont {Chakraborty}, \citenamefont {Rich}, \citenamefont
  {Caldwell}, \citenamefont {Williams}, \citenamefont {Fitch}, \citenamefont
  {Sauer}, \citenamefont {Frye}, \citenamefont {Hutson},\ and\ \citenamefont
  {Tarbutt}}]{Jurgilas2021}%
  \BibitemOpen
  \bibfield  {author} {\bibinfo {author} {\bibfnamefont {S.}~\bibnamefont
  {Jurgilas}}, \bibinfo {author} {\bibfnamefont {A.}~\bibnamefont
  {Chakraborty}}, \bibinfo {author} {\bibfnamefont {C.~J.~H.}\ \bibnamefont
  {Rich}}, \bibinfo {author} {\bibfnamefont {L.}~\bibnamefont {Caldwell}},
  \bibinfo {author} {\bibfnamefont {H.~J.}\ \bibnamefont {Williams}}, \bibinfo
  {author} {\bibfnamefont {N.~J.}\ \bibnamefont {Fitch}}, \bibinfo {author}
  {\bibfnamefont {B.~E.}\ \bibnamefont {Sauer}}, \bibinfo {author}
  {\bibfnamefont {M.~D.}\ \bibnamefont {Frye}}, \bibinfo {author}
  {\bibfnamefont {J.~M.}\ \bibnamefont {Hutson}},\ and\ \bibinfo {author}
  {\bibfnamefont {M.~R.}\ \bibnamefont {Tarbutt}},\ }\bibfield  {title}
  {\bibinfo {title} {Collisions between ultracold molecules and atoms in a
  magnetic trap},\ }\href {https://doi.org/10.1103/PhysRevLett.126.153401}
  {\bibfield  {journal} {\bibinfo  {journal} {Physical Review Letters}\
  }\textbf {\bibinfo {volume} {126}},\ \bibinfo {pages} {153401} (\bibinfo
  {year} {2021})}\BibitemShut {NoStop}%
\bibitem [{\citenamefont {Walker}\ and\ \citenamefont
  {Happer}(1997)}]{Walker:97}%
  \BibitemOpen
  \bibfield  {author} {\bibinfo {author} {\bibfnamefont {T.~G.}\ \bibnamefont
  {Walker}}\ and\ \bibinfo {author} {\bibfnamefont {W.}~\bibnamefont
  {Happer}},\ }\bibfield  {title} {\bibinfo {title} {Spin-exchange optical
  pumping of noble-gas nuclei},\ }\href
  {https://doi.org/10.1103/RevModPhys.69.629} {\bibfield  {journal} {\bibinfo
  {journal} {Review of Modern Physics}\ }\textbf {\bibinfo {volume} {69}},\
  \bibinfo {pages} {629} (\bibinfo {year} {1997})}\BibitemShut {NoStop}%
\bibitem [{\citenamefont {Tscherbul}\ \emph {et~al.}(2011)\citenamefont
  {Tscherbul}, \citenamefont {Zhang}, \citenamefont {Sadeghpour},\ and\
  \citenamefont {Dalgarno}}]{tscherbul2011anisotropic}%
  \BibitemOpen
  \bibfield  {author} {\bibinfo {author} {\bibfnamefont {T.~V.}\ \bibnamefont
  {Tscherbul}}, \bibinfo {author} {\bibfnamefont {P.}~\bibnamefont {Zhang}},
  \bibinfo {author} {\bibfnamefont {H.~R.}\ \bibnamefont {Sadeghpour}},\ and\
  \bibinfo {author} {\bibfnamefont {A.}~\bibnamefont {Dalgarno}},\ }\bibfield
  {title} {\bibinfo {title} {Anisotropic hyperfine interactions limit the
  efficiency of spin-exchange optical pumping of {$^{3}\mathrm{He}$} nuclei},\
  }\href {https://doi.org/10.1103/PhysRevLett.107.023204} {\bibfield  {journal}
  {\bibinfo  {journal} {Physical Review Letters}\ }\textbf {\bibinfo {volume}
  {107}},\ \bibinfo {pages} {023204} (\bibinfo {year} {2011})}\BibitemShut
  {NoStop}%
\bibitem [{\citenamefont {Wang}\ \emph {et~al.}(2021)\citenamefont {Wang},
  \citenamefont {Frye}, \citenamefont {Su}, \citenamefont {Cao}, \citenamefont
  {Liu}, \citenamefont {Zhang}, \citenamefont {Yang}, \citenamefont {Hutson},
  \citenamefont {Zhao}, \citenamefont {Bai} \emph {et~al.}}]{wang2021magnetic}%
  \BibitemOpen
  \bibfield  {author} {\bibinfo {author} {\bibfnamefont {X.-Y.}\ \bibnamefont
  {Wang}}, \bibinfo {author} {\bibfnamefont {M.~D.}\ \bibnamefont {Frye}},
  \bibinfo {author} {\bibfnamefont {Z.}~\bibnamefont {Su}}, \bibinfo {author}
  {\bibfnamefont {J.}~\bibnamefont {Cao}}, \bibinfo {author} {\bibfnamefont
  {L.}~\bibnamefont {Liu}}, \bibinfo {author} {\bibfnamefont {D.-C.}\
  \bibnamefont {Zhang}}, \bibinfo {author} {\bibfnamefont {H.}~\bibnamefont
  {Yang}}, \bibinfo {author} {\bibfnamefont {J.~M.}\ \bibnamefont {Hutson}},
  \bibinfo {author} {\bibfnamefont {B.}~\bibnamefont {Zhao}}, \bibinfo {author}
  {\bibfnamefont {C.-L.}\ \bibnamefont {Bai}}, \emph {et~al.},\ }\bibfield
  {title} {\bibinfo {title} {Magnetic feshbach resonances in collisions of
  $^{23}${Na}$^{40}${K} with $^{40}${K}},\ }\href
  {https://doi.org/10.1088/1367-2630/ac3318} {\bibfield  {journal} {\bibinfo
  {journal} {New Journal of Physics}\ }\textbf {\bibinfo {volume} {23}},\
  \bibinfo {pages} {115010} (\bibinfo {year} {2021})}\BibitemShut {NoStop}%
\bibitem [{\citenamefont {Park}\ \emph
  {et~al.}(2023{\natexlab{a}})\citenamefont {Park}, \citenamefont {Son},
  \citenamefont {Lu}, \citenamefont {Karman}, \citenamefont {Gronowski},
  \citenamefont {Tomza}, \citenamefont {Jamison},\ and\ \citenamefont
  {Ketterle}}]{park2023spectrum}%
  \BibitemOpen
  \bibfield  {author} {\bibinfo {author} {\bibfnamefont {J.~J.}\ \bibnamefont
  {Park}}, \bibinfo {author} {\bibfnamefont {H.}~\bibnamefont {Son}}, \bibinfo
  {author} {\bibfnamefont {Y.-K.}\ \bibnamefont {Lu}}, \bibinfo {author}
  {\bibfnamefont {T.}~\bibnamefont {Karman}}, \bibinfo {author} {\bibfnamefont
  {M.}~\bibnamefont {Gronowski}}, \bibinfo {author} {\bibfnamefont
  {M.}~\bibnamefont {Tomza}}, \bibinfo {author} {\bibfnamefont {A.~O.}\
  \bibnamefont {Jamison}},\ and\ \bibinfo {author} {\bibfnamefont
  {W.}~\bibnamefont {Ketterle}},\ }\bibfield  {title} {\bibinfo {title}
  {Spectrum of {Feshbach} resonances in $\mathrm{Na}\mathrm{Li}+\mathrm{Na}$
  collisions},\ }\href {https://doi.org/10.1103/PhysRevX.13.031018} {\bibfield
  {journal} {\bibinfo  {journal} {Physical Review X}\ }\textbf {\bibinfo
  {volume} {13}},\ \bibinfo {pages} {031018} (\bibinfo {year}
  {2023}{\natexlab{a}})}\BibitemShut {NoStop}%
\bibitem [{\citenamefont {Bird}\ \emph {et~al.}(2023)\citenamefont {Bird},
  \citenamefont {Tarbutt},\ and\ \citenamefont {Hutson}}]{bird2023tunable}%
  \BibitemOpen
  \bibfield  {author} {\bibinfo {author} {\bibfnamefont {R.~C.}\ \bibnamefont
  {Bird}}, \bibinfo {author} {\bibfnamefont {M.~R.}\ \bibnamefont {Tarbutt}},\
  and\ \bibinfo {author} {\bibfnamefont {J.~M.}\ \bibnamefont {Hutson}},\
  }\bibfield  {title} {\bibinfo {title} {Tunable feshbach resonances in
  collisions of ultracold molecules in {${}^{2}\mathrm{\ensuremath{\Sigma}}$}
  states with alkali-metal atoms},\ }\href
  {https://doi.org/10.1103/PhysRevResearch.5.023184} {\bibfield  {journal}
  {\bibinfo  {journal} {Physical Review Research}\ }\textbf {\bibinfo {volume}
  {5}},\ \bibinfo {pages} {023184} (\bibinfo {year} {2023})}\BibitemShut
  {NoStop}%
\bibitem [{\citenamefont {Park}\ \emph
  {et~al.}(2023{\natexlab{b}})\citenamefont {Park}, \citenamefont {Lu},
  \citenamefont {Jamison}, \citenamefont {Tscherbul},\ and\ \citenamefont
  {Ketterle}}]{park2023feshbach}%
  \BibitemOpen
  \bibfield  {author} {\bibinfo {author} {\bibfnamefont {J.~J.}\ \bibnamefont
  {Park}}, \bibinfo {author} {\bibfnamefont {Y.-K.}\ \bibnamefont {Lu}},
  \bibinfo {author} {\bibfnamefont {A.~O.}\ \bibnamefont {Jamison}}, \bibinfo
  {author} {\bibfnamefont {T.~V.}\ \bibnamefont {Tscherbul}},\ and\ \bibinfo
  {author} {\bibfnamefont {W.}~\bibnamefont {Ketterle}},\ }\bibfield  {title}
  {\bibinfo {title} {A {Feshbach resonance} in collisions between triplet
  ground-state molecules},\ }\href {https://doi.org/10.1038/s41586-022-05635-8}
  {\bibfield  {journal} {\bibinfo  {journal} {Nature}\ }\textbf {\bibinfo
  {volume} {614}},\ \bibinfo {pages} {54} (\bibinfo {year}
  {2023}{\natexlab{b}})}\BibitemShut {NoStop}%
\bibitem [{\citenamefont {Yang}\ \emph {et~al.}(2022)\citenamefont {Yang},
  \citenamefont {Cao}, \citenamefont {Su}, \citenamefont {Rui}, \citenamefont
  {Zhao},\ and\ \citenamefont {Pan}}]{yang2022creation}%
  \BibitemOpen
  \bibfield  {author} {\bibinfo {author} {\bibfnamefont {H.}~\bibnamefont
  {Yang}}, \bibinfo {author} {\bibfnamefont {J.}~\bibnamefont {Cao}}, \bibinfo
  {author} {\bibfnamefont {Z.}~\bibnamefont {Su}}, \bibinfo {author}
  {\bibfnamefont {J.}~\bibnamefont {Rui}}, \bibinfo {author} {\bibfnamefont
  {B.}~\bibnamefont {Zhao}},\ and\ \bibinfo {author} {\bibfnamefont {J.-W.}\
  \bibnamefont {Pan}},\ }\bibfield  {title} {\bibinfo {title} {Creation of an
  ultracold gas of triatomic molecules from an atom--diatomic molecule
  mixture},\ }\href {https://doi.org/10.1126/science.ade6307} {\bibfield
  {journal} {\bibinfo  {journal} {Science}\ }\textbf {\bibinfo {volume}
  {378}},\ \bibinfo {pages} {1009} (\bibinfo {year} {2022})}\BibitemShut
  {NoStop}%
\bibitem [{\citenamefont {Tizniti}\ \emph {et~al.}(2014)\citenamefont
  {Tizniti}, \citenamefont {Picard}, \citenamefont {Lique}, \citenamefont
  {Berteloite}, \citenamefont {Canosa}, \citenamefont {Alexander},\ and\
  \citenamefont {Sims}}]{Tizniti:14}%
  \BibitemOpen
  \bibfield  {author} {\bibinfo {author} {\bibfnamefont {M.}~\bibnamefont
  {Tizniti}}, \bibinfo {author} {\bibfnamefont {S.~D.~L.}\ \bibnamefont
  {Picard}}, \bibinfo {author} {\bibfnamefont {F.}~\bibnamefont {Lique}},
  \bibinfo {author} {\bibfnamefont {C.}~\bibnamefont {Berteloite}}, \bibinfo
  {author} {\bibfnamefont {A.}~\bibnamefont {Canosa}}, \bibinfo {author}
  {\bibfnamefont {M.~H.}\ \bibnamefont {Alexander}},\ and\ \bibinfo {author}
  {\bibfnamefont {I.~R.}\ \bibnamefont {Sims}},\ }\bibfield  {title} {\bibinfo
  {title} {The rate of the {F~+~H$_2$} reaction at very low temperatures},\
  }\href {https://doi.org/https://doi.org/10.1038/nchem.1835} {\bibfield
  {journal} {\bibinfo  {journal} {Nature Chemistry}\ }\textbf {\bibinfo
  {volume} {6}},\ \bibinfo {pages} {141} (\bibinfo {year} {2014})}\BibitemShut
  {NoStop}%
\bibitem [{\citenamefont {Yang}\ \emph {et~al.}(2019)\citenamefont {Yang},
  \citenamefont {Huang}, \citenamefont {Xiao}, \citenamefont {Chen},
  \citenamefont {Wang}, \citenamefont {Dai}, \citenamefont {Lique},
  \citenamefont {Alexander}, \citenamefont {Sun}, \citenamefont {Zhang},
  \citenamefont {Yang},\ and\ \citenamefont {Neumark}}]{Yang:19}%
  \BibitemOpen
  \bibfield  {author} {\bibinfo {author} {\bibfnamefont {T.}~\bibnamefont
  {Yang}}, \bibinfo {author} {\bibfnamefont {L.}~\bibnamefont {Huang}},
  \bibinfo {author} {\bibfnamefont {C.}~\bibnamefont {Xiao}}, \bibinfo {author}
  {\bibfnamefont {J.}~\bibnamefont {Chen}}, \bibinfo {author} {\bibfnamefont
  {T.}~\bibnamefont {Wang}}, \bibinfo {author} {\bibfnamefont {D.}~\bibnamefont
  {Dai}}, \bibinfo {author} {\bibfnamefont {F.}~\bibnamefont {Lique}}, \bibinfo
  {author} {\bibfnamefont {M.~H.}\ \bibnamefont {Alexander}}, \bibinfo {author}
  {\bibfnamefont {Z.}~\bibnamefont {Sun}}, \bibinfo {author} {\bibfnamefont
  {D.~H.}\ \bibnamefont {Zhang}}, \bibinfo {author} {\bibfnamefont
  {X.}~\bibnamefont {Yang}},\ and\ \bibinfo {author} {\bibfnamefont {D.~M.}\
  \bibnamefont {Neumark}},\ }\bibfield  {title} {\bibinfo {title} {Enhanced
  reactivity of fluorine with para-hydrogen in cold interstellar clouds by
  resonance-induced quantum tunnelling},\ }\href
  {https://doi.org/10.1038/s41557-019-0280-3} {\bibfield  {journal} {\bibinfo
  {journal} {Nature Chemistry}\ }\textbf {\bibinfo {volume} {11}},\ \bibinfo
  {pages} {744} (\bibinfo {year} {2019})}\BibitemShut {NoStop}%
\bibitem [{\citenamefont {Mayle}\ \emph {et~al.}(2012)\citenamefont {Mayle},
  \citenamefont {Ruzic},\ and\ \citenamefont {Bohn}}]{mayle2012statistical}%
  \BibitemOpen
  \bibfield  {author} {\bibinfo {author} {\bibfnamefont {M.}~\bibnamefont
  {Mayle}}, \bibinfo {author} {\bibfnamefont {B.~P.}\ \bibnamefont {Ruzic}},\
  and\ \bibinfo {author} {\bibfnamefont {J.~L.}\ \bibnamefont {Bohn}},\
  }\bibfield  {title} {\bibinfo {title} {Statistical aspects of ultracold
  resonant scattering},\ }\href {https://doi.org/10.1103/PhysRevA.85.062712}
  {\bibfield  {journal} {\bibinfo  {journal} {Physical Review A}\ }\textbf
  {\bibinfo {volume} {85}},\ \bibinfo {pages} {062712} (\bibinfo {year}
  {2012})}\BibitemShut {NoStop}%
\bibitem [{\citenamefont {Mayle}\ \emph {et~al.}(2013)\citenamefont {Mayle},
  \citenamefont {Qu{\'e}m{\'e}ner}, \citenamefont {Ruzic},\ and\ \citenamefont
  {Bohn}}]{mayle2013scattering}%
  \BibitemOpen
  \bibfield  {author} {\bibinfo {author} {\bibfnamefont {M.}~\bibnamefont
  {Mayle}}, \bibinfo {author} {\bibfnamefont {G.}~\bibnamefont
  {Qu{\'e}m{\'e}ner}}, \bibinfo {author} {\bibfnamefont {B.~P.}\ \bibnamefont
  {Ruzic}},\ and\ \bibinfo {author} {\bibfnamefont {J.~L.}\ \bibnamefont
  {Bohn}},\ }\bibfield  {title} {\bibinfo {title} {Scattering of ultracold
  molecules in the highly resonant regime},\ }\href@noop {} {\bibfield
  {journal} {\bibinfo  {journal} {Physical Review A}\ }\textbf {\bibinfo
  {volume} {87}},\ \bibinfo {pages} {012709} (\bibinfo {year}
  {2013})}\BibitemShut {NoStop}%
\bibitem [{\citenamefont {Christianen}\ \emph {et~al.}(2019)\citenamefont
  {Christianen}, \citenamefont {Karman},\ and\ \citenamefont
  {Groenenboom}}]{Christianen2019}%
  \BibitemOpen
  \bibfield  {author} {\bibinfo {author} {\bibfnamefont {A.}~\bibnamefont
  {Christianen}}, \bibinfo {author} {\bibfnamefont {T.}~\bibnamefont
  {Karman}},\ and\ \bibinfo {author} {\bibfnamefont {G.~C.}\ \bibnamefont
  {Groenenboom}},\ }\bibfield  {title} {\bibinfo {title} {Quasiclassical method
  for calculating the density of states of ultracold collision complexes},\
  }\href {https://doi.org/10.1103/PhysRevA.100.032708} {\bibfield  {journal}
  {\bibinfo  {journal} {Physical Review A}\ }\textbf {\bibinfo {volume}
  {100}},\ \bibinfo {pages} {032708} (\bibinfo {year} {2019})}\BibitemShut
  {NoStop}%
\bibitem [{\citenamefont {Liu}\ \emph {et~al.}(2020{\natexlab{a}})\citenamefont
  {Liu}, \citenamefont {Hu}, \citenamefont {Nichols}, \citenamefont {Grimes},
  \citenamefont {Karman}, \citenamefont {Guo},\ and\ \citenamefont
  {Ni}}]{liu2020photo}%
  \BibitemOpen
  \bibfield  {author} {\bibinfo {author} {\bibfnamefont {Y.}~\bibnamefont
  {Liu}}, \bibinfo {author} {\bibfnamefont {M.-G.}\ \bibnamefont {Hu}},
  \bibinfo {author} {\bibfnamefont {M.~A.}\ \bibnamefont {Nichols}}, \bibinfo
  {author} {\bibfnamefont {D.~D.}\ \bibnamefont {Grimes}}, \bibinfo {author}
  {\bibfnamefont {T.}~\bibnamefont {Karman}}, \bibinfo {author} {\bibfnamefont
  {H.}~\bibnamefont {Guo}},\ and\ \bibinfo {author} {\bibfnamefont {K.-K.}\
  \bibnamefont {Ni}},\ }\bibfield  {title} {\bibinfo {title} {Photo-excitation
  of long-lived transient intermediates in ultracold reactions},\ }\href
  {https://doi.org/https://doi.org/10.1038/s41567-020-0968-8} {\bibfield
  {journal} {\bibinfo  {journal} {Nature Physics}\ }\textbf {\bibinfo {volume}
  {16}},\ \bibinfo {pages} {1132} (\bibinfo {year}
  {2020}{\natexlab{a}})}\BibitemShut {NoStop}%
\bibitem [{\citenamefont {Gregory}\ \emph {et~al.}(2020)\citenamefont
  {Gregory}, \citenamefont {Blackmore}, \citenamefont {Bromley},\ and\
  \citenamefont {Cornish}}]{gregory2020loss}%
  \BibitemOpen
  \bibfield  {author} {\bibinfo {author} {\bibfnamefont {P.~D.}\ \bibnamefont
  {Gregory}}, \bibinfo {author} {\bibfnamefont {J.~A.}\ \bibnamefont
  {Blackmore}}, \bibinfo {author} {\bibfnamefont {S.~L.}\ \bibnamefont
  {Bromley}},\ and\ \bibinfo {author} {\bibfnamefont {S.~L.}\ \bibnamefont
  {Cornish}},\ }\bibfield  {title} {\bibinfo {title} {Loss of ultracold
  $^{87}\mathrm{Rb}^{133}\mathrm{Cs}$ molecules via optical excitation of
  long-lived two-body collision complexes},\ }\href
  {https://doi.org/10.1103/PhysRevLett.124.163402} {\bibfield  {journal}
  {\bibinfo  {journal} {Physical Review Letters}\ }\textbf {\bibinfo {volume}
  {124}},\ \bibinfo {pages} {163402} (\bibinfo {year} {2020})}\BibitemShut
  {NoStop}%
\bibitem [{\citenamefont {Gersema}\ \emph {et~al.}(2021)\citenamefont
  {Gersema}, \citenamefont {Voges}, \citenamefont {Meyer~zum Alten~Borgloh},
  \citenamefont {Koch}, \citenamefont {Hartmann}, \citenamefont {Zenesini},
  \citenamefont {Ospelkaus}, \citenamefont {Lin}, \citenamefont {He},\ and\
  \citenamefont {Wang}}]{Gersema2021}%
  \BibitemOpen
  \bibfield  {author} {\bibinfo {author} {\bibfnamefont {P.}~\bibnamefont
  {Gersema}}, \bibinfo {author} {\bibfnamefont {K.~K.}\ \bibnamefont {Voges}},
  \bibinfo {author} {\bibfnamefont {M.}~\bibnamefont {Meyer~zum
  Alten~Borgloh}}, \bibinfo {author} {\bibfnamefont {L.}~\bibnamefont {Koch}},
  \bibinfo {author} {\bibfnamefont {T.}~\bibnamefont {Hartmann}}, \bibinfo
  {author} {\bibfnamefont {A.}~\bibnamefont {Zenesini}}, \bibinfo {author}
  {\bibfnamefont {S.}~\bibnamefont {Ospelkaus}}, \bibinfo {author}
  {\bibfnamefont {J.}~\bibnamefont {Lin}}, \bibinfo {author} {\bibfnamefont
  {J.}~\bibnamefont {He}},\ and\ \bibinfo {author} {\bibfnamefont
  {D.}~\bibnamefont {Wang}},\ }\bibfield  {title} {\bibinfo {title} {Probing
  photoinduced two-body loss of ultracold nonreactive bosonic
  $^{23}\mathrm{Na}^{87}\mathrm{Rb}$ and $^{23}\mathrm{Na}^{39}\mathrm{K}$
  molecules},\ }\href {https://doi.org/10.1103/PhysRevLett.127.163401}
  {\bibfield  {journal} {\bibinfo  {journal} {Physical Review Letters}\
  }\textbf {\bibinfo {volume} {127}},\ \bibinfo {pages} {163401} (\bibinfo
  {year} {2021})}\BibitemShut {NoStop}%
\bibitem [{\citenamefont {Bause}\ \emph {et~al.}(2021)\citenamefont {Bause},
  \citenamefont {Schindewolf}, \citenamefont {Tao}, \citenamefont {Duda},
  \citenamefont {Chen}, \citenamefont {Qu\'em\'ener}, \citenamefont {Karman},
  \citenamefont {Christianen}, \citenamefont {Bloch},\ and\ \citenamefont
  {Luo}}]{Bause2021collisions}%
  \BibitemOpen
  \bibfield  {author} {\bibinfo {author} {\bibfnamefont {R.}~\bibnamefont
  {Bause}}, \bibinfo {author} {\bibfnamefont {A.}~\bibnamefont {Schindewolf}},
  \bibinfo {author} {\bibfnamefont {R.}~\bibnamefont {Tao}}, \bibinfo {author}
  {\bibfnamefont {M.}~\bibnamefont {Duda}}, \bibinfo {author} {\bibfnamefont
  {X.-Y.}\ \bibnamefont {Chen}}, \bibinfo {author} {\bibfnamefont
  {G.}~\bibnamefont {Qu\'em\'ener}}, \bibinfo {author} {\bibfnamefont
  {T.}~\bibnamefont {Karman}}, \bibinfo {author} {\bibfnamefont
  {A.}~\bibnamefont {Christianen}}, \bibinfo {author} {\bibfnamefont
  {I.}~\bibnamefont {Bloch}},\ and\ \bibinfo {author} {\bibfnamefont {X.-Y.}\
  \bibnamefont {Luo}},\ }\bibfield  {title} {\bibinfo {title} {Collisions of
  ultracold molecules in bright and dark optical dipole traps},\ }\href
  {https://doi.org/10.1103/PhysRevResearch.3.033013} {\bibfield  {journal}
  {\bibinfo  {journal} {Physical Review Research}\ }\textbf {\bibinfo {volume}
  {3}},\ \bibinfo {pages} {033013} (\bibinfo {year} {2021})}\BibitemShut
  {NoStop}%
\bibitem [{\citenamefont {Nichols}\ \emph {et~al.}(2022)\citenamefont
  {Nichols}, \citenamefont {Liu}, \citenamefont {Zhu}, \citenamefont {Hu},
  \citenamefont {Liu},\ and\ \citenamefont {Ni}}]{nichols2022detection}%
  \BibitemOpen
  \bibfield  {author} {\bibinfo {author} {\bibfnamefont {M.~A.}\ \bibnamefont
  {Nichols}}, \bibinfo {author} {\bibfnamefont {Y.-X.}\ \bibnamefont {Liu}},
  \bibinfo {author} {\bibfnamefont {L.}~\bibnamefont {Zhu}}, \bibinfo {author}
  {\bibfnamefont {M.-G.}\ \bibnamefont {Hu}}, \bibinfo {author} {\bibfnamefont
  {Y.}~\bibnamefont {Liu}},\ and\ \bibinfo {author} {\bibfnamefont {K.-K.}\
  \bibnamefont {Ni}},\ }\bibfield  {title} {\bibinfo {title} {Detection of
  long-lived complexes in ultracold atom-molecule collisions},\ }\href
  {https://doi.org/10.1103/PhysRevX.12.011049} {\bibfield  {journal} {\bibinfo
  {journal} {Physical Review X}\ }\textbf {\bibinfo {volume} {12}},\ \bibinfo
  {pages} {011049} (\bibinfo {year} {2022})}\BibitemShut {NoStop}%
\bibitem [{\citenamefont {Bause}\ \emph {et~al.}(2023)\citenamefont {Bause},
  \citenamefont {Christianen}, \citenamefont {Schindewolf}, \citenamefont
  {Bloch},\ and\ \citenamefont {Luo}}]{bause2023ultracold}%
  \BibitemOpen
  \bibfield  {author} {\bibinfo {author} {\bibfnamefont {R.}~\bibnamefont
  {Bause}}, \bibinfo {author} {\bibfnamefont {A.}~\bibnamefont {Christianen}},
  \bibinfo {author} {\bibfnamefont {A.}~\bibnamefont {Schindewolf}}, \bibinfo
  {author} {\bibfnamefont {I.}~\bibnamefont {Bloch}},\ and\ \bibinfo {author}
  {\bibfnamefont {X.-Y.}\ \bibnamefont {Luo}},\ }\bibfield  {title} {\bibinfo
  {title} {Ultracold sticky collisions: Theoretical and experimental status},\
  }\href {https://doi.org/https://doi.org/10.1021/acs.jpca.2c08095} {\bibfield
  {journal} {\bibinfo  {journal} {The Journal of Physical Chemistry A}\
  }\textbf {\bibinfo {volume} {127}},\ \bibinfo {pages} {729} (\bibinfo {year}
  {2023})}\BibitemShut {NoStop}%
\bibitem [{\citenamefont {Jachymski}\ \emph {et~al.}(2022)\citenamefont
  {Jachymski}, \citenamefont {Gronowski},\ and\ \citenamefont
  {Tomza}}]{jachymski2022collisional}%
  \BibitemOpen
  \bibfield  {author} {\bibinfo {author} {\bibfnamefont {K.}~\bibnamefont
  {Jachymski}}, \bibinfo {author} {\bibfnamefont {M.}~\bibnamefont
  {Gronowski}},\ and\ \bibinfo {author} {\bibfnamefont {M.}~\bibnamefont
  {Tomza}},\ }\bibfield  {title} {\bibinfo {title} {Collisional losses of
  ultracold molecules due to intermediate complex formation},\ }\href
  {https://doi.org/10.1103/PhysRevA.106.L041301} {\bibfield  {journal}
  {\bibinfo  {journal} {Physical Review A}\ }\textbf {\bibinfo {volume}
  {106}},\ \bibinfo {pages} {L041301} (\bibinfo {year} {2022})}\BibitemShut
  {NoStop}%
\bibitem [{\citenamefont {Tscherbul}\ and\ \citenamefont
  {D'Incao}(2023)}]{Tscherbul:23}%
  \BibitemOpen
  \bibfield  {author} {\bibinfo {author} {\bibfnamefont {T.~V.}\ \bibnamefont
  {Tscherbul}}\ and\ \bibinfo {author} {\bibfnamefont {J.~P.}\ \bibnamefont
  {D'Incao}},\ }\bibfield  {title} {\bibinfo {title} {Ultracold molecular
  collisions in magnetic fields: {Efficient} incorporation of hyperfine
  structure in the total rotational angular momentum representation},\ }\href
  {https://doi.org/10.1103/PhysRevA.108.053317} {\bibfield  {journal} {\bibinfo
   {journal} {Physical Review A}\ }\textbf {\bibinfo {volume} {108}},\ \bibinfo
  {pages} {053317} (\bibinfo {year} {2023})}\BibitemShut {NoStop}%
\bibitem [{\citenamefont {Herzberg}(1950)}]{Herzberg}%
  \BibitemOpen
  \bibfield  {author} {\bibinfo {author} {\bibfnamefont {G.}~\bibnamefont
  {Herzberg}},\ }\href@noop {} {\emph {\bibinfo {title} {Spectra of Diatomic
  Molecules}}}\ (\bibinfo  {publisher} {D.Van Nostrand Company, Inc},\ \bibinfo
  {year} {1950})\BibitemShut {NoStop}%
\bibitem [{\citenamefont {Ni}\ \emph {et~al.}(2008)\citenamefont {Ni},
  \citenamefont {Ospelkaus}, \citenamefont {De~Miranda}, \citenamefont {Pe'Er},
  \citenamefont {Neyenhuis}, \citenamefont {Zirbel}, \citenamefont
  {Kotochigova}, \citenamefont {Julienne}, \citenamefont {Jin},\ and\
  \citenamefont {Ye}}]{ni2008high}%
  \BibitemOpen
  \bibfield  {author} {\bibinfo {author} {\bibfnamefont {K.-K.}\ \bibnamefont
  {Ni}}, \bibinfo {author} {\bibfnamefont {S.}~\bibnamefont {Ospelkaus}},
  \bibinfo {author} {\bibfnamefont {M.}~\bibnamefont {De~Miranda}}, \bibinfo
  {author} {\bibfnamefont {A.}~\bibnamefont {Pe'Er}}, \bibinfo {author}
  {\bibfnamefont {B.}~\bibnamefont {Neyenhuis}}, \bibinfo {author}
  {\bibfnamefont {J.}~\bibnamefont {Zirbel}}, \bibinfo {author} {\bibfnamefont
  {S.}~\bibnamefont {Kotochigova}}, \bibinfo {author} {\bibfnamefont
  {P.}~\bibnamefont {Julienne}}, \bibinfo {author} {\bibfnamefont
  {D.}~\bibnamefont {Jin}},\ and\ \bibinfo {author} {\bibfnamefont
  {J.}~\bibnamefont {Ye}},\ }\bibfield  {title} {\bibinfo {title} {A high
  phase-space-density gas of polar molecules},\ }\href
  {https://doi.org/10.1126/science.1163861} {\bibfield  {journal} {\bibinfo
  {journal} {Science}\ }\textbf {\bibinfo {volume} {322}},\ \bibinfo {pages}
  {231} (\bibinfo {year} {2008})}\BibitemShut {NoStop}%
\bibitem [{\citenamefont {Hu}\ \emph {et~al.}(2019)\citenamefont {Hu},
  \citenamefont {Liu}, \citenamefont {Grimes}, \citenamefont {Lin},
  \citenamefont {Gheorghe}, \citenamefont {Vexiau}, \citenamefont
  {Bouloufa-Maafa}, \citenamefont {Dulieu}, \citenamefont {Rosenband},\ and\
  \citenamefont {Ni}}]{hu2019direct}%
  \BibitemOpen
  \bibfield  {author} {\bibinfo {author} {\bibfnamefont {M.-G.}\ \bibnamefont
  {Hu}}, \bibinfo {author} {\bibfnamefont {Y.}~\bibnamefont {Liu}}, \bibinfo
  {author} {\bibfnamefont {D.~D.}\ \bibnamefont {Grimes}}, \bibinfo {author}
  {\bibfnamefont {Y.-W.}\ \bibnamefont {Lin}}, \bibinfo {author} {\bibfnamefont
  {A.~H.}\ \bibnamefont {Gheorghe}}, \bibinfo {author} {\bibfnamefont
  {R.}~\bibnamefont {Vexiau}}, \bibinfo {author} {\bibfnamefont
  {N.}~\bibnamefont {Bouloufa-Maafa}}, \bibinfo {author} {\bibfnamefont
  {O.}~\bibnamefont {Dulieu}}, \bibinfo {author} {\bibfnamefont
  {T.}~\bibnamefont {Rosenband}},\ and\ \bibinfo {author} {\bibfnamefont
  {K.-K.}\ \bibnamefont {Ni}},\ }\bibfield  {title} {\bibinfo {title} {Direct
  observation of bimolecular reactions of ultracold {KRb} molecules},\ }\href
  {https://doi.org/10.1126/science.aay9531} {\bibfield  {journal} {\bibinfo
  {journal} {Science}\ }\textbf {\bibinfo {volume} {366}},\ \bibinfo {pages}
  {1111} (\bibinfo {year} {2019})}\BibitemShut {NoStop}%
\bibitem [{\citenamefont {Schweiger}\ and\ \citenamefont
  {Jeschke}(2001)}]{SchweigerBook}%
  \BibitemOpen
  \bibfield  {author} {\bibinfo {author} {\bibfnamefont {A.}~\bibnamefont
  {Schweiger}}\ and\ \bibinfo {author} {\bibfnamefont {G.}~\bibnamefont
  {Jeschke}},\ }\href@noop {} {\emph {\bibinfo {title} {Principles of Pulse
  Electron Paramagnetic Resonance}}}\ (\bibinfo  {publisher} {Cambridge
  University Press},\ \bibinfo {year} {2001})\BibitemShut {NoStop}%
\bibitem [{\citenamefont {Aldegunde}\ and\ \citenamefont
  {Hutson}(2017)}]{AldegundePRA2017}%
  \BibitemOpen
  \bibfield  {author} {\bibinfo {author} {\bibfnamefont {J.}~\bibnamefont
  {Aldegunde}}\ and\ \bibinfo {author} {\bibfnamefont {J.~M.}\ \bibnamefont
  {Hutson}},\ }\bibfield  {title} {\bibinfo {title} {Hyperfine structure of
  alkali-metal diatomic molecules},\ }\href
  {https://doi.org/10.1103/PhysRevA.96.042506} {\bibfield  {journal} {\bibinfo
  {journal} {Physical Review A}\ }\textbf {\bibinfo {volume} {96}},\ \bibinfo
  {pages} {042506} (\bibinfo {year} {2017})}\BibitemShut {NoStop}%
\bibitem [{\citenamefont {Croft}\ \emph {et~al.}(2017)\citenamefont {Croft},
  \citenamefont {Makrides}, \citenamefont {Li}, \citenamefont {Petrov},
  \citenamefont {Kendrick}, \citenamefont {Balakrishnan},\ and\ \citenamefont
  {Kotochigova}}]{croft2017universality}%
  \BibitemOpen
  \bibfield  {author} {\bibinfo {author} {\bibfnamefont {J.}~\bibnamefont
  {Croft}}, \bibinfo {author} {\bibfnamefont {C.}~\bibnamefont {Makrides}},
  \bibinfo {author} {\bibfnamefont {M.}~\bibnamefont {Li}}, \bibinfo {author}
  {\bibfnamefont {A.}~\bibnamefont {Petrov}}, \bibinfo {author} {\bibfnamefont
  {B.}~\bibnamefont {Kendrick}}, \bibinfo {author} {\bibfnamefont
  {N.}~\bibnamefont {Balakrishnan}},\ and\ \bibinfo {author} {\bibfnamefont
  {S.}~\bibnamefont {Kotochigova}},\ }\bibfield  {title} {\bibinfo {title}
  {Universality and chaoticity in ultracold {K} + {KRb} chemical reactions},\
  }\href {https://doi.org/10.1038/ncomms15897} {\bibfield  {journal} {\bibinfo
  {journal} {Nature communications}\ }\textbf {\bibinfo {volume} {8}},\
  \bibinfo {pages} {15897} (\bibinfo {year} {2017})}\BibitemShut {NoStop}%
\bibitem [{\citenamefont {Kendrick}\ \emph {et~al.}(2021)\citenamefont
  {Kendrick}, \citenamefont {Li}, \citenamefont {Li}, \citenamefont
  {Kotochigova}, \citenamefont {Croft},\ and\ \citenamefont
  {Balakrishnan}}]{Kendrick:21}%
  \BibitemOpen
  \bibfield  {author} {\bibinfo {author} {\bibfnamefont {B.~K.}\ \bibnamefont
  {Kendrick}}, \bibinfo {author} {\bibfnamefont {H.}~\bibnamefont {Li}},
  \bibinfo {author} {\bibfnamefont {M.}~\bibnamefont {Li}}, \bibinfo {author}
  {\bibfnamefont {S.}~\bibnamefont {Kotochigova}}, \bibinfo {author}
  {\bibfnamefont {J.~F.~E.}\ \bibnamefont {Croft}},\ and\ \bibinfo {author}
  {\bibfnamefont {N.}~\bibnamefont {Balakrishnan}},\ }\bibfield  {title}
  {\bibinfo {title} {Non-adiabatic quantum interference in the ultracold
  {Li~+~LiNa~$\to$~Li$_2$~+~Na} reaction},\ }\href
  {https://doi.org/10.1039/D0CP05499B} {\bibfield  {journal} {\bibinfo
  {journal} {Physical Chemistry Chemical Physics}\ }\textbf {\bibinfo {volume}
  {23}},\ \bibinfo {pages} {5096} (\bibinfo {year} {2021})}\BibitemShut
  {NoStop}%
\bibitem [{\citenamefont {Morita}\ \emph {et~al.}(2019)\citenamefont {Morita},
  \citenamefont {Krems},\ and\ \citenamefont {Tscherbul}}]{Morita:19}%
  \BibitemOpen
  \bibfield  {author} {\bibinfo {author} {\bibfnamefont {M.}~\bibnamefont
  {Morita}}, \bibinfo {author} {\bibfnamefont {R.~V.}\ \bibnamefont {Krems}},\
  and\ \bibinfo {author} {\bibfnamefont {T.~V.}\ \bibnamefont {Tscherbul}},\
  }\bibfield  {title} {\bibinfo {title} {Universal probability distributions of
  scattering observables in ultracold molecular collisions},\ }\href
  {https://doi.org/10.1103/PhysRevLett.123.013401} {\bibfield  {journal}
  {\bibinfo  {journal} {Physical Review Letters}\ }\textbf {\bibinfo {volume}
  {123}},\ \bibinfo {pages} {013401} (\bibinfo {year} {2019})}\BibitemShut
  {NoStop}%
\bibitem [{\citenamefont {Majorana}(1932)}]{Majorana32}%
  \BibitemOpen
  \bibfield  {author} {\bibinfo {author} {\bibfnamefont {E.}~\bibnamefont
  {Majorana}},\ }\bibfield  {title} {\bibinfo {title} {Atomi orientati in campo
  magnetico variabile},\ }\href {https://doi.org/10.1007/BF02960953} {\bibfield
   {journal} {\bibinfo  {journal} {Il Nuovo Cimento}\ ,\ \bibinfo {pages} {43}}
  (\bibinfo {year} {1932})}\BibitemShut {NoStop}%
\bibitem [{\citenamefont {Morita}\ \emph {et~al.}(2023)\citenamefont {Morita},
  \citenamefont {Kendrick}, \citenamefont {Kłos}, \citenamefont {Kotochigova},
  \citenamefont {Brumer},\ and\ \citenamefont {Tscherbul}}]{morita2023non}%
  \BibitemOpen
  \bibfield  {author} {\bibinfo {author} {\bibfnamefont {M.}~\bibnamefont
  {Morita}}, \bibinfo {author} {\bibfnamefont {B.~K.}\ \bibnamefont
  {Kendrick}}, \bibinfo {author} {\bibfnamefont {J.}~\bibnamefont {Kłos}},
  \bibinfo {author} {\bibfnamefont {S.}~\bibnamefont {Kotochigova}}, \bibinfo
  {author} {\bibfnamefont {P.}~\bibnamefont {Brumer}},\ and\ \bibinfo {author}
  {\bibfnamefont {T.~V.}\ \bibnamefont {Tscherbul}},\ }\bibfield  {title}
  {\bibinfo {title} {Signatures of non-universal quantum dynamics of ultracold
  chemical reactions of polar alkali dimer molecules with alkali metal atoms:
  {Li($^2$S) + NaLi($a^3\Sigma^+$) → Na($^2$S) + Li$_2$($a^3\Sigma_u^+$)}},\
  }\href {https://doi.org/10.1021/acs.jpclett.3c00159} {\bibfield  {journal}
  {\bibinfo  {journal} {The Journal of Physical Chemistry Letters}\ }\textbf
  {\bibinfo {volume} {14}},\ \bibinfo {pages} {3413} (\bibinfo {year}
  {2023})}\BibitemShut {NoStop}%
\bibitem [{\citenamefont {Liu}\ \emph {et~al.}(2020{\natexlab{b}})\citenamefont
  {Liu}, \citenamefont {Grimes}, \citenamefont {Hu},\ and\ \citenamefont
  {Ni}}]{liu2020probing}%
  \BibitemOpen
  \bibfield  {author} {\bibinfo {author} {\bibfnamefont {Y.}~\bibnamefont
  {Liu}}, \bibinfo {author} {\bibfnamefont {D.~D.}\ \bibnamefont {Grimes}},
  \bibinfo {author} {\bibfnamefont {M.-G.}\ \bibnamefont {Hu}},\ and\ \bibinfo
  {author} {\bibfnamefont {K.-K.}\ \bibnamefont {Ni}},\ }\bibfield  {title}
  {\bibinfo {title} {Probing ultracold chemistry using ion spectrometry},\
  }\href {https://doi.org/10.1039/C9CP07015J} {\bibfield  {journal} {\bibinfo
  {journal} {Physical Chemistry Chemical Physics}\ }\textbf {\bibinfo {volume}
  {22}},\ \bibinfo {pages} {4861} (\bibinfo {year}
  {2020}{\natexlab{b}})}\BibitemShut {NoStop}%
\bibitem [{\citenamefont {Borsalino}\ \emph {et~al.}(2014)\citenamefont
  {Borsalino}, \citenamefont {Londo\~no Flor\`ez}, \citenamefont {Vexiau},
  \citenamefont {Dulieu}, \citenamefont {Bouloufa-Maafa},\ and\ \citenamefont
  {Luc-Koenig}}]{borsalino2014efficient}%
  \BibitemOpen
  \bibfield  {author} {\bibinfo {author} {\bibfnamefont {D.}~\bibnamefont
  {Borsalino}}, \bibinfo {author} {\bibfnamefont {B.}~\bibnamefont {Londo\~no
  Flor\`ez}}, \bibinfo {author} {\bibfnamefont {R.}~\bibnamefont {Vexiau}},
  \bibinfo {author} {\bibfnamefont {O.}~\bibnamefont {Dulieu}}, \bibinfo
  {author} {\bibfnamefont {N.}~\bibnamefont {Bouloufa-Maafa}},\ and\ \bibinfo
  {author} {\bibfnamefont {E.}~\bibnamefont {Luc-Koenig}},\ }\bibfield  {title}
  {\bibinfo {title} {Efficient optical schemes to create ultracold {KRb}
  molecules in their rovibronic ground state},\ }\href
  {https://doi.org/10.1103/PhysRevA.90.033413} {\bibfield  {journal} {\bibinfo
  {journal} {Physical Review A}\ }\textbf {\bibinfo {volume} {90}},\ \bibinfo
  {pages} {033413} (\bibinfo {year} {2014})}\BibitemShut {NoStop}%
\bibitem [{\citenamefont {Liu}\ \emph {et~al.}(2021)\citenamefont {Liu},
  \citenamefont {Hu}, \citenamefont {Nichols}, \citenamefont {Yang},
  \citenamefont {Xie}, \citenamefont {Guo},\ and\ \citenamefont
  {Ni}}]{liu2021precision}%
  \BibitemOpen
  \bibfield  {author} {\bibinfo {author} {\bibfnamefont {Y.}~\bibnamefont
  {Liu}}, \bibinfo {author} {\bibfnamefont {M.-G.}\ \bibnamefont {Hu}},
  \bibinfo {author} {\bibfnamefont {M.~A.}\ \bibnamefont {Nichols}}, \bibinfo
  {author} {\bibfnamefont {D.}~\bibnamefont {Yang}}, \bibinfo {author}
  {\bibfnamefont {D.}~\bibnamefont {Xie}}, \bibinfo {author} {\bibfnamefont
  {H.}~\bibnamefont {Guo}},\ and\ \bibinfo {author} {\bibfnamefont {K.-K.}\
  \bibnamefont {Ni}},\ }\bibfield  {title} {\bibinfo {title} {Precision test of
  statistical dynamics with state-to-state ultracold chemistry},\ }\href
  {https://doi.org/10.1038/s41586-021-03459-6} {\bibfield  {journal} {\bibinfo
  {journal} {Nature}\ }\textbf {\bibinfo {volume} {593}},\ \bibinfo {pages}
  {379} (\bibinfo {year} {2021})}\BibitemShut {NoStop}%
\bibitem [{\citenamefont {Hansson}\ and\ \citenamefont
  {Watson}(2005)}]{hansson2005comment}%
  \BibitemOpen
  \bibfield  {author} {\bibinfo {author} {\bibfnamefont {A.}~\bibnamefont
  {Hansson}}\ and\ \bibinfo {author} {\bibfnamefont {J.~K.}\ \bibnamefont
  {Watson}},\ }\bibfield  {title} {\bibinfo {title} {A comment on
  {H{\"o}nl-London} factors},\ }\href
  {https://doi.org/10.1016/j.jms.2005.06.009} {\bibfield  {journal} {\bibinfo
  {journal} {Journal of Molecular Spectroscopy}\ }\textbf {\bibinfo {volume}
  {233}},\ \bibinfo {pages} {169} (\bibinfo {year} {2005})}\BibitemShut
  {NoStop}%
\end{thebibliography}

\begin{thebibliography}{76}%

\makeatletter
\providecommand \@ifxundefined [1]{%
 \@ifx{#1\undefined}
}%
\providecommand \@ifnum [1]{%
 \ifnum #1\expandafter \@firstoftwo
 \else \expandafter \@secondoftwo
 \fi
}%
\providecommand \@ifx [1]{%
 \ifx #1\expandafter \@firstoftwo
 \else \expandafter \@secondoftwo
 \fi
}%
\providecommand \natexlab [1]{#1}%
\providecommand \enquote  [1]{``#1''}%
\providecommand \bibnamefont  [1]{#1}%
\providecommand \bibfnamefont [1]{#1}%
\providecommand \citenamefont [1]{#1}%
\providecommand \href@noop [0]{\@secondoftwo}%
\providecommand \href [0]{\begingroup \@sanitize@url \@href}%
\providecommand \@href[1]{\@@startlink{#1}\@@href}%
\providecommand \@@href[1]{\endgroup#1\@@endlink}%
\providecommand \@sanitize@url [0]{\catcode `\\12\catcode `\$12\catcode
  `\&12\catcode `\#12\catcode `\^12\catcode `\_12\catcode `\%12\relax}%
\providecommand \@@startlink[1]{}%
\providecommand \@@endlink[0]{}%
\providecommand \url  [0]{\begingroup\@sanitize@url \@url }%
\providecommand \@url [1]{\endgroup\@href {#1}{\urlprefix }}%
\providecommand \urlprefix  [0]{URL }%
\providecommand \Eprint [0]{\href }%
\providecommand \doibase [0]{https://doi.org/}%
\providecommand \selectlanguage [0]{\@gobble}%
\providecommand \bibinfo  [0]{\@secondoftwo}%
\providecommand \bibfield  [0]{\@secondoftwo}%
\providecommand \translation [1]{[#1]}%
\providecommand \BibitemOpen [0]{}%
\providecommand \bibitemStop [0]{}%
\providecommand \bibitemNoStop [0]{.\EOS\space}%
\providecommand \EOS [0]{\spacefactor3000\relax}%
\providecommand \BibitemShut  [1]{\csname bibitem#1\endcsname}%
\let\auto@bib@innerbib\@empty


\bibitem [{\citenamefont {Pashov}\ \emph {et~al.}(2007)\citenamefont {Pashov},
  \citenamefont {Docenko}, \citenamefont {Tamanis}, \citenamefont {Ferber},
  \citenamefont {Kn\"ockel},\ and\ \citenamefont {Tiemann}}]{PashovPRA2007}%
  \BibitemOpen
  \bibfield  {author} {\bibinfo {author} {\bibfnamefont {A.}~\bibnamefont
  {Pashov}}, \bibinfo {author} {\bibfnamefont {O.}~\bibnamefont {Docenko}},
  \bibinfo {author} {\bibfnamefont {M.}~\bibnamefont {Tamanis}}, \bibinfo
  {author} {\bibfnamefont {R.}~\bibnamefont {Ferber}}, \bibinfo {author}
  {\bibfnamefont {H.}~\bibnamefont {Kn\"ockel}},\ and\ \bibinfo {author}
  {\bibfnamefont {E.}~\bibnamefont {Tiemann}},\ }\bibfield  {title} {\bibinfo
  {title} {Coupling of the ${X}^{1}{\Sigma}^{+}$ and $a^{3}{\Sigma}^{+}$ states
  of {KRb}},\ }\href {https://doi.org/10.1103/PhysRevA.76.022511} {\bibfield
  {journal} {\bibinfo  {journal} {Physical Review A}\ }\textbf {\bibinfo
  {volume} {76}},\ \bibinfo {pages} {022511} (\bibinfo {year}
  {2007})}\BibitemShut {NoStop}%
\bibitem [{\citenamefont {Knowles}\ and\ \citenamefont
  {Werner}(1992)}]{KnowlesTCA1992}%
  \BibitemOpen
  \bibfield  {author} {\bibinfo {author} {\bibfnamefont {P.~J.}\ \bibnamefont
  {Knowles}}\ and\ \bibinfo {author} {\bibfnamefont {H.-J.}\ \bibnamefont
  {Werner}},\ }\bibfield  {title} {\bibinfo {title} {Internally contracted
  multiconfiguration-reference configuration interaction calculations for
  excited states},\ }\href {https://doi.org/10.1007/BF01117405} {\bibfield
  {journal} {\bibinfo  {journal} {Theoretica Chimica Acta}\ }\textbf {\bibinfo
  {volume} {84}},\ \bibinfo {pages} {95} (\bibinfo {year} {1992})}\BibitemShut
  {NoStop}%
\bibitem [{\citenamefont {{Werner}}\ and\ \citenamefont
  {{Knowles}}(1988)}]{WernerJCP1988}%
  \BibitemOpen
  \bibfield  {author} {\bibinfo {author} {\bibfnamefont {H.-J.}\ \bibnamefont
  {{Werner}}}\ and\ \bibinfo {author} {\bibfnamefont {P.~J.}\ \bibnamefont
  {{Knowles}}},\ }\bibfield  {title} {\bibinfo {title} {{An efficient
  internally contracted multiconfiguration-reference configuration interaction
  method}},\ }\href {https://doi.org/10.1063/1.455556} {\bibfield  {journal}
  {\bibinfo  {journal} {Journal of Chemical Physics}\ }\textbf {\bibinfo
  {volume} {89}},\ \bibinfo {pages} {5803} (\bibinfo {year}
  {1988})}\BibitemShut {NoStop}%
\bibitem [{\citenamefont {Werner}\ \emph {et~al.}(2012)\citenamefont {Werner},
  \citenamefont {Knowles}, \citenamefont {Knizia}, \citenamefont {Manby},\ and\
  \citenamefont {Schütz}}]{WernerWIREs2012}%
  \BibitemOpen
  \bibfield  {author} {\bibinfo {author} {\bibfnamefont {H.-J.}\ \bibnamefont
  {Werner}}, \bibinfo {author} {\bibfnamefont {P.~J.}\ \bibnamefont {Knowles}},
  \bibinfo {author} {\bibfnamefont {G.}~\bibnamefont {Knizia}}, \bibinfo
  {author} {\bibfnamefont {F.~R.}\ \bibnamefont {Manby}},\ and\ \bibinfo
  {author} {\bibfnamefont {M.}~\bibnamefont {Schütz}},\ }\bibfield  {title}
  {\bibinfo {title} {Molpro: a general-purpose quantum chemistry program
  package},\ }\href {https://doi.org/10.1002/wcms.82} {\bibfield  {journal}
  {\bibinfo  {journal} {Wiley Interdisciplinary Reviews: Computational
  Molecular Science}\ }\textbf {\bibinfo {volume} {2}},\ \bibinfo {pages} {242}
  (\bibinfo {year} {2012})}\BibitemShut {NoStop}%
\bibitem [{\citenamefont {{Werner}}\ \emph {et~al.}(2020)\citenamefont
  {{Werner}}, \citenamefont {{Knowles}}, \citenamefont {{Manby}}, \citenamefont
  {{Black}}, \citenamefont {{Doll}}, \citenamefont {{He{\ss}elmann}},
  \citenamefont {{Kats}}, \citenamefont {{K{\"o}hn}}, \citenamefont {{Korona}},
  \citenamefont {{Kreplin}}, \citenamefont {{Ma}}, \citenamefont {{Miller}},
  \citenamefont {{Mitrushchenkov}}, \citenamefont {{Peterson}}, \citenamefont
  {{Polyak}}, \citenamefont {{Rauhut}},\ and\ \citenamefont
  {{Sibaev}}}]{WernerJCP2020}%
  \BibitemOpen
  \bibfield  {author} {\bibinfo {author} {\bibfnamefont {H.-J.}\ \bibnamefont
  {{Werner}}}, \bibinfo {author} {\bibfnamefont {P.~J.}\ \bibnamefont
  {{Knowles}}}, \bibinfo {author} {\bibfnamefont {F.~R.}\ \bibnamefont
  {{Manby}}}, \bibinfo {author} {\bibfnamefont {J.~A.}\ \bibnamefont
  {{Black}}}, \bibinfo {author} {\bibfnamefont {K.}~\bibnamefont {{Doll}}},
  \bibinfo {author} {\bibfnamefont {A.}~\bibnamefont {{He{\ss}elmann}}},
  \bibinfo {author} {\bibfnamefont {D.}~\bibnamefont {{Kats}}}, \bibinfo
  {author} {\bibfnamefont {A.}~\bibnamefont {{K{\"o}hn}}}, \bibinfo {author}
  {\bibfnamefont {T.}~\bibnamefont {{Korona}}}, \bibinfo {author}
  {\bibfnamefont {D.~A.}\ \bibnamefont {{Kreplin}}}, \bibinfo {author}
  {\bibfnamefont {Q.}~\bibnamefont {{Ma}}}, \bibinfo {author} {\bibfnamefont
  {I.}~\bibnamefont {{Miller}}, \bibfnamefont {Thomas~F.}}, \bibinfo {author}
  {\bibfnamefont {A.}~\bibnamefont {{Mitrushchenkov}}}, \bibinfo {author}
  {\bibfnamefont {K.~A.}\ \bibnamefont {{Peterson}}}, \bibinfo {author}
  {\bibfnamefont {I.}~\bibnamefont {{Polyak}}}, \bibinfo {author}
  {\bibfnamefont {G.}~\bibnamefont {{Rauhut}}},\ and\ \bibinfo {author}
  {\bibfnamefont {M.}~\bibnamefont {{Sibaev}}},\ }\bibfield  {title} {\bibinfo
  {title} {{The Molpro quantum chemistry package}},\ }\href
  {https://doi.org/10.1063/5.0005081} {\bibfield  {journal} {\bibinfo
  {journal} {The Journal of Chemical Physics}\ }\textbf {\bibinfo {volume}
  {152}},\ \bibinfo {eid} {144107} (\bibinfo {year} {2020})}\BibitemShut
  {NoStop}%
\bibitem [{\citenamefont {Werner}\ \emph {et~al.}(2019)\citenamefont {Werner},
  \citenamefont {Knowles} \emph {et~al.}}]{MOLPROv2019v2}%
  \BibitemOpen
  \bibfield  {author} {\bibinfo {author} {\bibfnamefont {H.-J.}\ \bibnamefont
  {Werner}}, \bibinfo {author} {\bibfnamefont {P.~J.}\ \bibnamefont {Knowles}},
  \emph {et~al.},\ }\href@noop {} {\bibinfo {title} {Molpro, 2019.2 , a package
  of ab initio programs}} (\bibinfo {year} {2019}),\ \bibinfo {note} {see
  https://www.molpro.net}\BibitemShut {NoStop}%
\bibitem [{\citenamefont {Stevens}\ \emph {et~al.}(1992)\citenamefont
  {Stevens}, \citenamefont {Krauss}, \citenamefont {Basch},\ and\ \citenamefont
  {Jasien}}]{StevensCJCh1992}%
  \BibitemOpen
  \bibfield  {author} {\bibinfo {author} {\bibfnamefont {W.~J.}\ \bibnamefont
  {Stevens}}, \bibinfo {author} {\bibfnamefont {M.}~\bibnamefont {Krauss}},
  \bibinfo {author} {\bibfnamefont {H.}~\bibnamefont {Basch}},\ and\ \bibinfo
  {author} {\bibfnamefont {P.~G.}\ \bibnamefont {Jasien}},\ }\bibfield  {title}
  {\bibinfo {title} {Relativistic compact effective potentials and efficient,
  shared-exponent basis sets for the third-, fourth-, and fifth-row atoms},\
  }\href {https://doi.org/10.1139/v92-085} {\bibfield  {journal} {\bibinfo
  {journal} {Canadian Journal of Chemistry}\ }\textbf {\bibinfo {volume}
  {70}},\ \bibinfo {pages} {612} (\bibinfo {year} {1992})}\BibitemShut
  {NoStop}%
\bibitem [{\citenamefont {{Ladjimi}}\ \emph {et~al.}(2020)\citenamefont
  {{Ladjimi}}, \citenamefont {{Zrafi}}, \citenamefont {{Allouche}},\ and\
  \citenamefont {{Berriche}}}]{LadjimiJQSRT2020}%
  \BibitemOpen
  \bibfield  {author} {\bibinfo {author} {\bibfnamefont {H.}~\bibnamefont
  {{Ladjimi}}}, \bibinfo {author} {\bibfnamefont {W.}~\bibnamefont {{Zrafi}}},
  \bibinfo {author} {\bibfnamefont {A.-R.}\ \bibnamefont {{Allouche}}},\ and\
  \bibinfo {author} {\bibfnamefont {H.}~\bibnamefont {{Berriche}}},\ }\bibfield
   {title} {\bibinfo {title} {{Ab-initio study of the ground and low-lying
  excited states including the spin-orbit effect of RbBa molecule and laser
  cooling feasibility}},\ }\href {https://doi.org/10.1016/j.jqsrt.2020.107069}
  {\bibfield  {journal} {\bibinfo  {journal} {Journal of Quantitative
  Spectroscopy \& Radiative Transfer}\ }\textbf {\bibinfo {volume} {252}},\
  \bibinfo {eid} {107069} (\bibinfo {year} {2020})}\BibitemShut {NoStop}%
\bibitem [{\citenamefont {{Ganyushin}}\ and\ \citenamefont
  {{Neese}}(2013)}]{GanyushinJCP2013}%
  \BibitemOpen
  \bibfield  {author} {\bibinfo {author} {\bibfnamefont {D.}~\bibnamefont
  {{Ganyushin}}}\ and\ \bibinfo {author} {\bibfnamefont {F.}~\bibnamefont
  {{Neese}}},\ }\bibfield  {title} {\bibinfo {title} {{A fully variational
  spin-orbit coupled complete active space self-consistent field approach:
  Application to electron paramagnetic resonance g-tensors}},\ }\href
  {https://doi.org/10.1063/1.4793736} {\bibfield  {journal} {\bibinfo
  {journal} {The Journal of Chemical Physics}\ }\textbf {\bibinfo {volume}
  {138}},\ \bibinfo {pages} {104113} (\bibinfo {year} {2013})}\BibitemShut
  {NoStop}%
\bibitem [{\citenamefont {{Neese}}\ \emph {et~al.}(2020)\citenamefont
  {{Neese}}, \citenamefont {{Wennmohs}}, \citenamefont {{Becker}},\ and\
  \citenamefont {{Riplinger}}}]{NeeseJCP2020}%
  \BibitemOpen
  \bibfield  {author} {\bibinfo {author} {\bibfnamefont {F.}~\bibnamefont
  {{Neese}}}, \bibinfo {author} {\bibfnamefont {F.}~\bibnamefont {{Wennmohs}}},
  \bibinfo {author} {\bibfnamefont {U.}~\bibnamefont {{Becker}}},\ and\
  \bibinfo {author} {\bibfnamefont {C.}~\bibnamefont {{Riplinger}}},\
  }\bibfield  {title} {\bibinfo {title} {{The ORCA quantum chemistry program
  package}},\ }\href {https://doi.org/10.1063/5.0004608} {\bibfield  {journal}
  {\bibinfo  {journal} {The Journal of Chemical Physics}\ }\textbf {\bibinfo
  {volume} {152}},\ \bibinfo {eid} {224108} (\bibinfo {year}
  {2020})}\BibitemShut {NoStop}%
\bibitem [{\citenamefont {Neese}(2018)}]{NeeseWIREs2018}%
  \BibitemOpen
  \bibfield  {author} {\bibinfo {author} {\bibfnamefont {F.}~\bibnamefont
  {Neese}},\ }\bibfield  {title} {\bibinfo {title} {Software update: the orca
  program system, version 4.0},\ }\href {https://doi.org/10.1002/wcms.1327}
  {\bibfield  {journal} {\bibinfo  {journal} {Wiley Interdisciplinary Reviews:
  Computational Molecular Science}\ }\textbf {\bibinfo {volume} {8}},\ \bibinfo
  {pages} {e1327} (\bibinfo {year} {2018})}\BibitemShut {NoStop}%
\bibitem [{\citenamefont {{Sandhoefer}}\ and\ \citenamefont
  {{Neese}}(2012)}]{SandhoeferJCP2012}%
  \BibitemOpen
  \bibfield  {author} {\bibinfo {author} {\bibfnamefont {B.}~\bibnamefont
  {{Sandhoefer}}}\ and\ \bibinfo {author} {\bibfnamefont {F.}~\bibnamefont
  {{Neese}}},\ }\bibfield  {title} {\bibinfo {title} {{One-electron
  contributions to the g-tensor for second-order Douglas-Kroll-Hess theory}},\
  }\href {https://doi.org/10.1063/1.4747454} {\bibfield  {journal} {\bibinfo
  {journal} {The Journal of Chemical Physics}\ }\textbf {\bibinfo {volume}
  {137}},\ \bibinfo {pages} {094102} (\bibinfo {year} {2012})}\BibitemShut
  {NoStop}%
\bibitem [{\citenamefont {Franzke}\ \emph {et~al.}(2020)\citenamefont
  {Franzke}, \citenamefont {Spiske}, \citenamefont {Pollak},\ and\
  \citenamefont {Weigend}}]{FranzkeJCTC2020}%
  \BibitemOpen
  \bibfield  {author} {\bibinfo {author} {\bibfnamefont {Y.~J.}\ \bibnamefont
  {Franzke}}, \bibinfo {author} {\bibfnamefont {L.}~\bibnamefont {Spiske}},
  \bibinfo {author} {\bibfnamefont {P.}~\bibnamefont {Pollak}},\ and\ \bibinfo
  {author} {\bibfnamefont {F.}~\bibnamefont {Weigend}},\ }\bibfield  {title}
  {\bibinfo {title} {Segmented contracted error-consistent basis sets of
  quadruple-{$\zeta$} valence quality for one- and two-component relativistic
  all-electron calculations},\ }\href
  {https://doi.org/10.1021/acs.jctc.0c00546} {\bibfield  {journal} {\bibinfo
  {journal} {Canadian Journal of Chemistry}\ }\textbf {\bibinfo {volume}
  {16}},\ \bibinfo {pages} {5658} (\bibinfo {year} {2020})}\BibitemShut
  {NoStop}%
\bibitem [{\citenamefont {Curl}(1965)}]{CurlMolPhys1965}%
  \BibitemOpen
  \bibfield  {author} {\bibinfo {author} {\bibfnamefont {R.}~\bibnamefont
  {Curl}},\ }\bibfield  {title} {\bibinfo {title} {The relationship between
  electron spin rotation coupling constants and g-tensor components},\ }\href
  {https://doi.org/10.1080/00268976500100761} {\bibfield  {journal} {\bibinfo
  {journal} {Molecular Physics}\ }\textbf {\bibinfo {volume} {9}},\ \bibinfo
  {pages} {585} (\bibinfo {year} {1965})}\BibitemShut {NoStop}%
\bibitem [{\citenamefont {Saitow}\ and\ \citenamefont
  {Neese}(2018)}]{SaitowJCP2018}%
  \BibitemOpen
  \bibfield  {author} {\bibinfo {author} {\bibfnamefont {M.}~\bibnamefont
  {Saitow}}\ and\ \bibinfo {author} {\bibfnamefont {F.}~\bibnamefont {Neese}},\
  }\bibfield  {title} {\bibinfo {title} {{Accurate spin-densities based on the
  domain-based local pair-natural orbital coupled-cluster theory}},\ }\href
  {https://doi.org/10.1063/1.5027114} {\bibfield  {journal} {\bibinfo
  {journal} {Journal of Chemical Physics}\ }\textbf {\bibinfo {volume} {149}},\
  \bibinfo {pages} {034104} (\bibinfo {year} {2018})}\BibitemShut {NoStop}%
\bibitem [{\citenamefont {Aldegunde}\ \emph {et~al.}(2008)\citenamefont
  {Aldegunde}, \citenamefont {Rivington}, \citenamefont {\ifmmode~\dot{Z}\else
  \.{Z}\fi{}uchowski},\ and\ \citenamefont {Hutson}}]{Aldegunde:08}%
  \BibitemOpen
  \bibfield  {author} {\bibinfo {author} {\bibfnamefont {J.}~\bibnamefont
  {Aldegunde}}, \bibinfo {author} {\bibfnamefont {B.~A.}\ \bibnamefont
  {Rivington}}, \bibinfo {author} {\bibfnamefont {P.~S.}\ \bibnamefont
  {\ifmmode~\dot{Z}\else \.{Z}\fi{}uchowski}},\ and\ \bibinfo {author}
  {\bibfnamefont {J.~M.}\ \bibnamefont {Hutson}},\ }\bibfield  {title}
  {\bibinfo {title} {Hyperfine energy levels of alkali-metal dimers:
  Ground-state polar molecules in electric and magnetic fields},\ }\href
  {https://doi.org/10.1103/PhysRevA.78.033434} {\bibfield  {journal} {\bibinfo
  {journal} {Physical Review A}\ }\textbf {\bibinfo {volume} {78}},\ \bibinfo
  {pages} {033434} (\bibinfo {year} {2008})}\BibitemShut {NoStop}%
\bibitem [{\citenamefont {Brown}\ and\ \citenamefont
  {Carrington}(2003)}]{Carrington:03}%
  \BibitemOpen
  \bibfield  {author} {\bibinfo {author} {\bibfnamefont {J.~M.}\ \bibnamefont
  {Brown}}\ and\ \bibinfo {author} {\bibfnamefont {A.}~\bibnamefont
  {Carrington}},\ }\href@noop {} {\emph {\bibinfo {title} {Rotational
  spectroscopy of diatomic molecules}}}\ (\bibinfo  {publisher} {Cambridge
  university press},\ \bibinfo {year} {2003})\BibitemShut {NoStop}%
\bibitem [{\citenamefont {Johnson}(1973)}]{Johnson:73}%
  \BibitemOpen
  \bibfield  {author} {\bibinfo {author} {\bibfnamefont {B.~R.}\ \bibnamefont
  {Johnson}},\ }\bibfield  {title} {\bibinfo {title} {The multichannel
  log-derivative method for scattering calculations},\ }\href
  {https://doi.org/10.1016/0021-9991(73)90049-1} {\bibfield  {journal}
  {\bibinfo  {journal} {Journal of Computational Physics}\ }\textbf {\bibinfo
  {volume} {13}},\ \bibinfo {pages} {445} (\bibinfo {year} {1973})}\BibitemShut
  {NoStop}%
\bibitem [{\citenamefont {Manolopoulos}(1986)}]{Manolopoulos:86}%
  \BibitemOpen
  \bibfield  {author} {\bibinfo {author} {\bibfnamefont {D.~E.}\ \bibnamefont
  {Manolopoulos}},\ }\bibfield  {title} {\bibinfo {title} {An improved log
  derivative method for inelastic scattering},\ }\href
  {https://doi.org/10.1063/1.451472} {\bibfield  {journal} {\bibinfo  {journal}
  {Journal of Chemical Physics}\ }\textbf {\bibinfo {volume} {85}},\ \bibinfo
  {pages} {6425} (\bibinfo {year} {1986})}\BibitemShut {NoStop}%
\bibitem [{\citenamefont {Zare}(1988)}]{Zare:88}%
  \BibitemOpen
  \bibfield  {author} {\bibinfo {author} {\bibfnamefont {R.~N.}\ \bibnamefont
  {Zare}},\ }\href@noop {} {\emph {\bibinfo {title} {Angular Momentum}}}\
  (\bibinfo  {publisher} {Wiley, NY},\ \bibinfo {year} {1988})\BibitemShut
  {NoStop}%
\bibitem [{\citenamefont {Idziaszek}\ and\ \citenamefont
  {Julienne}(2010)}]{Idziaszek:10}%
  \BibitemOpen
  \bibfield  {author} {\bibinfo {author} {\bibfnamefont {Z.}~\bibnamefont
  {Idziaszek}}\ and\ \bibinfo {author} {\bibfnamefont {P.~S.}\ \bibnamefont
  {Julienne}},\ }\bibfield  {title} {\bibinfo {title} {Universal rate constants
  for reactive collisions of ultracold molecules},\ }\href
  {https://doi.org/https://doi.org/10.1103/PhysRevLett.104.113202} {\bibfield
  {journal} {\bibinfo  {journal} {Physical Review Letters}\ }\textbf {\bibinfo
  {volume} {104}},\ \bibinfo {pages} {113202} (\bibinfo {year}
  {2010})}\BibitemShut {NoStop}%
\bibitem [{\citenamefont {Gorshkov}\ \emph {et~al.}(2011)\citenamefont
  {Gorshkov}, \citenamefont {Manmana}, \citenamefont {Chen}, \citenamefont
  {Demler}, \citenamefont {Lukin},\ and\ \citenamefont {Rey}}]{Gorshkov:11}%
  \BibitemOpen
  \bibfield  {author} {\bibinfo {author} {\bibfnamefont {A.~V.}\ \bibnamefont
  {Gorshkov}}, \bibinfo {author} {\bibfnamefont {S.~R.}\ \bibnamefont
  {Manmana}}, \bibinfo {author} {\bibfnamefont {G.}~\bibnamefont {Chen}},
  \bibinfo {author} {\bibfnamefont {E.}~\bibnamefont {Demler}}, \bibinfo
  {author} {\bibfnamefont {M.~D.}\ \bibnamefont {Lukin}},\ and\ \bibinfo
  {author} {\bibfnamefont {A.~M.}\ \bibnamefont {Rey}},\ }\bibfield  {title}
  {\bibinfo {title} {Quantum magnetism with polar alkali-metal dimers},\ }\href
  {https://doi.org/10.1103/PhysRevA.84.033619} {\bibfield  {journal} {\bibinfo
  {journal} {Physical Review A}\ }\textbf {\bibinfo {volume} {84}},\ \bibinfo
  {pages} {033619} (\bibinfo {year} {2011})}\BibitemShut {NoStop}%
\bibitem [{\citenamefont {Kotochigova}(2010)}]{Kotochigova:10}%
  \BibitemOpen
  \bibfield  {author} {\bibinfo {author} {\bibfnamefont {S.}~\bibnamefont
  {Kotochigova}},\ }\bibfield  {title} {\bibinfo {title} {Dispersion
  interactions and reactive collisions of ultracold polar molecules},\ }\href
  {http://dx.doi.org/10.1088/1367-2630/12/7/073041} {\bibfield  {journal}
  {\bibinfo  {journal} {New Journal of Physics}\ }\textbf {\bibinfo {volume}
  {12}},\ \bibinfo {pages} {073041} (\bibinfo {year} {2010})}\BibitemShut
  {NoStop}%
\bibitem [{\citenamefont {Karman}\ \emph {et~al.}(2023)\citenamefont {Karman},
  \citenamefont {Gronowski}, \citenamefont {Tomza}, \citenamefont {Park},
  \citenamefont {Son}, \citenamefont {Lu}, \citenamefont {Jamison},\ and\
  \citenamefont {Ketterle}}]{karman2023ab}%
  \BibitemOpen
  \bibfield  {author} {\bibinfo {author} {\bibfnamefont {T.}~\bibnamefont
  {Karman}}, \bibinfo {author} {\bibfnamefont {M.}~\bibnamefont {Gronowski}},
  \bibinfo {author} {\bibfnamefont {M.}~\bibnamefont {Tomza}}, \bibinfo
  {author} {\bibfnamefont {J.~J.}\ \bibnamefont {Park}}, \bibinfo {author}
  {\bibfnamefont {H.}~\bibnamefont {Son}}, \bibinfo {author} {\bibfnamefont
  {Y.-K.}\ \bibnamefont {Lu}}, \bibinfo {author} {\bibfnamefont {A.~O.}\
  \bibnamefont {Jamison}},\ and\ \bibinfo {author} {\bibfnamefont
  {W.}~\bibnamefont {Ketterle}},\ }\bibfield  {title} {\bibinfo {title} {Ab
  initio calculation of the spectrum of feshbach resonances in {NaLi} $+$ {Na}
  collisions},\ }\href {https://doi.org/10.1103/PhysRevA.108.023309} {\bibfield
   {journal} {\bibinfo  {journal} {Physical Review A}\ }\textbf {\bibinfo
  {volume} {108}},\ \bibinfo {pages} {023309} (\bibinfo {year}
  {2023})}\BibitemShut {NoStop}%
\bibitem [{\citenamefont {Ospelkaus}\ \emph {et~al.}(2010)\citenamefont
  {Ospelkaus}, \citenamefont {Ni}, \citenamefont {Wang}, \citenamefont
  {De~Miranda}, \citenamefont {Neyenhuis}, \citenamefont {Qu{\'e}m{\'e}ner},
  \citenamefont {Julienne}, \citenamefont {Bohn}, \citenamefont {Jin},\ and\
  \citenamefont {Ye}}]{ospelkaus2010quantum}%
  \BibitemOpen
  \bibfield  {author} {\bibinfo {author} {\bibfnamefont {S.}~\bibnamefont
  {Ospelkaus}}, \bibinfo {author} {\bibfnamefont {K.-K.}\ \bibnamefont {Ni}},
  \bibinfo {author} {\bibfnamefont {D.}~\bibnamefont {Wang}}, \bibinfo {author}
  {\bibfnamefont {M.}~\bibnamefont {De~Miranda}}, \bibinfo {author}
  {\bibfnamefont {B.}~\bibnamefont {Neyenhuis}}, \bibinfo {author}
  {\bibfnamefont {G.}~\bibnamefont {Qu{\'e}m{\'e}ner}}, \bibinfo {author}
  {\bibfnamefont {P.}~\bibnamefont {Julienne}}, \bibinfo {author}
  {\bibfnamefont {J.}~\bibnamefont {Bohn}}, \bibinfo {author} {\bibfnamefont
  {D.}~\bibnamefont {Jin}},\ and\ \bibinfo {author} {\bibfnamefont
  {J.}~\bibnamefont {Ye}},\ }\bibfield  {title} {\bibinfo {title}
  {Quantum-state controlled chemical reactions of ultracold potassium-rubidium
  molecules},\ }\href {https://doi.org/10.1126/science.1184121} {\bibfield
  {journal} {\bibinfo  {journal} {Science}\ }\textbf {\bibinfo {volume}
  {327}},\ \bibinfo {pages} {853} (\bibinfo {year} {2010})}\BibitemShut
  {NoStop}%
\bibitem [{\citenamefont {Li}\ \emph {et~al.}(2019)\citenamefont {Li},
  \citenamefont {Li}, \citenamefont {Makrides}, \citenamefont {Petrov},\ and\
  \citenamefont {Kotochigova}}]{li2019universal}%
  \BibitemOpen
  \bibfield  {author} {\bibinfo {author} {\bibfnamefont {H.}~\bibnamefont
  {Li}}, \bibinfo {author} {\bibfnamefont {M.}~\bibnamefont {Li}}, \bibinfo
  {author} {\bibfnamefont {C.}~\bibnamefont {Makrides}}, \bibinfo {author}
  {\bibfnamefont {A.}~\bibnamefont {Petrov}},\ and\ \bibinfo {author}
  {\bibfnamefont {S.}~\bibnamefont {Kotochigova}},\ }\bibfield  {title}
  {\bibinfo {title} {Universal scattering of ultracold atoms and molecules in
  optical potentials},\ }\href {https://doi.org/10.3390/atoms7010036}
  {\bibfield  {journal} {\bibinfo  {journal} {Atoms}\ }\textbf {\bibinfo
  {volume} {7}},\ \bibinfo {pages} {36} (\bibinfo {year} {2019})}\BibitemShut
  {NoStop}%
\bibitem [{\citenamefont {Frye}\ and\ \citenamefont
  {Hutson}(2021)}]{frye2021complexes}%
  \BibitemOpen
  \bibfield  {author} {\bibinfo {author} {\bibfnamefont {M.~D.}\ \bibnamefont
  {Frye}}\ and\ \bibinfo {author} {\bibfnamefont {J.~M.}\ \bibnamefont
  {Hutson}},\ }\bibfield  {title} {\bibinfo {title} {Complexes formed in
  collisions between ultracold alkali-metal diatomic molecules and atoms},\
  }\href {https://doi.org/10.1088/1367-2630/ac3ff8} {\bibfield  {journal}
  {\bibinfo  {journal} {New Journal of Physics}\ }\textbf {\bibinfo {volume}
  {23}},\ \bibinfo {pages} {125008} (\bibinfo {year} {2021})}\BibitemShut
  {NoStop}%
\end{thebibliography}

%

\clearpage
\onecolumngrid

\section*{Supplementary Materials}

\renewcommand{\thefigure}{S\arabic{figure}}

\renewcommand{\theequation}{S\arabic{equation}}

\setcounter{subsection}{0}
\setcounter{figure}{0}
\setcounter{equation}{0}
\setcounter{table}{0}

In these supplementary materials, we present detailed information on the theoretical calculations. 
We begin by describing the details of electronic structure calculations of the Rb-KRb  PES and spin-rotation interactions in Sec. A. These interactions are used as a basis for the quantum theory of Rb~+~KRb collisions in the rigid-rotor approximation, which we present in Sec.~B. This section contains the derivations of the matrix elements of the relevant intermolecular spin-dependent interactions in the Rb-KRb complex (Sec.~B1) such as the isotropic (Fermi contact) and tensor hyperfine interactions, as well as the spin-rotation  interaction. The matrix elements of the intraatomic (K) and intramolecular (KRb) hyperfine interactions are derived in Sec.~B2. Section~B3 discusses the parametrization of the {\it ab initio} Rb-KRb interactions used in CC calculations. Section B4 presents CC calculations of rotational and hyperfine branching ratios as a function of the PES scaling parameter. Finally, in Sec.~B5, we  present a mechanism for breaking  the conservation of the total rotational angular momentum $J_r$ by the spin-rotation interaction in the Rb-KRb complex.

Additionally, we include further details on the characterization of Rb-KRb inelastic collision rate, and the KRb$_2$ complex lifetime measurement.

\subsection{Details of electronic structure calculations}

We use state-of-the-art quantum-chemical methods to calculate the electronic structure of the Rb-KRb complex.

\subsubsection{Jacobi coordinates}

To describe the complex geometry, we use Jacobi coordinates $R$, $r$ and $\theta$, which are defined as follows:
\begin{itemize}
\item $R$ is the distance between the center of mass of the KRb molecule and the Rb atom,
\item $r$ is the intramonomer distance in KRb. For the ground vibrational level of KRb, its vibrationally averaged values is $r_0=7.6986$~bohr calculated from the experimental potential~\cite{PashovPRA2007}. The corresponding inner and outer classical turning points are $r_\text{in}=7.452$~bohr and $r_\text{out}=7.937$~bohr.
\item $\theta$ is the angle between the axis of the KRb molecule (oriented from Rb to K) and the axis
connecting the Rb with the center of mass of the molecular (oriented from the molecule to the atom). $\theta=0^\circ$ gives the Rb-K-Rb linear arrangement, while $\theta=180^\circ$ gives the K-Rb-Rb one.
\end{itemize}

\subsubsection{Potential energy surfaces}

The three-dimensional potential energy surfaces $V(R,r,\theta)$ for the Rb-KRb complex are computed using the multiconfiguration reference internally contracted configuration interaction method incuding single and double excitations~\cite{KnowlesTCA1992,WernerJCP1988}, as implemented in the Molpro 2019.2 packages of \textit{ab initio}
programs~\cite{WernerWIREs2012,WernerJCP2020,MOLPROv2019v2}. We obtain the orbitals in the state-averaged complete active space method with a minimal active space composed of 3 electrons and 3 orbitals. We use the large-core effective pseudopotentials SBKJC ECP18 for K and SBKJC ECP36 for Rb \cite{StevensCJCh1992} with appropriate Gaussian basis sets composed of (8s,8p,5d,3f) and (5s,5p,4d,3f) functions for Rb and K, respectively. We took the basis set for Rb from \cite{LadjimiJQSRT2020} and optimized a new basis set for K.

\begin{figure*}
\includegraphics[width=0.7\textwidth]{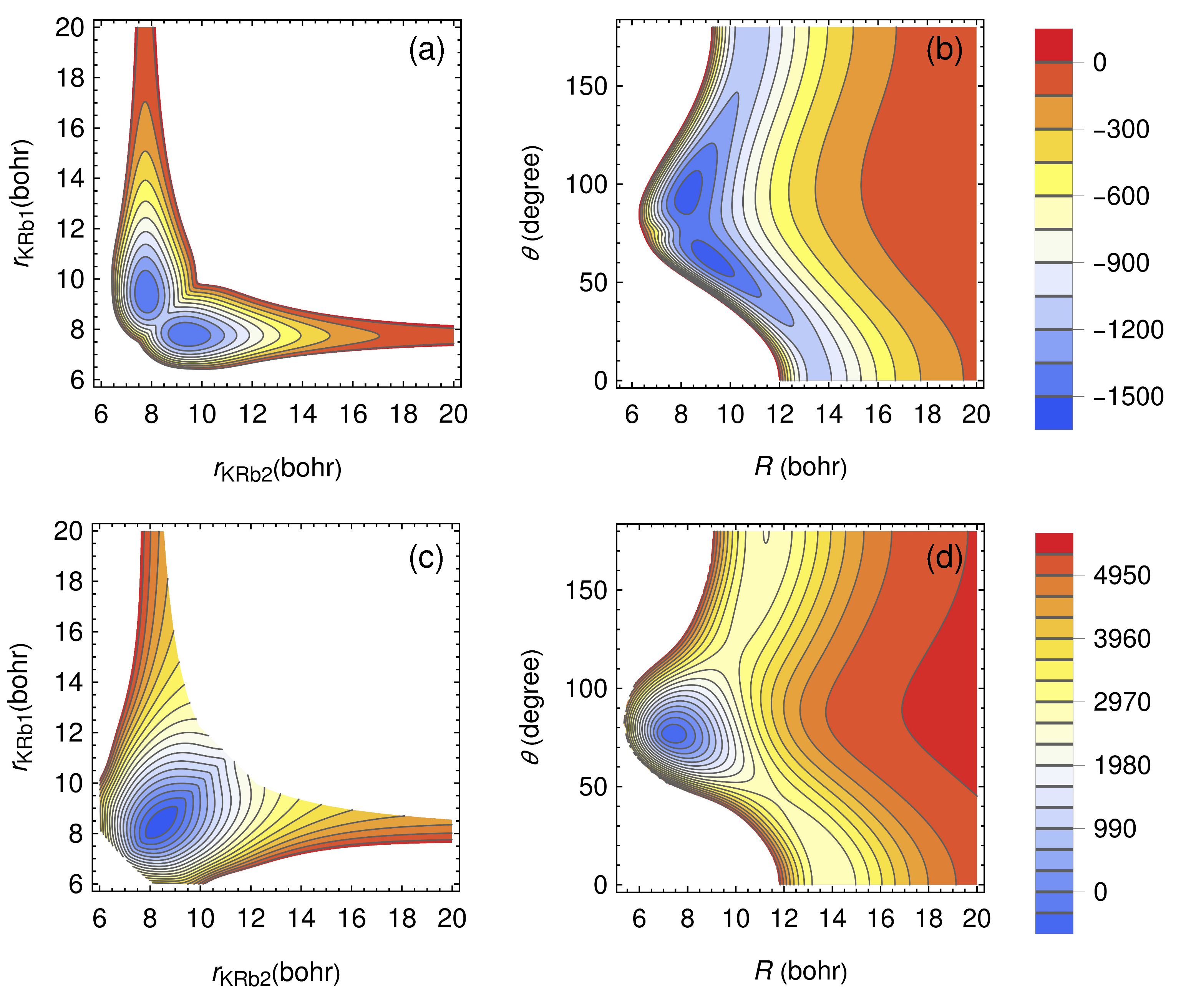}
\caption{ Two-dimensional cuts of the three-dimensional potential energy surfaces of the ground (panels (a), (b)) and first excited (panels (c), (d)) electronic states. The energy scales are in the cm$^{-1}$ and kept the same for all plots in the same row. Panels (a) and (c) show PESs as a function of two K-Rb distances in KRb$_2$ for the Rb-K-Rb angle equal to 62.2$^\circ$ (for this angle conical intersection occurs for $r_\text{KRb1}=r_\text{KRb2}=r_{0}$). Panels (b) and (d) show PESs in the Jacobi coordinates for intramonomer KRb distance fixed at $r_0$. }\label{fig:2DPES}
\end{figure*}

The full three-dimensional potential energy surfaces are computed for the ground and first excited electronic states. Their two-dimensional cuts are presented in Fig.~\ref{fig:2DPES}. 
Panel (b) shows a large anisotropy of atom-molecule interactions in the rigid rotor approximation. In panels (b) and (d), the location of the conical intersection at $\theta_\text{CI}=76.2^\circ$ is visible. Panel (a) shows that the exchange of Rb atoms can proceed without any electronic barrier.

\subsubsection{Electron spin resonance g-tensor and electron spin-rotation coupling}

We employ the all-electron state-averaged complete active space self-consistent field method \cite{GanyushinJCP2013} as implemented in the ORCA 5.0.4 program~\cite{NeeseJCP2020,NeeseWIREs2018} to compute electron spin resonance g-tensor $\mathbf{g}(R,r,\theta)$. The relativistic effects are described by the second-order Douglas-Kroll-Hess Hamiltonian taking into account picture-change effects \cite{SandhoeferJCP2012}. We decided not to include the magnetic field in the Foldy-Wouthuysen transformation due to numerical problems. The active space includes 3 electrons and 30 orbitals. We perform state-average over 40 doublet states and 40 quartet states. We use all-electron Karlsruhe-type quadruple-$\zeta$ valence relativistic basis set extended by additional $p$ and $d$ core functions and diffuse function \cite{FranzkeJCTC2020} (x2c-QZVPPDall-s), which we additionally decontracted. 

The coupling between electron spin and the rotation of whole atom-molecule complex can be estimated from the Curl's relation \cite{CurlMolPhys1965}, which has the following general form:
\begin{equation}
	g_{ij}-g_e \delta_{i,j}=\frac{1}{\hbar^2} \sum_k \epsilon_{ik} I_{kj}\,,
\end{equation}
where $g_e$ is the free electron g-factor, $\epsilon_{ik}$ are the components of spin-rotation tensor, $\bm{\epsilon}$, $I_{kj}$ are the components of inertial tensor, $\bm{I}$ (defined with respect of center of mass of the whole Rb-KRb system), and $g_{ij}$ are the components of electron spin resonance g-tensor, $\bm{g}$. We calculate $\bm{\epsilon}(R,r,\theta)$ using following relation:
\begin{equation}
	\bm{\epsilon}=\left( \bm{P}^{-1}\bm{g}\bm{P} - g_e \bm{1} \right) \left( \bm{P}^{-1} \bm{I} \bm{P} \right)^{-1}\,,
\end{equation}
where $\bm{1}$ is $3 \times 3$ identity matrix, and $\bm{P}$ matrix transforms $\bm{g}$ into diagonal form by operation: $\bm{P}^{-1}\bm{g}\bm{P}$. The scalar  isotropic electronic spin-rotation coupling constant is computed as:
\begin{equation}
\epsilon_{iso}=\frac{\text{Tr}{(\bm{\epsilon})}}{3}.
\end{equation}

In our calculations, we observed that the g-tensor reaches smoothly zero at the conical intersection. This results in a drastic enhancement of the spin-rotation coupling, which reaches a large but finite value, as presented in Fig.~\ref{fig:PESs}(b).

\subsubsection{Hyperfine couplings}

The hyperfine and nuclear electric quadrupole interaction are computed using domain-based local pair-natural orbital coupled-cluster singles and doubles \cite{SaitowJCP2018}.

\subsection{Details of coupled-channel calculations}

Here, we provide the details of our CC calculations of ultracold Rb~+~KRb collisions in a magnetic field, including the matrix elements and selection rules for key interactions. We also explore the question of whether the total rotational angular momentum of the Rb-KRb complex is conserved. Finally, we present the rotational and hyperfine state distributions of the collision  products and analyze their    sensitivity to  the interaction potential.

 The Hamiltonian of the atom-molecule collision complex is given by
\begin{equation}\label{eq:H}
\hat{{H}} = - \frac{1}{2\mu R}  \frac{\partial^2}{\partial R^2} R
                     + \frac{\hat{{L}}^2}{2\mu R^2} 
                     + \hat{{H}}_{\mathrm{mol}} + \hat{{H}}_\text{at} + 
                     \hat{V}(R,\theta),
\end{equation}
where the first two terms describe the kinetic energy of the Rb-KRb system in the center of mass frame, $\mu$ is the reduced mass of  Rb-KRb, $\hat{H}_\text{mol}$ is the Hamiltonian of the isolated KRb molecule \cite{Aldegunde:08,Carrington:03}, $\hat{H}_\text{at}$ is that of the Rb atom, $R$ is the magnitude of the atom-molecule separation vector $\mathbf{R}$ and $\theta$ is the Jacobi angle  between  $\mathbf{R}$ and the internuclear distance vector $\mathbf{r}$. The Rb-KRb interaction $\hat{V}(R,\theta)=\hat{V}_\text{si}(R,\theta) +  \hat{V}_\text{sd}(R,\theta)$ includes the scalar (spin-independent) interaction potential and the spin-dependent term (see below) with the KRb internuclear distance $r$ fixed at its equilibrium value. While the rigid-rotor approximation neglects the vibrational motion of KRb, it is necessary to employ in the present context to make the CC calculations computationally feasible.

The wavefunction of the Rb-KRb collision complex is expanded in a space-fixed total rotational angular momentum basis (see the main text) 
\begin{equation}
\label{eq:psi}
|\Psi\rangle = \frac{1}{R} \sum_{n}F_{n}(R) |\Phi_n\rangle, 
\end{equation}
with the basis functions given by 
\begin{equation}
\label{eq:basis}
|\Phi_n\rangle =  |(NL)J_rM_r\rangle |n_s^\text{mol}\rangle |n_s^\text{at}\rangle
=|(NL)J_r M_r\rangle |M_{I_1} M_{I_2}\rangle |F_a M_{F_a}\rangle.
\end{equation}

Here,  $|(NL)J_rM_r\rangle$ are  the eigenstates  of $\hat{J}_r^2$ and $\hat{J}_{r_z}$, and $|n_s^\text{mol}\rangle$ and $|n_s^\text{at}\rangle$ are the spin  basis functions for the KRb molecule and the Rb atom, respectively.  We choose the uncoupled nuclear spin basis for the KRb molecule, $|n_s^\text{mol}\rangle = |I_1 M_{I_1}\rangle |I_2 M_{I_2}\rangle = |M_{I_1} M_{I_2}\rangle$ as a direct product of eigenstates of ${\hat{I}}_m^2$ and $\hat{{I}}_{m_z}$, where $\mathbf{\hat{I}}_m$ is the nuclear spin angular momentum operator  for the $m$-th nucleus in KRb ($m=1$ for K, $m=2$ for Rb) and   $\hat{{I}}_{m_z}$ is its projection  
on the magnetic field vector $\mathbf{B}$, which defines the space-fixed quantization axis.
 For the Rb atom, we take $|n_s^\text{at}\rangle=|F_a M_{F_a}\rangle$, where $F_a$ and $M_{F_a}$ are the values of the total angular momentum $\hat{\mathbf{F}}_a=\hat{\mathbf{I}}_a+\hat{\mathbf{S}}$ and its space-fixed projection.

Substituting this expansion into the Schr\"odinger equation $\hat{H}{\Psi}=E\Psi$, where $\hat{H}$ the Hamiltonian of the atom-molecule collision complex  and $E$ is the total energy, leads to a system of CC equations
\begin{equation}\label{eq:CC}
 \Big[ \frac{d^2}{dR^2}+2\mu E \Big] F_{n}(R)=2\mu
 \sum_{n'} \langle \Phi_n | \hat{V} (R,\theta)+ \frac{\hat{{L}}^2}{2\mu R^2} + \hat{H}_\text{mol} + \hat{H}_\text{at}   | \Phi_{n'}\rangle  
  F_{n'}(R).
\end{equation}
We solve these equations numerically on a radial grid extending from $R=4\,a_0$ to $R_m=35\, a_0$ with the grid spacing of 0.02~$a_0$ and from  $R_m=35\, a_0$ to $R_f=400\,a_0$ with the grid spacing of 0.1~$a_0$ using the log-derivative algorithm \cite{Johnson:73,Manolopoulos:86}.  

At $R=R_f$ the radial solutions $F_{n}(R)$ are matched to the Riccati-Bessel functions to yield the scattering $S$-matrix \cite{Johnson:73}, from which the integral cross sections, collision rates, and rotational and hyperfine branching ratios are obtained following the standard procedures \cite{Tscherbul:23}.

\subsubsection{Matrix elements and selection rules: Intermolecular interactions}

The matrix elements of $\hat{V} (R,\theta)$ on the right-hand side of Eq.~\eqref{eq:CC}  determine the selection rules for the spin-independent and spin-dependent  intermolecular interactions in Rb-KRb, $\hat{V}(R,\theta)=\hat{V}_\text{sd}(R,\theta) + \hat{V}_\text{si}(R,\theta)$ (\TT{see above}). 
 Spin-dependent terms are also present in the Hamiltonians of the colliding fragments $\hat{H}_\text{mol}$ and $\hat{H}_\text{at}$, see the following section.  These  can be evaluated using the Wigner-Eckart theorem \cite{Zare:88} as described below. We expand the angular dependence of all Rb-KRb interactions  in Legendre polynomials, such as $V(R,\theta) = \sum_\lambda V_{\lambda}(R) P_\lambda(\cos\theta)$ for the interaction PES, $A^\text{FC}_k(R,\theta) = \sum_\lambda A^\text{FC}_{k,\lambda}(R) P_\lambda(\cos\theta)$ for the Fermi contact couplings, etc.

{Through these expansions, the spin-dependent and spin-independent Rb-KRb interactions are rigorously included at all values of $R$ in our CC calculations  (at the level of the RR approximation). This is necessary to account for the intrinsically multichannel, spin-dependent nature of the observed hyperfine-to-rotational energy transfer in Rb-KRb collisions, including the final rotational state distributions of KRb products.

The rigorous treatment of short-range interactions is an essential new aspect of the theory developed here, and distinguish our work from previous theoretical studies based on single-channel universal models, which model short-range physics by adding an artificial loss channel \cite{Idziaszek:10} or quantum reactive scattering computations, which do account for the short-range interactions and vibrational modes, but neglect the electronic and nuclear spins \cite{croft2017universality}.}

The matrix elements of the spin-independent Rb-KRb interaction in the total rotational angular momentum basis \eqref{eq:basis} are given by
\begin{multline}\label{Eq:interaction_pot}
 \langle (NL)J_rM_r| \langle  M_{I_1} M_{I_2} | \langle F_a M_{F_a}| V_\text{si}(R,\theta) | (N'L')J_r'M_r'\rangle
    | M_{I_1}' M_{I_2}'\rangle |F_a' M_{F_a}'\rangle
    = \delta_{M_{I_1} M_{I_1}'} \delta_{M_{I_2} M_{I_2}'} \delta_{F_a F_a'}\delta_{M_{F_a} M_{F_a'}}   \delta_{J_rJ_r'}\delta_{M_r M_r'} 
    \\ \times
     (-1)^{J_r+N+N'} [(2N+1)(2N'+1)(2L+1)(2L'+1)]^{1/2}\sum_\lambda V_\lambda(R) \sixj{N}{L'}{L}{N'}{J_r}{\lambda} \threejm{N}{0}{\lambda}{0}{N'}{0} 
     \threejm{L}{0}{\lambda}{0}{L'}{0}. 
\end{multline}
This expression shows that while the spin-independent part of the Rb-KRb interaction 
can change the rotational state of KRb without affecting the internal hyperfine states of either KRb or Rb it cannot drive  hyperfine-to-rotational energy transfer in Rb$|2,2\rangle$~+~KRb$|0,0,-4,1/2\rangle$ collisions.

We now proceed to analyze the spin-dependent part of the Rb-KRb interaction, which is responsible for the hyperfine-to-rotational energy transfer.
We begin with the Fermi contact coupling [the first term in Eq.~\eqref{V_sd}], which can be conveniently recast in the following form
\begin{equation}\label{Eq:FC}
\hat{V}_\text{sd}^\text{FC}  = \sum_{m=1,2}  A^{\text{FC}}_m(R,\theta)  \hat{\mathbf{S}} \cdot \hat{\mathbf{I}}_m
+  A^\text{FC}_a (R,\theta) \hat{\mathbf{S}} \cdot \hat{\mathbf{I}}_a,
\end{equation}
where the sum over $m$ now runs over the KRb nuclear spins ($m=1,2$), and the index $a$ refers to the colliding Rb atom.  The first term in Eq.~\eqref{Eq:FC} describes the interaction of $\hat{\mathbf{S}}$ with the nuclear spins inside KRb, and thus $A^\text{FC}_m (R,\theta)\to 0$ as $R\to \infty$.

The second term in Eq.~\eqref{Eq:FC} describes the modification of the hyperfine constant of the Rb atom by the approaching KRb molecule, and is defined in such a way that $A_{a}^\text{FC} (R,\theta)\to 0$ in the limit $R\to \infty$.

We note that this term is diagonal in the atomic quantum numbers $F_a$ and $M_{F_a}$ because $\hat{\mathbf{S}} \cdot \hat{\mathbf{I}}_a=\frac{1}{2}[\hat{F}_a^2 - \hat{I}_a^2 - S^2]$, where $\hat{\mathbf{F}}_a=\hat{\mathbf{I}}_a + \hat{\mathbf{S}}$ is the total angular momentum of the atom. The term $A^\text{FC}_a (R,\theta) \hat{\mathbf{S}} \cdot \hat{\mathbf{I}}_a$ therefore acts as a small $F_a$-dependent perturbation to the Rb-KRb interaction PES that is diagonal in $M_{F_a}$. As a result, it cannot couple the initial $|22\rangle$ hyperfine state of Rb to the lower hyperfine states, and we will  neglect it in what follows.  

The matrix elements of the Fermi contact interaction involving the K nucleus in KRb  can be obtained by expanding the first term in Eq.~\eqref{Eq:FC} in Legendre polynomials as  $A^\text{FC}_m(R,\theta) = \sum_\lambda A^\text{FC}_{m,\lambda}(R) P_\lambda(\cos\theta)$ with $m=1$ and evaluating the angular matrix elements in the basis \eqref{eq:basis} using the expressions from Ref.~\cite{Tscherbul:23} 
\begin{multline}\label{eq:FC1mxel1}
\langle \Phi_n | A^{\text{FC}}_m (R,\theta) \hat{\mathbf{S}} \cdot \hat{\mathbf{I}}_m | \Phi_{n'}\rangle
= \sum_\lambda A^\text{FC}_{m,\lambda}(R)    
        \langle (NL)J_rM_r | P_\lambda(\cos\theta) | (N'L')J_r'M_r' \rangle 
        \langle  M_{I_1} M_{I_2} | \langle F_a M_{F_a} |   \hat{\mathbf{S}}\cdot \hat{\mathbf{I}}_m | M_{I_1}' M_{I_2}'\rangle |F_a' M_{F_a}'\rangle
  \\ =  \delta_{J_rJ_r'}\delta_{M_r M_r'} 
\delta_{\bar{M}_{I_m} \bar{M}_{I_m}'}
  \sum_\lambda A^\text{FC}_{m,\lambda}(R)  (-1)^{J_r+N+N'}
  [(2N+1)(2N'+1)]^{1/2}[(2L+1)(2L'+1)]^{1/2} \sixj{N}{L'}{L}{N'}{J_r}{\lambda} \threejm{N}{0}{\lambda}{0}{N'}{0} 
   \\ \times   \threejm{L}{0}{\lambda}{0}{L'}{0} 
         (-1)^{M_{F_a} - M_{F_a}' } (-1)^{F_a-M_{F_a}} \threejm{F_a}{-M_{F_a}}{1}{ M_{F_a} - M_{F_a}'  }{F_a'}{M_{F_a}'} (-1)^{I_a+S+F_a+1} 
 [(2F_a+1)(2F_a'+1)]^{1/2} 
  \\ \times
  [(2S+1)S(S+1)]^{1/2}
\sixj{S}{F_a'}{F_a}{S}{I_a}{1} (-1)^{I_m - M_{I_m}}  [(2I_m+1)I_m(I_m+1)]^{1/2} \threejm{I_m}{-M_{I_m}}{1}{M_{I_m} - M_{I_m}'}{I_m}{M_{I_m}'},
\end{multline}
where the notation $\bar{M}_{I_1}=M_{I_2}$ and $\bar{M}_{I_2}=M_{I_1}$ indicates that the matrix elements for the first nucleus in KRb $(m=1)$ are diagonal in $M_{I_2}$ and those for the second nucleus $(m=2)$ are diagonal in $M_{I_1}$.
It follows from Eq.~\eqref{eq:FC1mxel1} that the FC interaction due to the $m$-th nucleus of KRb conserves the value of $J_r$ and $M_r+M_{I_m}+M_{F_a}$.

Further, Eq.~\eqref{eq:FC1mxel1} immediately establishes that {\it the FC interaction has the selection rule  $\Delta M_{F_a}=\pm 1,\,0$, which is consistent with the experimental observations (see the main text).}  This can be seen from either the structure of the FC interaction, which contains the term $\hat{\mathbf{S}}\cdot \hat{\mathbf{I}}_m = \frac{1}{2} [\hat{S}_-\hat{I}_{m+} + \hat{S}_+\hat{I}_{m-}] + \hat{S_z}\hat{I}_z$ (where $\hat{S}_\pm$ and $\hat{I}_{m\pm}$ are the raising and lowering operators) or from the 3-$j$ symbol $ \threejm{F_a}{-M_{F_a}}{1}{q}{F_a'}{M_{F_a}'} $, which vanishes  unless $\Delta M_{F_a}=\pm 1$ or 0.

Importantly, this selection rule holds for transition probabilities only in the weak-coupling limit (i.e., within the range of validity of the Fermi Golden Rule).

Equation \eqref{eq:FC1mxel1} also shows that {\it the FC interactions couple not only the different hyperfine states of the incident Rb atom but also the different rotational states of KRb.} While  the incoming Rb's hyperfine state is flipped by the $\hat{S}_-\hat{I}_{m+} $ coupling (see above), the  anisotropy of the FC interaction, represented by the terms $A^\text{FC}_{m,\lambda}(R)$ with $\lambda>0$, couples the different rotational states of KRb, driving transitions from the ground ($N=0$) to rotationally excited ($N=1$ and $N=2$) states observed experimentally.  As shown below, CC calculations including only the FC contact interactions and a model Rb-KRb interaction PES can reproduce the experimental branching ratios for both the final rotational states of KRb molecules and hyperfine states of the Rb atom.

We now turn to the tensor hyperfine interactions between the electron spin of the Rb-KRb complex $\mathbf{\hat{S}}$ and the nuclear spins of KRb  $\mathbf{\hat{I}}_m$ ($m=1,2$).  We approximate this   interaction by its long-range part, which is given by the magnetic dipole-dipole interaction between $\hat{\mathbf{S}}$ and $\hat{\mathbf{I}}_m$: 
\begin{equation}\label{Vm_dipolar2}
\hat{V}^{(m)}_\text{dip} = -\left(\frac{24\pi}{5}\right)^{1/2} \frac{g_m\mu_N}{g_S\mu_B}   \frac{\alpha^2}{R^3} \sum_q (-1)^q Y_{2,-q}(\hat{R})[\hat{\mathbf{S}} \otimes \hat{\mathbf{I}}_m ]^{(2)}_q,
\end{equation}
where $\mu_N$ and $\mu_B$ are the nuclear and electron Bohr magnetons,  $g_m$ are the gyromagnetic ratios for the $m$-th nuclear spin of KRb \cite{Aldegunde:08}, $g_S\simeq 2.002$  is the electron spin $g$-factor, and $\alpha$ is the fine structure constant. 

The matrix elements of the tensor hyperfine interactions in the total rotational angular momentum basis \eqref{eq:basis} can be evaluated by taking into account the factorized structure of the basis functions 
\cite{Tscherbul:23}  

\begin{multline}\label{mxel_aniso1tram}
  \langle (NL)J_rM_r | \langle {M_{I_1} M_{I_2}} | \langle {F_a M_{F_a}} | \hat{V}^{(m)}_\text{dip}
|(N'L')J_r'M_r'\rangle  |{M_{I_1}' M_{I_2}'}\rangle |{F_a' M_{F_a}'} \rangle
= -\delta_{NN'} \delta_{\bar{M}_{I_m} \bar{M}_{I_m}'}
\sqrt{30} \frac{g_m\mu_N}{g_S\mu_B}   \frac{\alpha^2}{R^3} 
\\ \times 
(-1)^{J_r-M_r} \threejm{J_r}{-M_r}{2}{M_r-M_r'}{J_r'}{M_r'}
 (-1)^{N+L'+J_r} 
[(2J_r+1)(2J_r'+1)]^{1/2}
\sixj{L}{J_r'}{J_r}{L'}{N}{2}  
(-1)^l [(2L+1)(2L'+1)]^{1/2}
\\ \times
 \threejm{L}{0}{2}{0}{L'}{0}
  \threejm{1}{M_{F_a} - M_{F_a}'}{1}{ M_{I_m} - M_{I_m}' }{2}{M_r-M_r'}(-1)^{I_m-M_{I_m}} 
  \threejm{I_m}{-M_{I_m}}{1}{ M_{I_m} - M_{I_m}' }{I_m}{M_{I_m}'}
  \\ \times
   [(2I_m+1)I_m(I_m+1)]^{1/2} (-1)^{I_a+S+2F_a-M_{F_a}+1}
[(2F_a+1)(2F_a'+1)]^{1/2}   [(2S+1)S(S+1)]^{1/2} 
\\ \times
 \threejm{F_a}{-M_{F_a}}{1}{M_{F_a} - M_{F_a}'}{F_a'}{M_{F_a}'}
\sixj{S}{F_a'}{F_a}{S}{I_a}{1},
\end{multline}
which is valid for $m=1$ and 2.
Importantly, this expression shows that {\it tensor hyperfine interactions have exactly the same atomic hyperfine selection rule as the FC  interactions,  namely, $|\Delta M_{F_a}|=1$}. As a result, the  initial state of Rb  $|{2,2}\rangle$ is directly coupled only to the final hyperfine states $|1,1\rangle$ and $|{2,1}\rangle$.

However, unlike the FC interactions, {\it the tensor hyperfine interactions couple the different values of  the total rotational angular momentum $J_r$}. The selection rule $\Delta J_r = 0, \pm1, \pm 2$ is imposed by  the 6-$j$ symbol  $\sixj{L}{J_r'}{J_r}{L'}{N}{2}$.  
 In addition, the asymptotic form of the tensor hyperfine interactions \eqref{Vm_dipolar2} does not couple the different rotational states of KRb.

\begin{figure}
\includegraphics[scale=0.4,trim = 0 0 0 0]{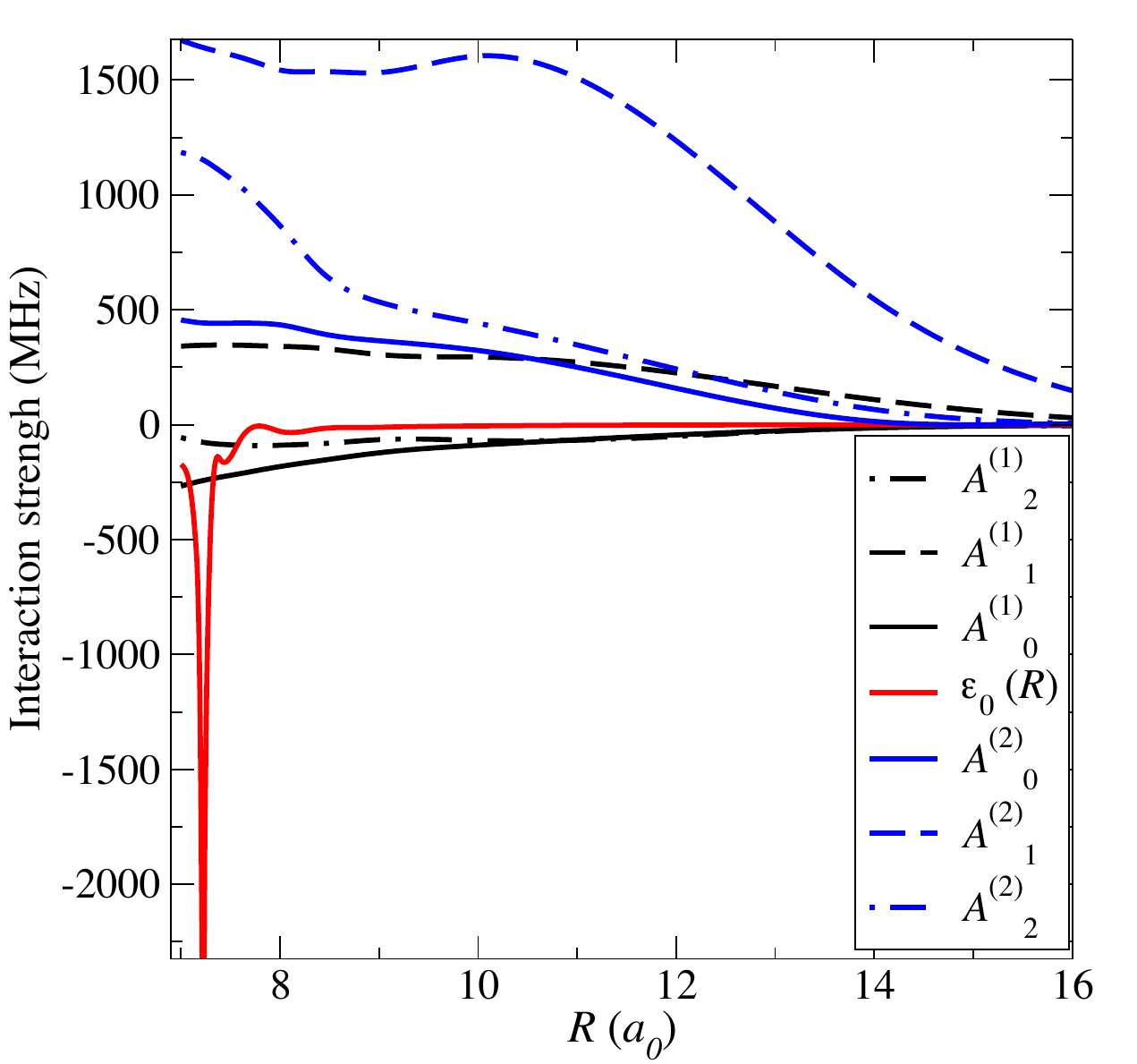}
\caption{Spin-dependent interactions in Rb-KRb as a function of $R$. Black lines: Legendre moments of the Fermi contact interaction involving the K nucleus in KRb $A^{(1)}_\lambda(R)$ with $\lambda=0$ (full line), $\lambda=1$ (dashed line), $\lambda=2$ (dash-dotted line). Blue lines:  Legendre moments of the Fermi contact interaction involving the Rb nucleus in KRb $A^{(2)}_\lambda(R)$ with $\lambda=0$ (full line), $\lambda=1$ (dashed line), $\lambda=2$ (dash-dotted line). Red line: The isotropic spin-rotation interaction $\epsilon_0(R)$.}
\label{fig:interactions_vsR}
\end{figure}

Finally, we consider the matrix elements of the intermolecular spin-rotation interaction 
\begin{equation}\label{eq:SpinRotInt}
\hat{V}_\text{sd}^\text{SR} = -\hat{\mathbf{S}}\, \bm{\epsilon}(R,\theta)\, \hat{\mathbf{J}}_r = -\hat{\mathbf{S}}\, \bm{\epsilon}(R,\theta)\, (\hat{\mathbf{N}} + \hat{\mathbf{L}}),
\end{equation}
where $\bm{\epsilon}$ is the spin-rotation tensor. We focus on the  isotropic part  of the spin-rotation tensor $\bm{\epsilon}\simeq {\epsilon}_0(R,\theta)\mathbf{1}$, where $\mathbf{1}$ is a $3\times 3$ unit matrix, which is the dominant contribution to the spin-rotation interaction in Rb-KRb at short range according to {\it ab initio} calculations. For qualitative estimates of the role of the spin-rotation interaction in Rb~+~KRb collisions, we also neglect the angular anisotropy of the isotropic spin-rotation tensor.   Hence, Eq.~\eqref{eq:SpinRotInt} becomes
\begin{equation}\label{eq:SpinRotInt_iso}
\hat{V}_\text{sd}^\text{SR}  =  -{\epsilon}_0(R) \hat{\mathbf{S}} \cdot (\hat{\mathbf{N}} + \hat{\mathbf{L}}),
\end{equation}
A plot of the scalar spin-rotation coupling as a function of $R$ as obtained from {\it ab initio} calculations is shown in Fig.~\ref{fig:interactions_vsR}.

We observe that the coupling is strongly peaked in a narrow range of $R$ near the CI between the two lowest PES.

The matrix elements of the spin-rotation interaction in the TRAM basis are
\begin{multline}
 \langle (NL)J_rM_r | \langle {M_{I_1} M_{I_2}} | \langle {F_a M_{F_a}} | \hat{V}_\text{sd}^\text{SR} 
|(N'L')J_r'M_r'\rangle  |{M_{I_1}' M_{I_2}'}\rangle |{F_a' M_{F_a}'} \rangle 
\\ = -\epsilon_0(R) \sum_q (-1)^q 
\langle {M_{I_1} M_{I_2}} | \langle {F_a M_{F_a}} | \hat{S}^{(1)}_q |{M_{I_1}' M_{I_2}'}\rangle |{F_a' M_{F_a}'} \rangle 
\langle (NL)J_rM_r |(\hat{N}^{(1)}_{-q} + \hat{L}^{(1)}_{-q}) |(N'L')J_r'M_r'\rangle.
\end{multline}
Evaluating the rotational, orbital, and spin matrix elements on the right-hand side as described above and in Ref.~\cite{Tscherbul:23} and setting $q=M_r'-M_r=M_{F_a} - M_{F_a}'$, we find
\begin{multline}\label{eq:SpinRotInt_mxel}
 \langle (NL)J_rM_r | \langle {M_{I_1} M_{I_2}} | \langle {F_a M_{F_a}} | \hat{V}_\text{sd}^\text{SR} 
|(N'L')J_r'M_r'\rangle  |{M_{I_1}' M_{I_2}'}\rangle |{F_a' M_{F_a}'} \rangle 
 = -\epsilon_0(R) \delta_{NN'} \delta_{LL'}  \delta_{M_{I_1} M_{I_1}'} \delta_{M_{I_2} M_{I_2}'} (-1)^{J_r - M_r}
 \\ \times
 (-1)^{F_a - M_{F_a}} \threejm{F_a}{-M_{F_a}}{1}{M_{F_a}-M_{F_a}'}{F_a'}{M_{F_a}'} (-1)^{I_a+S+F_a+1}[(2F_a+1)(2F_a'+1)]^{1/2}
  \sixj{S}{F_a'}{F_a}{S}{I_a}{1} p_3(S)
 \\ \times
  [(2J_r+1)(2J_r+1)]^{1/2} \threejm{J_r}{-M_r}{1}{M_r-M_r'}{J_r'}{M_r'} \biggl{[}
(-1)^{N+L+J_r'+1}  p_3(N) \sixj{N}{J_r'}{J_r}{N'}{L}{1} 
\\ 
+ (-1)^{N+L'+J_r+1}  p_3(L) \sixj{L}{J_r'}{J_r}{L'}{N}{1}   \biggr{]},
\end{multline}
where $p_3(X)=[(2X+1)X(X+1)]^{1/2}$.
The spin-rotation interaction couples basis states with $J_r'-J_r = \pm 1$ and $M_{F_a}=\pm 1$, but conserves the value of $M_r+M_{F_a}$.

\subsubsection{Matrix elements and selection rules: Intramolecular interactions}

Finally, we consider the intramolecular interactions in the KRb molecule described by the Hamiltonian 
\begin{equation}\label{eq:Hmol}
\hat{H}_\text{mol}=\hat{H}^\text{rot}_\text{mol}+\hat{H}^\text{NEQ}_\text{mol}+\hat{H}^\text{ss}_\text{mol}+\hat{H}_\text{mol}^Z.
\end{equation}
 The dominant interactions include the rotational Hamiltonian, the scalar spin-spin interactions, the Zeeman Hamiltonian, and the nuclear electric quadrupole (NEQ) interaction
 \begin{equation}\label{NEQint}
\hat{H}_{\textrm{mol}}^\text{NEQ}= \sum_{m=1,2} \sum_{p=-2}^2 (-1)^{p} C^{2}_{p}(\theta,\phi) \frac{(eqQ)_m\sqrt{6}}{4I_m(2I_m-1)} [\hat{\mathbf{I}}_m\otimes \hat{\mathbf{I}}_m]^{(2)}_{-p},
\end{equation}
where $C^{k}_{p}(\theta,\phi)=\sqrt{\frac{4\pi}{2k+1}}Y_{kq}(\theta,\phi)$ are the renormalized (Racah) spherical harmonics,  $[\hat{\mathbf{I}}_m\otimes \hat{\mathbf{I}}_m]^{(2)}_{-p}$ is a spherical tensor product of the $m$-th nuclear spin in KRb with itself, and $(eqQ)_m$ are the NEQ interaction constants, $(eqQ)_1=0.45$~MHz for $^{40}$K and $(eqQ)_2=1.41$~MHz for $^{87}$Rb.

The rotational Hamiltonian $\hat{H}^\text{rot}_\text{mol}=B_e\hat{N}^2$, where $B_e$ is the rotational constant, is diagonal in the total rotational angular momentum basis, with the diagonal elements given by $N(N+1)$.
  The scalar spin-spin interaction term $\hat{H}_\text{ss}=c_4 \hat{\mathbf{I}}_1\cdot\hat{\mathbf{I}}_2$ \cite{Aldegunde:08,Gorshkov:11}, which is important for $N=0$, does not depend on the rotational degrees of freedom, and has the following matrix elements:
 \begin{multline}\label{Hss}
 \langle (NL)J_rM_r| \langle  M_{I_1} M_{I_2} | \langle F_a M_{F_a}| c_4 \hat{\mathbf{I}}_1\cdot\hat{\mathbf{I}}_2 | (N'L')J_r'M_r' \rangle    | M_{I_1}' M_{I_2}'\rangle |F_a' M_{F_a}'\rangle =  \delta_{F_a F_a'}\delta_{M_{F_a} M_{F_a'}}\delta_{NN'}  \delta_{LL'} \delta_{J_r J_r'} \delta_{M_r M_r'}
 c_4  
\\  \times (-1)^{I_1 - M_{I_1} + I_2 - M_{I_2}} [(2I_1+1)I_1(I_1+1)]^{1/2} [(2I_2+1)I_2(I_2+1)]^{1/2} (-1)^{M_{I_1}-M_{I_1}'}
\\  \times 
\threejm{I_1}{-M_{I_1}}{1}{M_{I_1}-M_{I_1}'}{I_1}{M_{I_1}'}
\threejm{I_2}{-M_{I_2}}{1}{M_{I_2}-M_{I_2}'}{I_2}{M_{I_2}'}.
\end{multline}

The Zeeman Hamiltonian of KRb is given by \cite{Aldegunde:08}

\begin{equation}
\hat{H}_\text{mol}^Z=- g_N \mu_N \hat{{N}}_z B- \sum_{m=1,2} g_{I_m} \mu_N \hat{{I}}_{m_z} B (1-\sigma_m), \\
\end{equation}
where $g_N$ is the rotational $g$-factor, $g_{I_m}$ is the $g$-factor for the $m$-th nucleus ($m=1,2$),  $\hat{\textbf{I}}_m$ are the nuclear spin operators, $\mu_N$ is the nuclear Bohr magneton, and $\sigma_m$ are the nuclear spin shielding factors. 
Neglecting the small rotational Zeeman term, the Zeeman interaction has only diagonal matrix elements
 \begin{multline}\label{HZeeman}
 \langle (NL)J_rM_r| \langle  M_{I_1} M_{I_2} | \langle F_a M_{F_a}| \hat{H}_\text{mol}^Z | (N'L')J_r'M_r' \rangle    | M_{I_1}' M_{I_2}'\rangle |F_a' M_{F_a}'\rangle =  \delta_{F_a F_a'}\delta_{M_{F_a} M_{F_a'}}\delta_{NN'}  \delta_{LL'} \delta_{J_r J_r'} \delta_{M_r M_r'} 
 \\  \times
 \delta_{M_{I_1} M_{I_1}' }  \delta_{M_{I_2} M_{I_2}' } [-g_1 \mu_N (1-\sigma_1) M_{I_1} - g_2 \mu_N (1-\sigma_2) M_{I_2} ].
\end{multline}

We neglect the  nuclear spin-rotation and tensor  interactions, which are several orders of magnitude weaker than the dominant NEQ and scalar spin-spin interactions.

By contrast, the intramolecular nuclear electric quadrupole (NEQ) interaction  couples the different rotational and nuclear spin states of the isolated KRb molecules, and as a result,  breaks the conservation of the total rotational angular momentum. To see this,  we evaluate the matrix elements of Eq.~\eqref{NEQint} in the total rotational angular momentum basis 
\begin{multline}
 \langle (NL)J_rM_r| \langle  M_{I_1} M_{I_2} | \langle F_a M_{F_a}| \hat{H}_{\textrm{mol}}^{NEQ,1} | (N'L')J_r'M_r' \rangle
    | M_{I_1}' M_{I_2}'\rangle |F_a' M_{F_a}'\rangle = \delta_{F_a F_a'}\delta_{M_{F_a} M_{F_a'}} \delta_{LL'}
   \frac{(eqQ)_1}{4} \delta_{M_{I_2} M_{I_2}'}  
    \\ \times (-1)^{J_r-M_r+L+J_r'}
    \sum_{p=-2}^2 (-1)^{p}
       \threejm{J_r}{-M_r}{2}{p}{J_r'}{M_r'} 
[(2J_r+1)(2J_r'+1)]^{1/2}
\sixj{N}{J_r'}{J_r}{N'}{L}{2} 
[(2N+1)(2N'+1)]^{1/2}
 \threejm{N}{0}{2}{0}{N'}{0}
    \\ \times
     (-1)^{I_1-M_{I_1}} \threejm{I_1}{-M_{I_1}}{2}{-p}{I_1}{M_{I_1}'}
  \threejm{I_1}{-I_{1}}{2}{0}{I_1}{{I_1}}^{-1}.
\end{multline}
The 3-$j$ symbols vanish unless $p=M_r - M_r'=M_{I_1}'-M_{I_1}$, and thus $M_r+M_{I_1}$ is conserved.
Here, the 3-$j$ symbols vanish unless $p=M_r - M_r'=M_{I_2}'-M_{I_2}$, and thus $M_r+M_{I_2}$ is conserved.
We observe that the NEQ couples the values of $J_r$ and $J_r'=J_r\pm 2$. However, test calculations show that this coupling is too weak to break the conservation of the total rotational angular momentum in ultracold Rb-KRb collisions.

\subsubsection{Parametrization of the Rb-KRb interactions}

To fully specify the matrix elements, we need to parametrize the potential energy surface and hyperfine interactions of the Rb-KRb complex as a function of $R$ and $\theta$. 

To this end, we computed the lowest-order anisotropic expansion coefficients  $V_\lambda(R)$ of the Rb-KRb interaction potential from the {\it ab initio} data points \cite{jachymski2022collisional} evaluated in the linear and T-shaped geometries. We fit the  data points with cubic spline functions, which are smoothly merged with the long-range parts $V_0^{LR}(R)=-f(R)C_{6,0}/R^6$ and  $V_2^{LR}(R)=-f(R)C_{6,2}/R^6$ using a switching function $f(R)$. The values of the long-range dispersion coefficients $C_{6,n}$ for Rb-KRb are taken from Ref.~\cite{Kotochigova:10}.

The resulting PES, which we will refer to as PES2, has the correct long-range behavior, and reproduces all of the expected features of the full Rb-KRb potential (i.e., large depth and strong anisotropy) except for the higher-order anisotropic terms with $\lambda\ge3$. To assess the importance of these terms, we supplement PES2 with additional  anisotropic terms  $V_\lambda(R) = 0.5 V_{\lambda-1}(R)$ for $3 \le \lambda\le 7$. The resulting PES is referred to as PES7.

Following a similar procedure, we obtain the lowest-order anisotropic expansions of the Rb-KRb Fermi contact interactions $V_{\lambda,m}^\text{iso}$ for $m=1,2$ and $\lambda\le 2$. 

Since no {\it ab initio} calculations are available for the Rb-KRb anisotropic hyperfine interactions, we approximate them by the  magnetic dipole-dipole interactions between the electron spin of the colliding Rb atom and the nuclear spins of KRb \eqref{Vm_dipolar2}.
These expressions are valid at large $R$ and neglect the short-range anisotropy of the tensor hyperfine interaction, which, however, is small.

We note that because ultracold scattering observables are known to be highly sensitive to fine details of intermolecular interactions \cite{,Morita:19}, even the most accurate {\it ab initio} PESs invariably require modifications  to fit experimental observations \cite{karman2023ab}. We perform such  modifications by scaling our model PESs by a constant factor $\lambda$ as described below.

\begin{figure}
\centering

\includegraphics[width=0.40\textwidth]{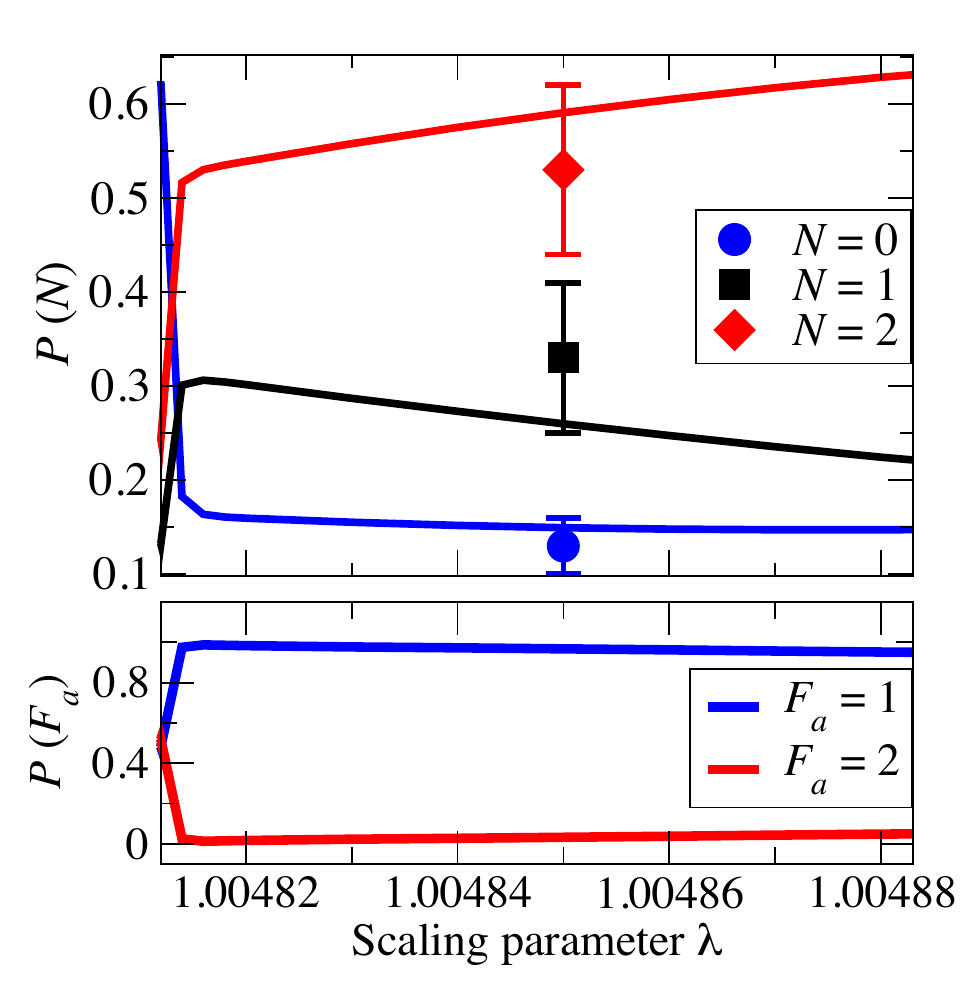}
  \caption{(Upper panel): Branching ratios for the final states of KRb produced in ultracold Rb($|2,2\rangle$)~+~KRb($|0,0,4,-1/2\rangle$) collisions computed using the CC approach on PES7 scaled by a constant parameter $\lambda$. (Bottom panel): Same as panel (a) but for the final hyperfine states of Rb. The collision energy is 1~$\mu$K and the magnetic field is 30~G. Experimental data are shown by symbols with error bars.}
\label{fig:PES_scaling}
\end{figure}

\begin{figure}
\centering
\includegraphics[width=0.40\textwidth]{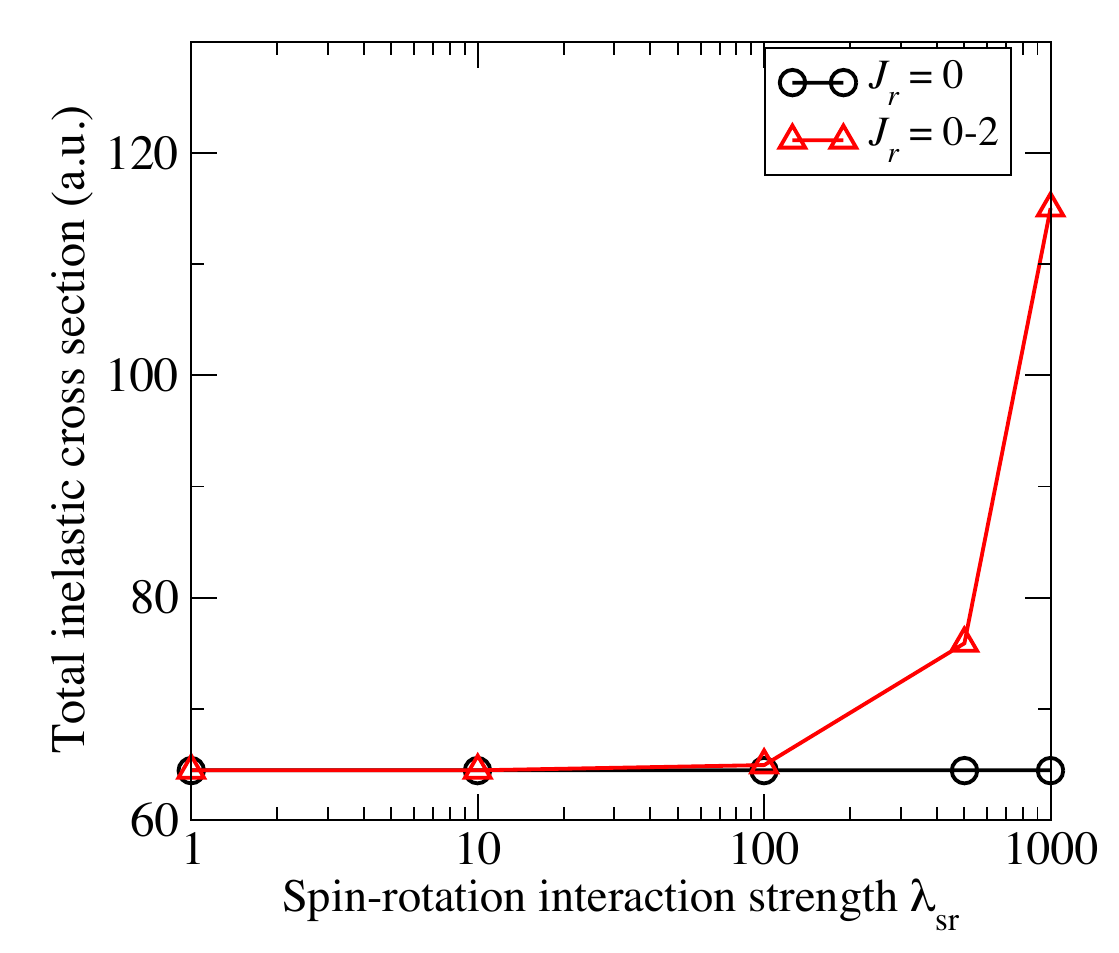}
  \caption{Total inelastic cross sections for Rb~+~KRb collisions computed using the CC approach with $N_\text{max}=4$ on the {\it ab initio} PES as a function of the scaling parameter of the spin-rotation interaction $\lambda_\text{sr}$. Circles: calculations including only the $J_r=0$ block. Diamonds: calculations including the three lowest $J_r$ blocks ($J_r=0{-}2$). The differences between the $J_r=0$ and $J_r=0{-}2$   results quantify the extent, to which $J_r$ is conserved.}
\label{fig:sri_scaling}
\end{figure}

\subsubsection{CC calculations of rotational and hyperfine branching ratios in ultracold Rb~+~KRb collisions}

As shown in Fig.~\ref{fig:hist} the rotational state distributions computed using the rigid-rotor CC approach on both PES2 and PES7 generally predict the preferential population of the $N=0$ final state of KRb in marked disagreement with experiment. However, there is a single  outlier PES corresponding to the value of $\lambda = 1.005$, which predicts an inverted rotational state distribution close to that observed experimentally. 
Scanning $\lambda$ in the vicinity of the value corresponding to the outlier in the upper panel of Fig.~\ref{fig:hist}, we find good agreement with experiment for all three final rotational states, as shown in Fig.~\ref{fig:PES_scaling}. However, the calculated {\it total inelastic cross section} corresponding to these values of $\lambda$ is  uncharacteristically small, much below the experimental values.

Therefore, even though our rigid-rotor CC calculations can explain the observed product state distributions by tuning the strength of the Rb-KRb interaction, this requires extensive tuning of the PES, and the resulting Rb-KRb interaction, for which theory matches experiment, is unlikely. In addition, the rigid-rotor calculations disagree with the full range of experimental observations, which include not only KRb product state distributions but also the total inelastic rates.
We note that our rigid rotor CC model rigorously accounts for couplings between all six angular momenta in the Rb-KRb collision complex, and only assumes that the total mechanical rotational angular momentum $J_r$  is conserved in a collision. Thus, its consistent inability to match the experimental observations strongly suggests the lack of  conservation of $J_r$  in the complex.  

This finding is consistent with the conclusion derived from the statistical model above, which suggested that the observed branching ratios only make sense in the absence of $J_r$ conservation.

\subsubsection{Mechanisms for the breakdown of $J_r$ conservation: The spin-rotation interaction near a conical intersection}

As a plausible candidate responsible for the breakdown of $J_r$ conservation in ultracold Rb~+~KRb collisions, we consider the spin-rotation interaction. As shown above, this interaction is strongly enhanced near the CI between the two lowest PESs of Rb-KRb in the classically allowed short-range region. 

This enhancement correlates with the electron g-factor reaching zero at the CI (see the main text). 
However, our CC calculations indicate that even the enhanced spin-rotation interaction is not strong enough to break the conservation of $J_r$ in the rigid-rotor approximation. In Fig.~\ref{fig:sri_scaling}, we plot the total inelastic cross sections for Rb~+~KRb collisions calculated with $J_r^\text{max}=0$ and $J_r^\text{max}=2$ as a function of the spin-rotation interaction strength quantified by the scaling parameter $\lambda_\text{sr}$. The difference between these cross sections quantifies the extent, to which $J_r$ is conserved. We observe that at $\lambda_\text{sr}=1$, which corresponds to the unscaled {\it ab initio} spin-rotation tensor (see above), the difference between the $J_r^\text{max}=0$ and $J_r^\text{max}=2$ results is less than 1\%, which indicates that $J_r$ is conserved to a very good approximation.

Because our rigid-rotor CC calculations do not account for the vibrational degree of freedom of KRb, they may underestimate the effective strength of the spin-rotation interaction within the complex. As shown in Refs.~\cite{mayle2012statistical,mayle2013scattering}, the $J_r$-breaking  interactions may be amplified when the lifetime of the complex is long compared to the typical collision time. To explore this scenario, we performed additional CC calculations with the spin-rotation interaction scaled by a constant factor $\lambda$. Figure~\ref{fig:sri_scaling} shows that appreciable differences between the $J_r^\text{max}=0$ and $J_r^\text{max}=2$ results occur only at $\lambda\ge 100$, which suggests that the effective spin-rotation interaction in the complex must be amplified by the factor of $\geq$100 to cause significant $J_r$ mixing. We further observe that the rotational state distributions of KRb products are far less sensitive to the spin-rotation interaction than the total integral cross sections and the hyperfine state distributions of Rb products.  This fact points even more strongly to the necessity of including the CI and vibrational degrees of freedom of KRb (in addition to the rotational and spin degrees of freedom considered here) in quantum scattering calculations on ultracold Rb+KRb collisions.

\subsection{Characterizing the inelastic collision rate}

\begin{figure*}
\centering
	\includegraphics[width=0.4\textwidth]{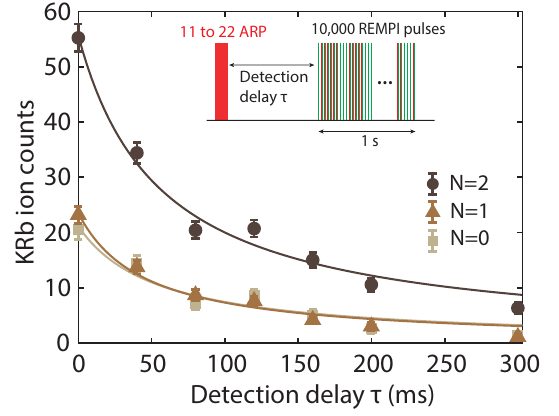}
  \caption{ KRb ion counts with 666 nm laser resonant with Q(2) (circles), Q(1) (triangles) and R(0) (squares) transitions as a function of ionization delay time. The solid lines are fitting to Eqn.~\eqref{eqn:delay}.  The $|1,1\rangle$ to $|2,2\rangle$ microwave ARP pulse length was 4 ms. The ionization block consisted of 10,000 REMPI pulses at 10 kHz repetition rate. 
  We plot the total KRb ion counts detected in the 10,000 REMPI events.  }
\label{fig:delay_scan}
\end{figure*}

We determine the rate coefficient of the hyperfine-to-rotational energy transfer process by monitoring the depletion rate of the reactants and quantifying the total number of molecules generated through this process.

Let $N_k(t)$ denote the number of molecules in the rotational state $N=k$ at time $t$, where $k=0,1$ or $2$. The collision product population $N_k(t)$ follows
\begin{align}\label{eq:dn_kdt}
\frac{d\,N_k(t)}{dt} &= \frac{\alpha_k}{V_\text{ov}} N_a(t) N_m(t),
\end{align}
where $\alpha_k$ is the formation rate coefficient of KRb molecules in rotational state $k$, $V_\text{ov}$ is the overlap volume of the atoms and molecules. $N_a(t), N_m(t)$ are the reactant atom and molecule number in the cloud at time $t$ respectively, with $t=0$ denoting the starting point of the collision. Note that we are able to distinguish $N=0$ collision products from KRb molecules before collisions, which are also at $N=0$ state, because our ionization pulses with the dark mask selectively detects molecules with sufficient kinetic energy to leave the masked region, i.e., only the collision products from the inelastic collision process.

On the other hand, the reactant loss follows
\begin{align}\label{eq:dNdt}
\frac{d\,N_a(t)}{d\,t} &= -\frac{k}{V_\text{ov}} N_a(t) N_m(t), \\
\frac{d\,N_m(t)}{d\,t} &= -\frac{k}{V_\text{ov}} N_a(t) N_m(t),
\end{align}
where $k$ is the loss rate coefficient.  In our experiment, we control the initial atom and molecule number to be the same $N_{a}(0) = N_{m}(0)$. In this case, the atom and molecule number can be approximately described by the two-body decay function
\begin{align}\label{eq:NaNm}
N_a(t) = N_m(t) = \frac{1}{N_{a,m}(0)^{-1}+\frac{k}{V_\text{ov}}t}. 
\end{align}
Combining Eq.~(\ref{eq:dn_kdt}) and (\ref{eq:NaNm}), we have
\begin{align}
\frac{d\,N_k(t)}{dt} &= \frac{\alpha_k}{V_\text{ov}} \left(\frac{1}{N_{a,m}(0)^{-1}+\frac{k}{V_\text{ov}}t}\right)^2.
\end{align}
Solving this differential equation we get
\begin{align}
N_k(t) = \frac{\alpha_k}{V_\text{ov}} N_{a,m}(0) \frac{t}{N_{a,m}(0)^{-1}+\frac{k}{V_\text{ov}}t},
\end{align}

Experimentally we can characterize $N_k(t) $ by varying the delay time $\tau$ between the $|2,2\rangle$ preparation pulse and the ionization detection block [see the inset of Fig.~\ref{fig:delay_scan}]. The total number of detected molecules at detection delay $\tau$ is effectively $c_k(\tau) = N_k(t=\infty)-N_k(\tau) + N_k(\tau+\delta_t)-N_k(\tau)$,
as the 1 s long detection sequence is enough to capture everything that remains in the detection volume and all the molecules generated after $\tau$. $N_k(\tau+\delta\tau)-N_k(\tau)$ represents all  the molecules remaining in the detection volume when we start the ionization sequence. Consider an infinite long cylindrical beam with radius $R = $ 1 mm, $\delta\tau = \frac{\pi}{2}\frac{R}{v}$. For the slowest molecule $N=2$, we get $\delta \tau = 2.1$ ms, which is negligible and $N_k(\tau+\delta\tau)-N_k(\tau)\approx 0$ considering the time scale of this experiment.  

Thus, at delay time $\tau$, the detected product population is
\begin{align}
c_k(\tau) &= N_k(t=\infty) - N_k(\tau) \nonumber \\
&=  N_{a,m}(0) \frac{\alpha_k}{k} \left( 1- \frac{\tau}{(N_{a,m}(0) k/V_{\text{ov}})^{-1}+\tau}\right).
\label{eqn:delay}
\end{align}
The measured KRb ion counts as a function of detection delay are shown in Fig.~\ref{fig:delay_scan}. The solid lines in Fig.~\ref{fig:delay_scan} are a fitting to Eq.~(\ref{eqn:delay}). The fitted half-lives $t_{1/2} = \frac{1}{N_{a,m}(0)k/V_{\text{ov}}}$ are 54.0(13.9) ms, 45.0(7.9) ms, and 56.1(6.9) ms for $N = 0,1,2$ respectively. The error bar here represents 68\% confidence interval. The three half-lives agree reasonably well and we take the mean value 51.7(9.6) to calculate the atom-molecule loss rate coefficient. 

To extract the loss rate coefficient we need the overlap volume $V_\text{ov}$ and initial atom and molecule number $N_{a,m}(0)$. The overlap volume is given by
\begin{align}
1/V_\text{ov} &=\int d^3\mathbf{r} \,n_a(\mathbf{r})n_m(\mathbf{r}) \nonumber \\
&= \int d^3\mathbf{r} \Pi_{i=x,y,z} \frac{1}{\sqrt{2\pi}\sigma_{\text{Rb},i}}\exp(-\frac{i^2}{2\sigma_{\text{Rb},i}^2})\cdot \frac{1}{\sqrt{2\pi}\sigma_{\text{KRb},i}}\exp(-\frac{i^2}{2\sigma_{\text{KRb},i}^2}),
\label{eqn:Vov}
\end{align}
with $\sigma_{\text{Rb(KRb)},i} = \sqrt{\frac{k_{\text{B}}T_{\text{Rb(KRb)}}}{m_{\text{Rb(KRb)}}\omega_{\text{Rb(KRb)},i}^2}}$, where $k_{\text{B}}$ is the Boltzmann constant, $T_{\text{Rb}}$ = 0.8 $\mu$K is the temperature for atom, and $T_{\text{KRb}}$ = 0.4 $\mu$K is the temperature for molecule respectively. $\omega_{\text{Rb}} = $ $2\pi\times$(212(5),212(5),63(3)) Hz and $\omega_{\text{KRb}} = $ $2\pi\times$(250(3),250(3),72(3)) Hz are the trapping frequencies for atoms and molecules respectively in the crossed optical dipole trap at 1064~nm.
Plugging in these numbers to Eq.~(\ref{eqn:Vov}), the overlap volume is 2.76$\times 10^{-8}$ cm$^3$.
The atom/molecule number at the start of the collision is about $N_{a,m}(0) \approx 8\times10^3$ (there is some amount of atom-molecule two-body loss during the field ramp). From the above numbers, we obtain an atom-molecule loss rate coefficient $k=\frac{V_{\text{ov}}}{N_{a,m}(0) t_{1/2}}= 6.7(1.4)\times10^{-11}$ s$^{-1}$cm$^3$. The extracted loss rate coefficient is similar to that of the system Rb ($|1,1\rangle$) + KRb in the ground rotational state~\cite{ospelkaus2010quantum,nichols2022detection}, which is close to the predicted universal loss rate for s-wave collisions between KRb and Rb ~\cite{li2019universal,frye2021complexes}. 

From the energy conservation point of view, all the inelastic collision product KRb molecules should be in N=0,1,2, and we can conclude that $\sum_{k=0,1,2} \alpha_k= k$, which implies that the hyperfine-to-rotational energy transfer process happens at near universal rate.

\subsection{Intermediate complex lifetime estimation}
In our previous work~\cite{nichols2022detection}, we characterized the KRb$_2$ intermediate complex lifetime in Rb ($|1,1\rangle$) + KRb (N=0) collision by direct photoionization.
In the present system, the complex lifetime is too short and we don't observe any KRb$_2$ ions using a similar measurement scheme as in~\cite{nichols2022detection}.

\begin{figure*}
\centering
	\includegraphics[width=0.5\textwidth]{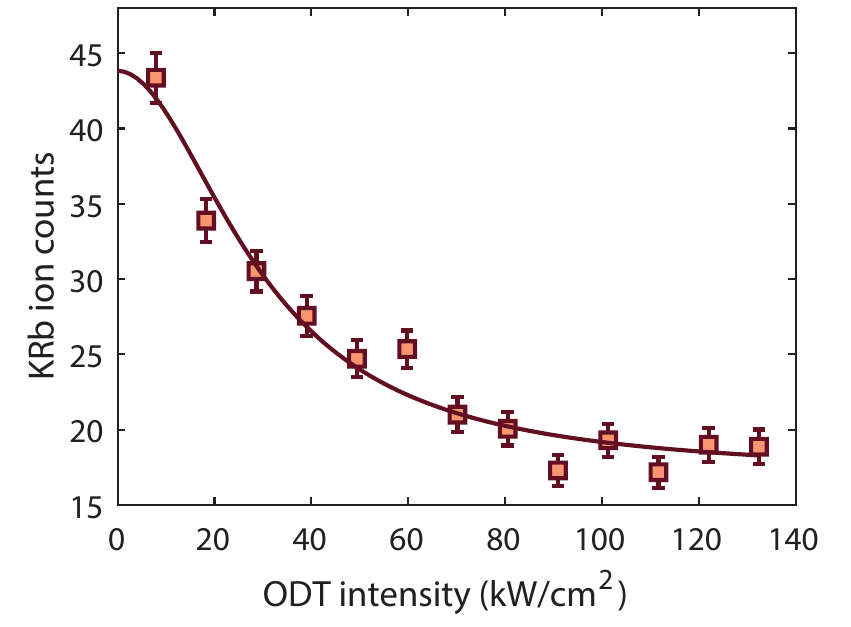}
  \caption{ Collision product KRb ion counts as a function of the static ODT intensity. REMPI beam was resonant with Q(2) transition.  Error bars represent shot noise. The solid line is fitted to Eq.~(\ref{eqn:lifetime}).}
\label{fig:Complex}
\end{figure*}

However, if we increase the ODT intensity, the photoexcitation of the intermediate complex KRb$_2$ will increase and less rotationally excited KRb molecules will form. Thus we can extract the intermediate complex lifetime by monitoring the product population as a function of ODT intensity. The measurement results are shown in Fig.~\ref{fig:Complex}. We assume that the KRb$_2$ complex formed in Rb $|2,2\rangle$ + KRb (N=0) collision has the same photoexcitation rate as in Rb $|1,1\rangle$ + KRb (N=0) collisions, which has been characterized in~\cite{nichols2022detection}. The first-order photoexcitation rate is $\beta_1$ = 0.50(3) $\mu$s$^{-1}$(kW/cm$^{2}$)$^{-1}$ and the second-order photoexcitation rate is $\beta_2$ = 2.1(8) $\mu$s$^{-1}$(kW/cm$^{2}$)$^{-2}$. 
We measure the product population at various ODT intensities and fit to the Equation (2) in Ref.~\cite{nichols2022detection}
\begin{align}\label{eqn:lifetime}
N & =  \frac{A}{1+B_1I+(\beta_{2}/\beta_{1})B_1I^2} + C.
\end{align}
Here, we fix the ratio between the second- and first-order photoexcitation rate to be $\beta_{2}/\beta_{1}$.  From the fitting, we can extract the complex lifetime $\tau_c = B_1/\beta_1 = 0.54(10)$ ns. Here, the error bar represents a 68\% confidence interval.

\end{document}